\documentclass[12pt]{iopart}

\usepackage{graphicx}

\newcommand {\be} {\begin{equation}}
\newcommand {\ee} {\end{equation}}
\newcommand {\bq} {\begin{eqnarray}}
\newcommand {\eq} {\end{eqnarray}}

\newcommand {\Fig} {Fig. \ref}
\newcommand {\Figs} {Figs. \ref}
\newcommand {\Eq} {Eq. \ref}

\def\ie{{\it i.e.}}

\def\etal{{\it et al.}}

\begin{document}
\bibliographystyle{plain}

\title{Modeling the mechanics of amorphous solids at different length and time scales}

\author{D. Rodney$^1$, A. Tanguy$^2$, D. Vandembroucq$^3$}

\address{$^1$Laboratoire Science et Ing\'enierie des Mat\'eriaux et Proc\'ed\'es, Grenoble INP, UJF, CNRS, Domaine Universitaire BP 46, F38402 Saint Martin d'H\`{e}res, France\\
$^2$Laboratoire de Physique de la Mati\`{e}re Condens\'ee et Nanostructures, Universit\'e Lyon 1, Domaine Scientifique de la Doua, F69622 Villeurbanne, France\\
$^3$Laboratoire Physique et M\'ecanique des Milieux H\'et\'erog\`{e}nes, CNRS/ESPCI/UPMC/Universit\'e Paris 7 Diderot, 10 rue Vauquelin, F75231 Paris, France}
\ead{david.rodney@grenoble-inp.fr}
\begin{abstract}
We review the recent literature on the simulation of the structure and deformation of amorphous glasses, including oxide and metallic glasses. We consider simulations at different length and time scales. At the nanometer scale, we review studies based on atomistic simulations, with a particular emphasis on the role of the potential energy landscape and of the temperature. At the micrometer scale, we present the different mesoscopic models of amorphous plasticity and show the relation between shear banding and the type of disorder and correlations (e.g. elastic) included in the models. At the macroscopic range, we review the different constitutive laws used in finite element simulations. We end the review by a critical discussion on the opportunities and challenges offered by multiscale modeling and transfer of information between scales to study amorphous plasticity.
\end{abstract}

\submitto{\MSMSE}
\maketitle

\section{Introduction}
\label{introduction}

Amorphous solids, or glasses, are not meant to deform. We all know that if we try to deform a piece of window panel, a compact disk or a sugar glass candy, the result is always the same: the glass is initially hard, deforms very little but breaks as soon as it starts to deform plastically. In mechanical terms, amorphous solids have a high strength and a low ductility. These traits also apply to the recently developed metallic glasses \cite{johnson-mrs1999,inoue-am2000,heilmaier-jmpt2001,zhang-am2003,johnson-prl2004,ashby-sm2006} while their crystalline counterparts are known for exactly the opposite: low elastic limit and large deformation before failure. The low ductility of glasses is due to the localization of the plastic deformation in thin shear bands where local heating leads to decohesion and catastrophic failure \cite{lewandowski-nmat2006}. Understanding and thus controlling shear band formation is the main challenge that has so far limited the use of glasses as structural materials \cite{schuh-am2007}.

\begin{figure}[tbp]
\begin{center}
\includegraphics[width=15cm]{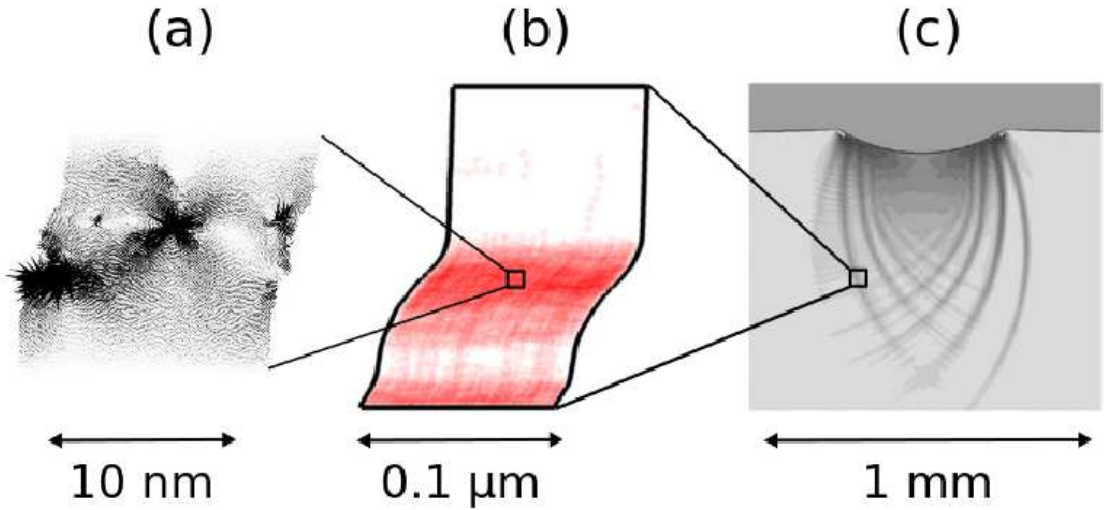}
\caption{\label{fig:multiscale}Illustration of a multiscale linking approach. The actual simulations were performed independently. In (a), atomic-scale simulations show localized atomic rearrangements, called shear transformations. Reproduced with permission from Ref. \cite{tanguy-epje2006}. The energy of shear transformations is used in (b) to perform kinetic Monte Carlo simulations (the actual scale of the simulation cell is smaller than written here, 50 nm). Reproduced with permission from Ref. \cite{homer-am2009}. In (c), a micromechanical constitutive law is used to simulate indentation with the finite element method. Reproduced with permission from Ref. \cite{su-am2006}.}
\end{center}
\end{figure}

Shear band formation is a typical multiscale phenomenon that occurs over three main length and time scales illustrated in \Fig{fig:multiscale}. The elementary plastic events are atomic rearrangements called shear transformations (STs) \cite{argon-am1979,argon-mse1979} that occur at the atomic scale and are localized both in space, involving only a few tens of atoms, and time, spanning a few picoseconds. At the micron scale, STs organize to form shear bands that appear within a few milliseconds and have a width on the order of a tenth of a micron \cite{schuh-am2007}. Finally, at the macroscopic scale, depending on the loading condition, either a single shear band forms as in tension test, or several shear bands form and interact, in case of confined plasticity as in indentation tests \cite{su-am2006}.

The study of plasticity in amorphous solids has greatly benefited from computer simulations; probably more than crystalline solids, because of the lack of experimental device able to identify plastic events in glasses, in the way electron microscopy images dislocations in crystals and also because the intrinsic brittleness of glasses forbids the use of standard mechanical tests. STs have been extensively studied using atomic-scale simulations (for a recent review, see \cite{falk-epj2010}). At the micron scale, rate-and-state models (for a recent review, see \cite{barrat-review2010}) based on the elastic interaction between STs have been used to study ST organization and its relation to shear banding. At the macroscopic scale, the finite element method (FEM) has been used to simulate indentation \cite{su-am2006,KBVD-ActaMat08}. We wish to note that the above studies were carried out independently, while in a multiscale approach, we would like to tie them together in order to exchange information between scales. Multiscale modeling has been successfully applied to crystalline plasticity, with two main types of approaches. The first is a coupling approach where several simulations at different scales are performed simultaneously in a single computation framework, such as the quasicontinuum method that couples atomistic and finite element simulations \cite{shenoy-jmps1999,phillips-msmse1999,miller-msmse2009}. The second possibility is a linking approach where simulations at different scales are done sequentially and the results obtained at one scale are modeled in the form of local rules or phenomenological parameters used as input in the simulation at the scale above. As an example, the linking approach has been applied to study the formation of shear bands in irradiated metals \cite{nogaret-jnm2008}.

Similar multiscale approaches would greatly benefit the study of amorphous solid mechanics. They have not yet been attempted, probably because our understanding of plasticity in amorphous solids is far less advanced than in crystals. In glasses, several very fundamental questions, starting at the atomic scale, should be answered before multiscale modeling can be performed:

\begin{enumerate}

\item At the atomic scale, what triggers a ST? Since glasses are disordered and contain large fluctuations in local structure, density and composition, flow defects equivalent to dislocations in crystals cannot be identified; only flow events or plastic rearrangements are observed. The question is then if there is a way to identify regions prone to plastic rearrangement?
\item Still at the atomic-scale, what is the role of temperature on plastic events? Plastic activity has been compared to an effective temperature, but what is the interplay between effective and real temperatures?
\item STs produce elastic stresses and strains through which they interact with one another. But how does disorder interfere with elasticity?

\end{enumerate}

Answering those questions would lead to the development of an equivalent of the dislocation theory of crystalline plasticity \cite{hirth-1982}. Multiscale modeling would then serve to answer equally fundamental questions at the micron and macroscopic scales:

\begin{enumerate}

\item At the micron scale, how do shear bands arise from the competition between ST interactions, local softening or hardening, and disorder in the glass structure and dynamics?
\item What happens during the millisecond or so before a shear band nucleates?
\item At the macroscopic scale, how are shear bands sensitive to the loading condition and how do they interact with one another?

\end{enumerate}

The underlying issue behind the above questions is whether or not the state of a glass can be described by a finite set of internal variables. And whether or not the evolution of these internal variables can be tied to the loading history of the glass in order to describe the evolution of the glass state under deformation and its influence on the mechanical response. In a linking approach, the internal variables can then be used at the micron-scale in rate-and-state models to determine constitutive laws used in FEM at the macroscopic scale. This multiscale approach would be the equivalent of the dislocation density-based crystal plasticity theories that have proved very efficient in modeling plastic deformation in metals and alloys \cite{madec-science2003,devincre-science2008}.

Several reviews have been published recently on the mechanical behavior of glasses (see Schuh \etal~\cite{schuh-am2007} for experiments, Falk and Maloney \cite{falk-epj2010} and Barrat and Lema\^{i}tre \cite{barrat-review2010} for simulations and modeling). In the present paper, we focus on the multiscale modeling aspect of the study of glasses and associated challenges, with a particular attention to plasticity. To review the numerical simulation methods at the different length- and timescales relevant for multiscale modeling, we start at the atomistic scale (Section \ref{at_dynamics} and \ref{at_def}), then address the mesoscopic scale (Section \ref{meso}) and finish by the macroscopic scale (Section \ref{macro}). At each scale, we emphasize on the main specificities, successes and limitations of the different numerical techniques used to model glasses. We finally discuss opportunities and challenges offered by a multiscale linking approach to the study of plasticity in glasses (Section \ref{perspective}).

\section{Atomic-scale dynamics in the absence of deformation}
\label{at_dynamics}

The dynamics of disordered solids and liquids at the atomic-scale is highly complex and is the subject of a very vast and ever growing literature that cannot be reviewed exhaustively. We will focus here on numerical studies by means of molecular dynamics (MD) and statics simulations, and we adopt a specific angle of approach with the help of the \emph{potential energy landscape} (PEL) picture. We start by presenting the elementary ingredients of atomistic simulations. We then review studies of the dynamics of liquids when cooled down to the glassy state.

\subsection{Practicalities}
\label{practicalities}

\textbf{Choice of interatomic potential.}
Accurately representing atomic bonding requires in principle electronic structure based methods, such as the density functional theory or self-consistent field calculations \cite{Tsuneyuki1988}. Electronic structure calculations have been applied to simulate liquids and glasses, but the tradeoff for increased accuracy is a decrease in system size (a few hundred atoms) and in simulated time (a few picoseconds). Monoatomic systems (for example \cite{Bazant1996,jakse-prl2003,ganesh-prb2006}), silica glasses (for example \cite{Tsuneyuki1988,Ispas2002}) and metal alloys (for example \cite{sheng-nature2006,ganesh-prb2008,kazimirov-prb2008,gu-apl2008}) have been simulated but semi-empirical potentials have also been used to access larger systems during longer times. For metallic alloys for example, the embedded atom method (EAM) formalism has been employed, in particular to model CuZr \cite{duan-prb2005,mendelev-jap2007a,mendelev-pm2009}, CuZrAl \cite{cheng-am2008} and CuMg \cite{bailey-prb2004} alloys. For liquid and glassy silica and silicon, several empirical potentials have been adjusted in order to fit, with a minimum number of parameters, ab-initio cohesive energy data \cite{Bazant1996,Tsuneyuki1988} and experimental characteristics such as the melting temperature, short-range order, diffusive motion, some characteristic vibrational frequencies and elastic moduli \cite{Stillinger1985,Tersoff1988,Beest1990,EDIP1997,Ispas2002,Huang2003,Ispas2005,Carre2008}. In the case of silicon, the difficulty is to find an empirical potential able to describe the profound structural breakdown that occurs upon melting, with an experimental average coordination of 4.0 in the solid phase (tetrahedral order due to sp3 covalent bonding in the crystal as well as in the amorphous phase) and  of 6.4 in the liquid state \cite{Tersoff1988}. Since the interatomic potentials depend on the full atomic configuration, they can be decomposed into two-, three-body potentials and above. The three-body potential serves to stabilize the ideal local bond angle, as in the famous Stillinger-Weber empirical potential \cite{Stillinger1985}, expressed as:
\be
V(1,...,N)=\sum_{i<j}v_2(i,j)+\sum_{i<j<k}v_3(i,j,k)
\ee
with
\be
v_2(i,j)=\epsilon.A.(Br_{ij}^{-p}-r_{ij}^{-q})\exp[(r_{ij}-a)^{-1}], \mbox{ r}_{ij}<a \mbox{ and =0 elsewhere}
\ee
and
\be
v_3(i,j,k)=h(r_{ij},r_{ik},\theta_{jik})+h(r_{ji},r_{jk},\theta_{ijk})+h(r_{ki},r_{kj},\theta_{ikj})
\ee
with
\be
h(r_{ij},r_{ik},\theta_{jik})=\epsilon\lambda \exp[\gamma (r_{ij}-a)^{-1}+\gamma (r_{ik}-a)^{-1}].(\cos\theta_{jik}+\frac{1}{3})^2,
\ee
where $r_{ij}$ is the bond length between atoms $i$ and $j$ and $\theta_{ijk}$ is the angle between the bonds $ji$ and $jk$. The experimental pair correlation however is better reproduced with more complex environment-dependent potentials, such as Tersoff potential \cite{Tersoff1988} or EDIP \cite{EDIP1997}. The latter is also able to represent the covalent-to-metallic transition through a parameter depending on the local coordination and describing the angular stiffness of the bond. These different empirical potentials have been tested on the nonlinear mechanical response of samples submitted to a large shear \cite{Pizzagalli2003} or crack propagation \cite{Hauch1999}. In the case of silica, the ionic character of the bond is usually described through a two-body potential with partial charges \cite{Beest1990, Carre2008, Micoulaut2006}, while the covalent character implies a three-body term \cite{Huang2003}. Surprisingly enough, the local tetrahedral structure of Si0$_2$ is well described by a simple two-body potential with different charges on Si and O, even with a small cut-off of the interaction range \cite{Carre2008}.

In the field of liquids and glasses, most studies have aimed at understanding generic properties and in consequence have used the simplest possible phenomenological potentials. For example, most numerical works employed a binary Lennard-Jones (LJ) potential, an additive pairwise potential that physically represents noble gasses (Ar, Kr, Xe) with van der Waals interactions.  Its expression is:
\begin{equation}
v_2(i,j) = 4\epsilon_{ij}\Big[\Big(\frac{\sigma_{ij}}{r_{ij}}\Big)^{12}-\Big(\frac{\sigma_{ij}}{r_{ij}}\Big)^6\Big]+ B_{ij}r+C_{ij}.
\end{equation}
The linear function in the r.h.s. is added to the original potential to ensure that the potential and interaction force go smoothly to zero at the cut-off radius, $R_C$. Sometimes, quadratic functions are used to additionally ensure smoothness of the second derivative. Table \ref{table:pot} summarizes the parameters of the 3 most common LJ potentials used in the literature.

\begin{table*}
\begin{tabular}{ccccccccccccc}
  \hline
  & $\epsilon_{AA}$ & $\epsilon_{BB}$ & $\epsilon_{AB}$ & $\sigma_{AA}$ & $\sigma_{BB}$ & $\sigma_{AB}$ & $R_C$ & $C_A$ & $C_B$ & $T_C$\\
  \hline
  Kob-Andersen \cite{kob-pre1995}& 1 & 0.5 & 1.5 & 1 & 0.88 & 0.8 & 2.5 & 0.8 & 0.2 & 0.435 \\
  Wahnstr\"{o}m \cite{wahnstrom-pra1991}& 1 & 1 & 1 & 1 & $\frac{5}{6}$ & $\frac{11}{12}$ & 2.5 & 0.5 & 0.5 & 0.57 \\
  Lan\c{c}on \etal~\cite{lancon-epl1986}& 0.5 & 0.5 & 1 & $2\sin(\frac{\pi}{10})$ & $2\sin(\frac{\pi}{5})$ & 1 & 5 & 0.45 & 0.55 & 0.485  \\
  \hline
\end{tabular}
\caption{\label{table:pot}Parameters of three common binary Lennard-Jones potentials. The cut-off radius is expressed in units of the corresponding $\sigma_{ij}$. $C_A$ and $C_B$ are the concentrations of the two species. $T_C$ is the mode-coupling temperature \cite{gotze-rpp1992} expressed in units of $\epsilon_{AA}/k_B$, where $k_B$ is Boltzmann constant.}
\end{table*}

By far, the most widely used potential was developed by Kob and Andersen \cite{kob-pre1995} to reproduce the properties of Ni$_{80}$P$_{20}$. One should note that this potential is highly non-additive, that is $\sigma_{AB}<(\sigma_{AA}+\sigma_{BB})/2$ and $\epsilon_{AB}>(\epsilon_{AA}+\epsilon_{BB})/2$. Similar non-additivity is present in Lan\c{c}on \etal~potential \cite{lancon-epl1986} (with the size ratio between $\sigma_{AB}$ and $(\sigma_{AA}+\sigma_{BB})/2$ reversed). The latter potential has been used in 2 dimensions because of its quasicrystalline ground state. By way of contrast, the second most common potential, developed by Wahnstr\"{o}m \cite{wahnstrom-pra1991}, is strictly additive. Thus, good glass-formability does not necessarily require non-additivity of the potential. Generalized Lennard-Jones potentials have also been employed, for instance to simulate CuZr metallic glasses \cite{kobayashi-am1980,kobayashi-jpsj1980} with exponents 4-8 instead of 6-12, which proved better suited to represent metallic bonding.

\textbf{Choice of boundary conditions.} The most popular boundary conditions are periodic. In most cases, deformations are applied using strain-controlled boundary conditions implemented by changing the simulation cell dimensions and shape: uniaxial traction and compression (with a pressure control in the transverse directions to maintain zero pressure transversally), pure and simple shear. In latter case, Lees-Edwards shifted periodic boundary conditions are applied, whereby periodicity across two opposite faces of the simulation cell are shifted with respect to each other in the shear direction (which is equivalent to applying periodic boundary conditions in a non-orthonormal cell \cite{allen-book}). Fixed and free boundary conditions have also been used.

The choice of boundary conditions is not inconsequential since fixed boundaries favor localization of the deformation in simple shear \cite{varnik-prl2003}. Similarly, uniaxial traction with free surfaces in the transverse directions also favor shear localization compared to simple shear with periodic boundary conditions \cite{cheng-am2009}.

\textbf{Molecular dynamics versus quasistatic simulations.} The first step of an atomistic simulation of a glass is to produce an initial configuration by quenching a liquid using MD. The procedure involves first equilibrating the liquid at an elevated temperature (typically several 1000 K) and then progressively decreasing the temperature below the glass transition at either fixed volume or fixed pressure. The main limitation of MD simulations is that the timescale is very limited, typically a few tens of nanoseconds. As a consequence, MD quench rates are on the order of 1000 K$/10^{-8}$ s, i.e. $10^{10}$ to $10^{12}$ K$/$s, which is enormously high compared to experimental quench rates that are rather on the order of $10^6$ K$/$s for the first synthesized metallic glasses \cite{klement-nature1960} to $0.1$ K$/$s for the most recent bulk metallic glasses \cite{johnson-mrs1999}. As a result, simulated glasses are inevitably far less relaxed than experimental glasses, with consequences on their propensity to form shear bands that will become apparent later in the text. Also, when simulating the deformation of glasses using MD, one wishes to reach a strain of order 1 in the timescale of the simulation, resulting in typical strain rates of $\dot{\gamma} = 1/10^{-8} = 10^8$ s$^{-1}$ compared to typically $10^{-3}$ s$^{-1}$ experimentally. In order to avoid this discrepancy, one option is to apply quasistatic (QS) deformations \cite{malandro-prl1998,maloney-prl2004a,maloney-pre2006}. The system is then sheared in small increments followed by energy minimizations at fixed applied strain. This protocol corresponds to the limit $T \rightarrow 0$, since the energy minimizations remove all thermal activation, and $\dot{\gamma} \rightarrow 0$, since the system is allowed to fully relax to a new energy minimum before a new strain increment is applied \cite{malandro-prl1998}.

The physical situation associated with the QS procedure involves two timescales characteristic of the dynamics of glassy materials \cite{kob-book}. The first timescale, $\tau_{diss}$, is the time it takes for a localized energy input to spread over the whole system and be dissipated as heat. The corresponding mechanisms can be viscous (in a soft material) or associated with phonon propagation or electron transfers (in a metallic glass). QS deformation corresponds to a typical time between consecutive strain increments much larger than $\tau_{diss}$, that is $\dot\gamma \ll \delta\gamma/\tau_{diss}$, where $\delta\gamma$ is the elementary strain increment. A second, much longer timescale is the structural relaxation time of the system, $\tau_{\alpha}$, associated with spontaneous aging processes that take place at finite temperature in the absence of external drive. By quenching after every displacement step, any such processes are suppressed and the time elapsed between consecutive increments is thus far smaller than $\tau_{\alpha}$, i.e. $\dot\gamma \gg \delta\gamma/\tau_{\alpha}$. Here, $\delta\gamma$, the elementary strain step, is chosen small enough to ensure that the system remains in its initial basin when starting from any equilibrium configuration ($\delta\gamma$ depends on the system size, probably logarithmically \cite{leonforte-phd} and is on the order of $10^{-5}$ for $L=100$ in LJ units in 2 and 3 dimensions \cite{Tanguy-PRB05}). This picture is of course oversimplified. Relaxations in glasses are in general stretched \cite{kob-book}, meaning that relaxation processes take place over a broad spectrum of timescales and the QS approach ignores the fast wing of the relaxation spectrum. Another drawback of the QS approach is that it ignores thermal effects that are unavoidable in experiments. The coupling between temperature and strain rate is highly non-trivial and progress in this area is discussed in Section \ref{temperature}. The first steps towards using accelerated dynamics have been taken \cite{kushima-jcp2009,rodney-prb2009} in order to access slow dynamics at the atomic scale, but such simulations are difficult because of the complexity of the configurational paths available to disordered systems.

\subsection{Cooling a liquid}
\label{cooling}

Understanding the glass transition and the associated dramatic slowing down and increasing heterogeneity of the liquid dynamics is a topic of intense research (for reviews, see for example \cite{ediger-jcp1996,debenedetti-nature2001,tarjus-jpcm2005,heuer-jpcm2008,tanaka-nmat2010,berthier-arXiv2010}). This field has greatly benefited from the recognition that the dynamics of a cooling liquid is intimately related to the topography of its underlying potential energy landscape \cite{widmer-prl2006}.

\textbf{What is the potential energy landscape?} An atomic configuration is represented in configuration space by its position vector, $R^{3N}$, a 3N-dimensional vector for a system of N particles in 3 dimensions. The potential energy of the system, $V(R^{3N})$, is then a 3N-dimensional surface, termed the \emph{potential energy landscape} (PEL), in the (3N+1)-dimensional space composed of configuration space and the energy axis \cite{goldstein-jcp1969,wales-book}. It is important to note that the PEL depends only on the interatomic potential and the boundary conditions. Thus, for a given potential and given boundary conditions, all configurations of a system, whether they are crystalline, amorphous or liquid, share the same PEL; only the region of configuration space visited by the system depends on the state of the system.

\begin{figure}
\begin{center}
\includegraphics[width=10cm]{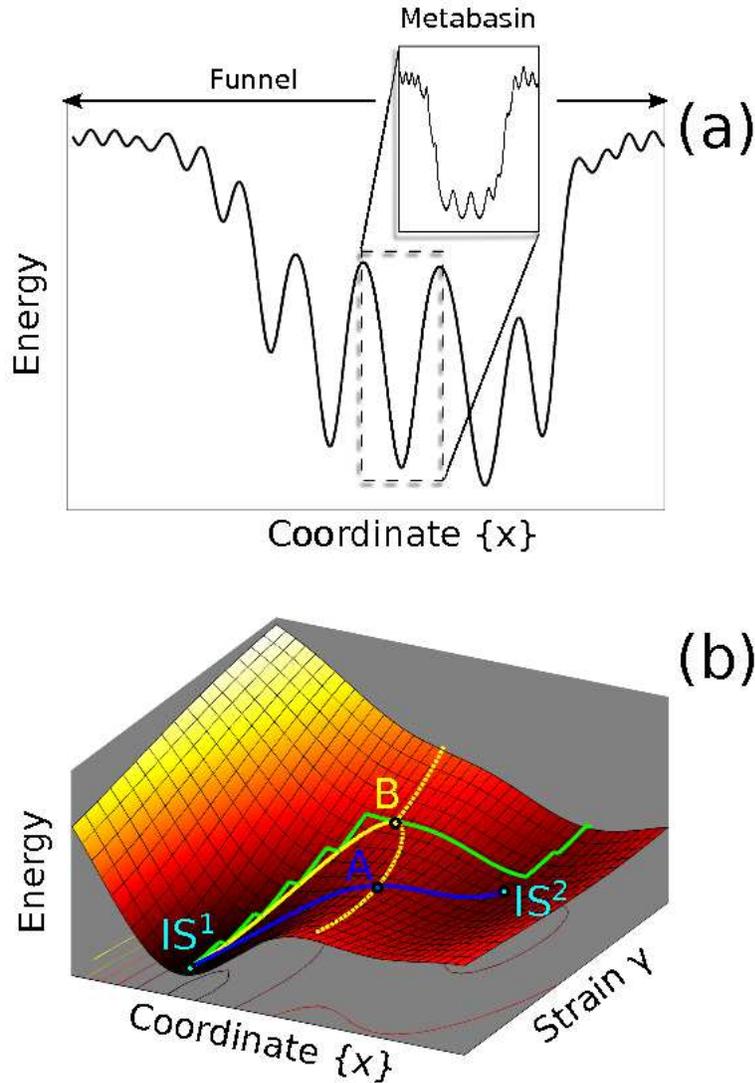}
\caption{\label{fig:PEL}Schematic illustration of potential energy landscapes in (a) two and (b) three dimensions. See text for details.}
\end{center}
\end{figure}

As sketched in \Fig{fig:PEL} in two and three dimensions, the PEL contains extrema, or \emph{stationary points}, which may be local maxima, local minima called \emph{inherent structures} (IS) \cite{stillinger-science1984} ($IS^1$ and $IS^2$ in \Fig{fig:PEL}(b)) and saddle points ($A$ in \Fig{fig:PEL}(b)). Stationary points are classified by their index, which is the number of negative eigenvalues of the Hessian matrix (matrix of second derivatives of the potential energy) computed at the stationary point \cite{wales-book}. The PEL can be partitioned into valleys, or \emph{basins}, that surround each local minima \cite{stillinger-science1984,stillinger-science1995}. More precisely, the basin (of attraction) of an IS is the region of configuration space where all configurations converge to the IS upon steepest-descent energy minimization. In \Fig{fig:PEL}(b), the basins of $IS^1$ and $IS^2$ are delimited by the yellow dashed line. Alternatively, the PEL can be partitioned using the basins of attraction of all stationary points and not only local minima by minimizing the objective function $\frac{1}{2}|\nabla V(R^{3N})|^2$ rather than the potential energy \cite{angelani-prl2000}, although one has to be careful because the objective function has more minima than stationary points on the PEL \cite{doye-jcp2002}.

The number of stationary points increases roughly as an exponential function of the number of atoms in the system \cite{stillinger-science1984,heuer-jpcm2008}. It is therefore impossible in practice to list exhaustively all stationary points in a PEL, except for very small systems (containing typically 32 atoms or less \cite{heuer-prl1997,angelani-pre2000}). On the other hand, statistical information, such as distributions of activation energies, can be computed accurately on tractable samples containing typically a few thousand events \cite{rodney-prl2009}.

\textbf{Hierarchical structure of the PEL.} The relation between liquid dynamics and PEL was probed by means of MD simulations (mostly binary Lennard-Jones liquids modeled using Kob and Andersen potential) with regular energy minimizations in order to determine the succession of inherent structures visited by the liquid during its time trajectory \cite{stillinger-science1984,jonsson-prl1988,sastry-nature1998}. Examples of quenches are shown in \Fig{fig:SHEAR}(a) using the Wahnstr\"{o}m LJ potential. The mode coupling temperature ($T_C$) \cite{gotze-rpp1992} plays the role of reference temperature for the dynamics of liquids at the atomic scale. Three regimes were identified \cite{sastry-nature1998}. At high temperature, above about $1.5 T_C$ for the system considered here, the system is fully liquid and the kinetic energy is high enough that the system hardly feels the underlying PEL.

At lower temperatures, the system becomes supercooled and enters the so-called \emph{landscape influenced regime}. Between $1.5 T_C$ and $T_C$, the liquid gets progressively trapped in basins of decreasing energy, i.e. it remains for longer periods of time in a given IS before hopping to the next IS with the help of thermal activation. A decoupling develops between the rapid motion of the supercooled liquid inside a basin (intrabasin vibration) and the infrequent transitions to a new basin (interbasin hoping) \cite{sciortino-prl1999}. Vibrations are however of large amplitude and the system spends most of its time in the basins of attraction of saddle points, i.e. stationary points with index 1 and above \cite{angelani-prl2000,broderix-prl2000}. A well-defined relation was found between the average stationary point index and the temperature or potential energy \cite{angelani-prl2000,broderix-prl2000} with the average index vanishing precisely at $T_C$. The progressive trapping of the system implies that the PEL has a multifunnel structure \cite{wales-nature1998, wales-book} where, as illustrated in \Fig{fig:PEL}(a), the basins are organized in pockets of inherent structures with increasing energy barriers for decreasing IS energy.

Funnels have a hierarchical structure. On short timescales, the system undergoes a back and forth motion between a limited number of basins forming a cluster, also called a \emph{metabasin} \cite{stillinger-science1995}, while on longer timescales, the system performs an irreversible Brownian diffusion between metabasins \cite{buchner-prl2000,doliwa-pre2003a,doliwa-pre2003b,denny-prl2003}. The above dynamics implies that, as illustrated in \Fig{fig:PEL}(a), metabasins are composed of basins connected by low-energy saddle points, while different metabasins are separated by higher energy barriers. Moreover, some metabasins are visited quickly whereas others are very long-lived \cite{doliwa-prl2003}. The overall structure of the PEL is therefore a multifunnel rough landscape \cite{wales-nature1998} with a hierarchy of energy barriers connecting basins and metabasins. Trapping of the supercooled liquid in this maze of interconnected basins and the multistep hopping process between long-lived metabasins which dictates the slow dynamics was put forward to explain the rapid slowing down and stretched exponential relaxations of liquids across the glass transition \cite{doliwa-prl2003,heuer-jpcm2008}.

Below $T_C$, the system enters the glassy state, or \emph{landscape dominated regime} \cite{schroder-jcp2000}, which is the true thermally-activated regime where the system spends most time vibrating near local minima and undergoes rare and thermally-activated transitions between ISs. As seen in \Fig{fig:SHEAR}(a), the energy of the final inherent structure decreases with decreasing quench rate, meaning that the glass is better relaxed. This is a consequence of the funnel structure of the PEL where decreasing quench rates give more time to the system to explore lower regions of the funnel before being trapped. When the temperature is lowered below $T_C$, the waiting time between transitions exceeds rapidly MD timescales and the PEL can no longer be probed by direct MD simulations. In this regime, saddle-point search methods can be used to identify activated states on a PEL. Examples of distributions of the activation energies of transition states (saddle-points of index 1) around the final ISs obtained after the above quenches are shown in \Fig{fig:SHEAR}(d). These distributions were obtained using the activation-relaxation technique (ART) \cite{malek-pre2000,cances-jcp2009,rodney-prb2009}. Distributions of activation energies have a characteristic shape with a broad energy spectrum and a maximum, i.e. a most likely activation energy. Similar distributions were obtained in LJ glasses \cite{angelani-prl2000,middleton-prb2001,rodney-prb2009} as well as amorphous silicon \cite{valiquette-prb2003}. Their exponential tail agrees with the master-equation approach developed by Dyre \cite{dyre-prl1987} in the case of rapidly quenched glasses. Also, the density of low activation saddles (below typically $5 \epsilon_{AA}$ for this system) decreases for configurations relaxed more slowly. The latter configurations are therefore more stable, both thermodynamically, because they have a lower energy, and kinetically, because they are surrounded by higher activation energies. In experimental glasses that are quenched much more slowly than in simulations, we expect activation energies to be shifted towards higher energies without low activation energies, although a quantitative study of this dependence remains to be done.

Below $T_C$, glasses are not in thermal equilibrium but keep evolving towards deeper regions of the PEL. This relaxation process is slow and referred to as structural relaxation or \emph{physical aging} \cite{utz-prl2000}. Two relaxation timescales have been identified, $\beta$-relaxations on short timescales and $\alpha$-relaxations on longer timescales \cite{johari-jcp1970}. In relation with the PEL picture, Stillinger \cite{stillinger-science1995} hypothesized that $\beta$-relaxations are transitions inside a given metabasin while $\alpha$-relaxations would be transitions between metabasins. MD simulations confirmed the first hypothesis but showed that the second is only partly true. A $\beta$-relaxation is a transition inside a given metabasin, termed \emph{nondiffusive} by Middleton and Wales \cite{middleton-prb2001} because it corresponds to a slight repositioning of atoms inside their shell of nearest neighbors (\emph{cage effect}). By way of contrast, transitions between metabasins involve bond switching and correspond to localized rearrangements in the form of string-like sequences of displacements whose size ($\sim$ 10 atoms) increases as the temperature decreases towards $T_C$ \cite{kob-prl1997,donati-prl1998,donati-pre1999,schroder-jcp2000,stevenson-natphys2006}. One such dynamical heterogeneity is however not an $\alpha$-relaxation in itself because it occurs on a timescale of the order of $1/10$ to $1/5$ of the $\alpha$-relaxation timescale \cite{appignanesi-prl2006}. An $\alpha$-relaxation is therefore made of 5 to 10 metabasin transitions. It remains localized in the microstructure but involves more atoms than string-like events ($\sim$ 40 atoms), is more globular and was termed \emph{democratic} \cite{appignanesi-prl2006}. Dynamical heterogeneities have been shown correlated to both localized soft modes \cite{widmer-naturephys2008} and fluctuations of static structural order \cite{kawasaki-prl2007,tanaka-nmat2010}.

\textbf{Atomic-scale glass structure.} Although glasses are disordered at long range, they exhibit short and medium range orders. There exists no general theory to predict the packing in a given glass but short and medium range orders have been studied in a number of metallic glasses using a combination of experiments and simulations (see for example Refs. \cite{kobayashi-jpsj1980,duan-prb2005,sheng-nature2006,chen-apl2006,park-sm2007,han-pre2007}). It was shown that solute atoms tend to remain separated from one another and their shell of first neighbors tends to form simple polyhedra. In several binary glasses, and in particular CuZr metallic glasses, the most abundant polyhedra are Cu-centered icosahedra, therefore atoms with a local fivefold atomic environment \cite{jonsson-prl1988,cheng-am2009,peng-apl2010}. Moreover, it was shown that in this system, the short-range order can be characterized by the density of these icosahedra, which density is strongly correlated with the level of relaxation of the glass. In particular, Cheng \etal~\cite{cheng-am2009} showed that the density of Cu-centered icosahedra increases rapidly in the landscape influenced regime and reaches a low-temperature value that increases with decreasing quench rate, i.e. increasing levels of relaxation. It is known that with the Wahnstr\"{o}m LJ potential used in \Fig{fig:SHEAR}(a), icosahedron-centered atoms characterize the short-range order \cite{coslovich-jcp2007,pedersen-prl2010}. We computed their fraction, noted ICO in \Fig{fig:SHEAR}(a), and found that the fraction of icosahedron-centered atoms increases with decreasing quench rate, which confirms the result mentioned above: the short-range order in glasses increases with decreasing quench rates. Icosahedron-centered atoms are clear markers of the level of relaxation of a glass but they concern unfortunately only a certain class of metallic glasses. For instance, they are not the most frequent local environment in monoatomic glasses and the addition of a three-body term to the interatomic potential changes drastically its atomic structure and ability to crystallize \cite{mokshin-pre2008,talati-epl2009,tanaka-natphys2006,fusco-pre2010}.

\section{Atomic-scale deformation of glasses}
\label{at_def}

\subsection{Quasi-static deformation}
\label{qs_def}

\begin{figure}
\begin{center}
\includegraphics[width=15cm]{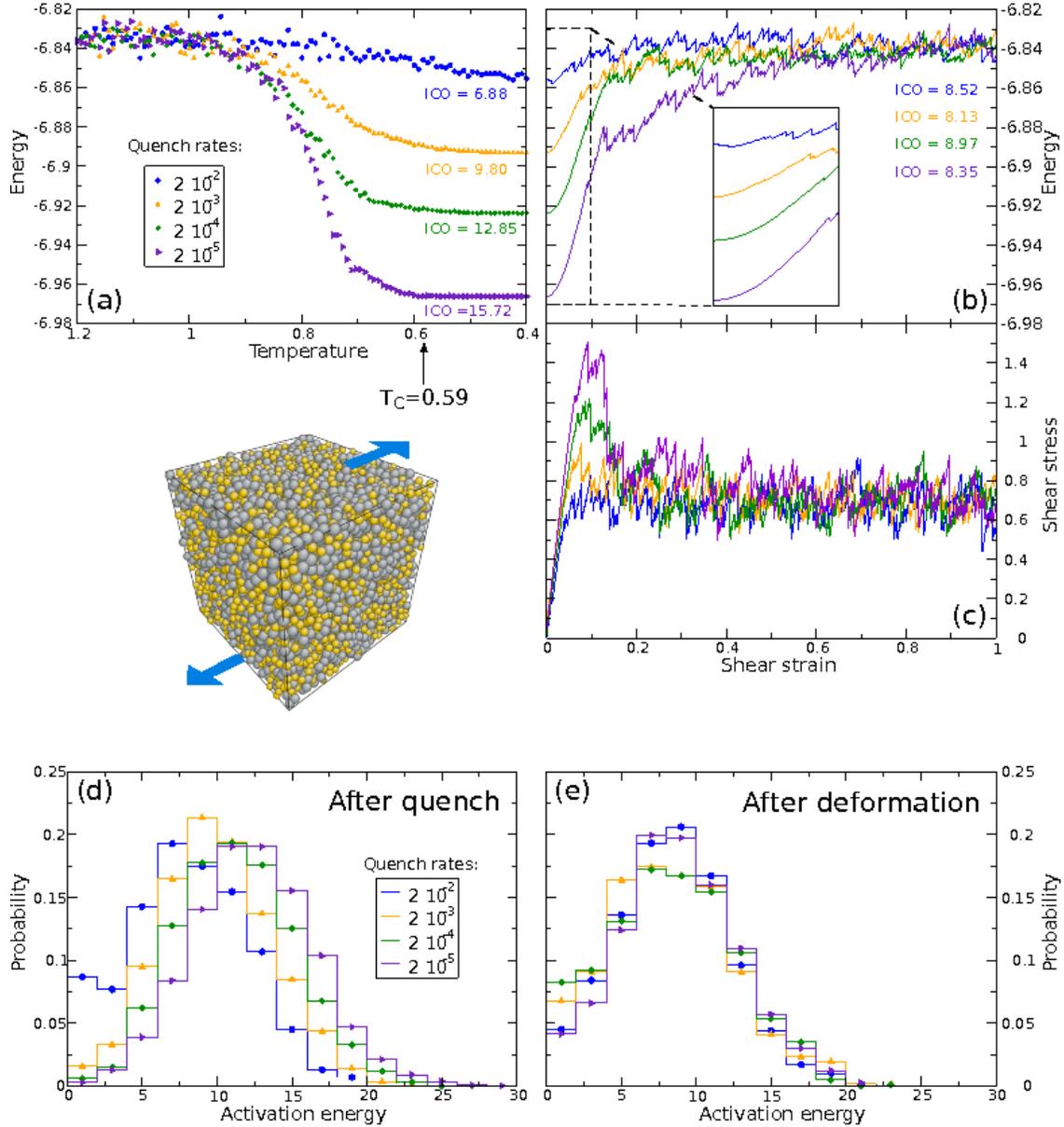}
\caption{\label{fig:SHEAR} Influence of quench rate on the cooling and deformation of glasses: (a) inherent structure energy during quenches at different rates, (b) energy-strain curve and (c) shear stress-shear strain curve during simple shear deformation, (d) distribution of activation energies around as-quenched configurations, (e) same distributions after plastic deformation. The system of 10000 atoms of a 50:50 binary mixture of LJ atoms modeled with Wahnstr\"{o}m potential, in a cubic cell of length 19.74 $\sigma_{AA}$. The mode-coupling temperature $T_C$ for this system is 0.59. Quenches were performed at fixed volume after equilibration at $k_BT=1.2 \epsilon_{AA}$ for $5\times10^3$ $t_0$. Quench rates are expressed in units of $\epsilon_{AA}/k_Bt_0$. The fraction ICO of icosahedron-centered atoms is computed using a Voronoi tessellation. In (b) and (c), shear deformation is applied quasistatically with a strain increment of $10^{-4}$ followed by energy minimization using Lees-Edwards boundary conditions. Activation energy distributions in (d) and (e) are obtained using the Activation-Relaxation Technique with samples containing a minimum of 1000 saddles.}
\end{center}
\end{figure}

\textbf{PEL and QS deformation.} The PEL picture also proved useful to rationalize the behavior of glasses under deformation, as first demonstrated by Malandro and Lacks \cite{malandro-prl1998,malandro-jcp1999,lacks-prl2001}. The picture is clearest during QS deformation where the glass evolution is exclusively strain activated and when rigid boundary conditions are used. In this case, a shear strain is applied by rigidly moving in opposite directions two slabs of atoms on opposite faces of the simulation cell. The PEL is then unchanged because the potential energy function is unchanged (the situation is different when stresses are applied because the PEL is then tilted by the work of the applied stress \cite{bacon-book}). The effect of an applied strain is then to force the system to visit new regions of the PEL. More precisely, adding a strain increment to a simulation cell amounts to moving the system in a given direction of configuration space, which we call the strain vector $\gamma$. The PEL may thus be represented in three dimensions as in \Fig{fig:PEL}(b), with the strain vector $\gamma$ decoupled from the other dimensions of configuration space, noted $\{x\}$. A possible path during QS deformation is shown in green in \Fig{fig:PEL}(b). Initially, the as-quenched glass is in an inherent structure, noted $IS^1$, at the bottom of an energy basin. The strain increments push the system away from this initial configuration. After each strain increment, the energy is minimized at fixed strain, i.e. minimized in the hyperplane perpendicular to the strain vector. The system then relaxes to the nearest local minimum in the hyperplane, which lies on the branch of stable equilibrium at the bottom of the basin, shown as a yellow solid line in \Fig{fig:PEL}(b). During the first few strain increments, the system remains in the initial basin and the deformation is purely elastic and reversible; if the strain is removed, the system relaxes back along the stable branch down to $IS^1$. The extension of this region of strict reversibility increases with decreasing system size and increasing level of relaxation of the quenched glass (see \Fig{fig:SHEAR}(b)).

With increasing strain, the system reaches the edge of the initial basin, noted $B$ in \Fig{fig:PEL}(b), a position of instability on the PEL where the stable branch at the bottom of the basin (yellow solid line) meets the unstable branch on the border of the basin (yellow dashed line). At this point, one eigenvalue of the Hessian matrix in the hyperplane of constant strain vanishes. Such instability is called a saddle-node or a fold bifurcation \cite{guckenheimer-book}. Near the instability, glasses have a universal behavior \cite{maloney-prl2004a,caroli-prl2010,procaccia-pre2010}: the vanishing eigenvalue and energy barrier between stable and unstable positions go to zero as $(\gamma_C-\gamma)^{1/2}$ and $(\gamma_C-\gamma)^{3/2}$ respectively, where $\gamma_C$ is the strain at the bifurcation. The subsequent energy minimization takes the system into a new basin, centered on $IS^2$ in \Fig{fig:PEL}(b). During the relaxation, the system is out-of-equilibrium and its trajectory depends on the energy minimization algorithm used. Across the transition, the position of the system on the PEL, as well as its energy and stress are discontinuous. The transition is irreversible because if the strain is decreased, the system remains in the new basin. Such transitions are the elementary events that produce dissipation and plastic strain during the deformation process. During a QS process the kinetic energy produced by the elementary events is assumed to be very efficiently dissipated in order to relax in the next visited basin.

\textbf{Shear transformations and avalanches.} \Fig{fig:SHEAR}(b) and (c) show examples of energy-strain and stress-strain curves during simple shear deformation of the quenched glasses obtained in \Fig{fig:SHEAR}(a). As expected from the above discussion, the curves are made of continuous elastic segments intersected by plastic events where stress and energy are discontinuous \cite{malandro-prl1998,malandro-jcp1999,maloney-prl2004a,tanguy-epje2006}. The deformation starts with a regime where the stress increases linearly with strain and the energy increases quadratically. This regime is larger than the region of strict reversibility mentioned above and contains small plastic events as evidenced by the discontinuities in the inset of \Fig{fig:SHEAR}(b). The latter correspond to localized rearrangements in the microstructure \cite{malandro-jcp1999,maloney-prl2004a,tanguy-epje2006,mayr-prl2006,delogu-prl2008}. They are the \emph{shear transformations} (STs) proposed by Argon \cite{argon-am1979,argon-mse1979} as elementary plastic events in amorphous solids. An example in two dimensions is shown in \Fig{fig:multiscale}(a). The rearrangement is composed of two regions: a plastic core region where atomic bonds are cut and reformed and a surrounding elastic region, which responds elastically to the plastic rearrangement. In \Fig{fig:multiscale}(a), mainly the elastic field is visible with a quadrupolar symmetry \cite{maloney-prl2004a,tanguy-epje2006}, characteristic of an Eshelby field (see Section \ref{ingredients} and Eq. \ref{eq:eshelby}). This field was shown to scale with the eigenmode whose eigenvalue vanishes at the instability \cite{maloney-prl2004b,lemaitre-pre2007,tanguy-epl2010}. Evaluating the size of a ST requires separating the plastic region, which is the true ST zone, from the elastic surrounding. Such separation is not straightforward, but usual estimates are 20 atoms in two dimensions \cite{tanguy-epje2006} and 100 in three dimensions \cite{zink-prb2006}. The corresponding local strain is about $0.1$, which satisfies the Lindemann instability criterion with a relative displacement of the particles of a tenth of the interatomic distance \cite{tanguy-epje2006,tsamados-pre2009}.

As strain increases, the system reaches its true elastic limit where plastic flow starts. Irreversible events are then dominated by large rearrangements that span the entire simulation cell \cite{maloney-prl2004a,tanguy-epje2006,bailey-prl2007}. Careful analysis of energy minimizations during such rearrangements \cite{maloney-prl2004a,demkowicz-prb2005} showed that, although the system is out-of-equilibrium and does not cross any equilibrium configuration (otherwise the energy minimization would stop), the path can be decomposed into a succession of localized unstable STs that trigger each other through their elastic strain and stress fields. The latter thus play the role in QS simulations of a mechanical noise that can push local regions beyond their stability limit leading to out-of-equilibrium cascades of STs \cite{lemaitre-pre2007}. Avalanches do not necessarily lead to persistent shear bands since, as discussed below, depending on the boundary conditions and on the interatomic interactions, the cumulated deformation can remain homogeneous in the avalanche regime for several 100 $\%$ strain.

Relating the elastic limit to the PEL picture, Harmon \etal~\cite{harmon-prl2007} proposed that the elastic limit occurs when the glass leaves its initial metabasin while STs are transitions between basins inside a given metabasin. However, from the PEL analysis of supercooled liquids presented in Section \ref{cooling}, we know that transitions inside a given metabasin correspond to small atomic adjustments that retain the nearest-neighbor shell while transitions between metabasins involve bond switching, as in STs. STs are thus elementary transitions between metabasins and are analogous to string-like events in supercooled liquids, while avalanches, made of a succession of STs, correspond to a transition over several metabasins and would be analogous to $\alpha$-relaxations.

\begin{figure}
\begin{center}
\includegraphics[width=15cm]{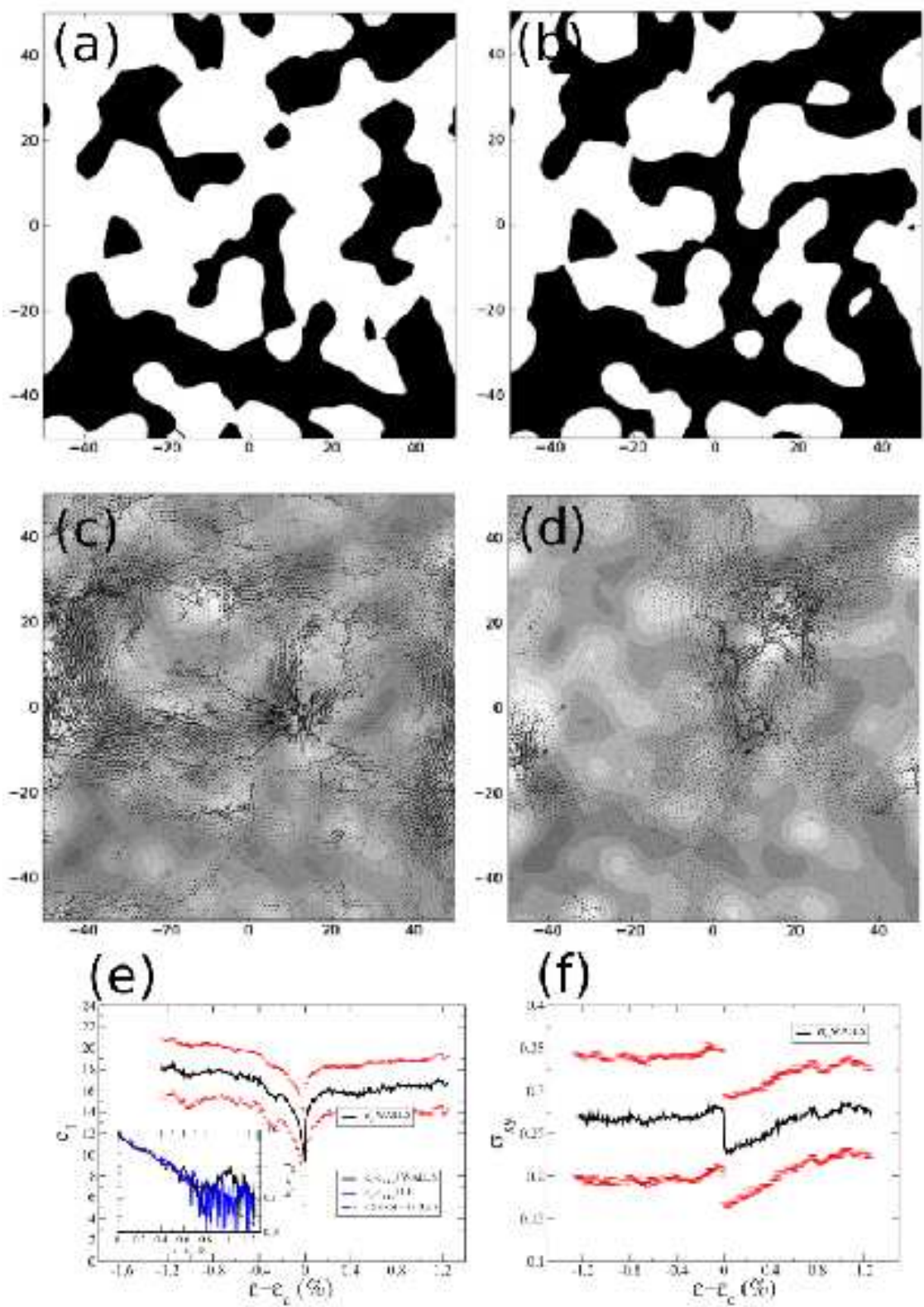}
\caption{\label{fig:ELASTICITY}Elasticity map and nonaffine displacements in a 2D shear LJ glass. In (a) and (b), the shear modulus is divided in rigid ($c_1 > \overline{c_1}$, black) and soft ($c_1 < \overline{c_1}$, white) zones for 2 strains: (a) 2.5 $\%$, (b) 2.55 $\%$. In (c) and (d), nonaffine displacements are superimposed on the local map of shear modulus for the same configurations as in (a) (for (c)) and (b) (for (d)). In (c), the nonaffine field is multiplied by 300 to illustrate the very strong correlation between elastic nonaffine field and elasticity map. In (e) and (f), are shown the variation of $c_1$ and shear stress, averaged over plastic events. Reproduced with permission from Ref. \cite{tsamados-pre2009}.}
\end{center}
\end{figure}

A question of prime importance then arises: can we predict the location of the next ST? Or said differently, is there a structural signature that indicates a region prone to plastic rearrangements, also called \emph{weak region}, or \emph{shear transformation zone} \cite{falk-pre1998}? Historically, the first proposed criterion was based on local atomic stresses. Srolovitz \etal~\cite{srolovitz-prb1981,srolovitz-am1983} in their early simulations of metallic glass plasticity identified three types of structural defects: atoms with high atomic tensile or compressive stress (n- and p-type defects respectively) and atoms with high atomic shear stress ($\tau$-defects). Plastic rearrangements were shown uncorrelated with n- and p-defects but a correlation was found with $\tau$-defects. A more recent study in a 2D sheared LJ glass \cite{tsamados-pre2009} showed however that the increase in local shear stress occurs only right before the plastic event, with an average relative increase of the local shear stress of only about $3 \%$, as shown in \Fig{fig:ELASTICITY}(f), very difficult to identify experimentally. Moreover, this small change is smaller than local stress fluctuations, and is not visible in the distribution of shear stresses for atoms in regions just about to rearrange plastically, which is the same as the overall distribution computed over all atoms \cite{tsamados-epje2008}. A criterion based on local stresses therefore appears not adequate at small scales, even if a definite yield stress appears at large scales (see discussion in Section \ref{ingredients}). A more promising route concerns local elastic moduli. Tsamados \etal~\cite{tsamados-pre2009,tanguy-epl2010} computed the local elasticity tensor through a coarse-graining procedure in a 2D sheared LJ glass and proposed as order parameter the lowest eigenvalue of the local elasticity tensor, $c_1$. Strong elastic heterogeneities were found at the nanometer scale that decay for larger coarse-graining scale. The nanometer scale (between 5 and 20 $\sigma_{AA}$) appears the most appropriate scale to describe elastic heterogeneities, since linear elasticity appears to be valid at that scale, and local anisotropy and spatial heterogeneities are measurable. As illustrated in \Figs{fig:ELASTICITY}(a) and (b), the glass is composed at this scale of a rigid scaffolding (black regions where $c_1 > \overline{c_1}$, where $\overline{c_1}$ is the average order parameter) and soft zones (white regions where $c_1 < \overline{c_1}$). Note that rigid and soft zones are not fixed in the microstructure but evolve dynamically during the deformation (as seen by comparing \Figs{fig:ELASTICITY}(a) and (b)). Also, as shown in \Figs{fig:ELASTICITY}(c) and (d), a strong correlation was found between plastic activity and soft zones at the nanometer scale: $c_1$ vanishes locally prior to a plastic event, with a marked $37 \%$ exponential decay of its amplitude, over a characteristic strain range of $0.2 \%$. Elastic constants being related to the Hessian matrix, this result shows that the global Hessian matrix, which has a vanishing eigenvalue at the plastic event, can be computed locally over finite-size regions with the same property: the local Hessian matrix has a vanishing eigenvalue in the region of plastic activity. But noticeably, the decay of the heterogeneous elastic constants measured on the local Hessian matrix \cite{tsamados-pre2009}, that is at the nanometer scale, appears far before ($\delta\gamma\approx 0.2\%$) the decay of the eigenvalues of the global Hessian (which occurs only within $10^{-3}\%$ of the transition \cite{maloney-prl2004b} with possibly an influence of the simulation cell size). Prediction of plastic activity thus combines the local measurement (at the nanometer scale) of a global quantity (the coefficients of the Hessian matrix). The same result was obtained for dislocation nucleation in crystals \cite{miller-jmps2008}. However, as shown in \Fig{fig:ELASTICITY}(e), $c_1$ decreases only near the instability and the next plastic event cannot be predicted until the glass is brought close to the instability. The question of the existence of identifiable structural defects in metallic glasses is therefore still largely open. The situation is different in amorphous silicon where weak (or liquidlike) and strong (solidlike) regions can be identified by looking at their radial and angular distribution functions  \cite{demkowicz-prl2004,talati-epl2009,fusco-pre2010}.

\textbf{Influence of initial configuration.} As seen in \Figs{fig:SHEAR}(b) and (c), plastic yielding strongly depends on the level of relaxation of the initial glass. Slowly quenched glasses that are more relaxed show a marked \emph{upper yield point} followed by softening, while less relaxed glasses have no upper yield point \cite{utz-prl2000,shi-prl2005,shi-prb2006,cheng-am2008, fusco-pre2010}. Also, the extension of the purely elastic regime increases strongly with the level of relaxation of the glass. After yielding, when simple shear is applied as in \Fig{fig:SHEAR}, the glass enters a steady state regime, called a \emph{flow state}, which is independent of the initial configuration in the sense that the energy, stress and fraction of icosahedron-centered atoms reach steady state values independent of the initial configuration, as shown in \Fig{fig:SHEAR}(b). One may say that the glass has lost the memory of its initial configuration \cite{utz-prl2000}. Persistent localization has been observed in this regime, but depends on the level of relaxation of the initial quenched glass, the boundary conditions as well as the interatomic potential. Persistent localization occurs only in slowly quenched glasses \cite{shi-prl2005} and is favored in simple shear by fixed boundaries \cite{varnik-prl2003} and in uniaxial straining by free surfaces \cite{shi-prl2005,shi-prb2006,cheng-am2008,cheng-am2009}. A strong three-body term in the interatomic potential also favors plastic localization, as well as a low pressure \cite{talati-epl2009,fusco-pre2010}. Sampling of the PEL around deformed configurations taken from the flow state, as shown in \Fig{fig:SHEAR}(e), shows a marked increase in the density of low activation energies compared to the initial quenched glasses except for the most rapidly quenched glass \cite{rodney-prl2009,rodney-prb2009}. This effect has also been measured experimentally \cite{Khonik-jap2009}. Similarly, as seen in \Fig{fig:SHEAR}(b), the fraction of icosahedron-centered atoms decreases after deformation \cite{cheng-am2009}, again with the exception of the most rapidly quenched glass, which was so far from equilibrium initially that it evolved towards a slightly more relaxed glass during deformation, a process called \emph{overaging} \cite{viasnoff-prl2002}.

\subsection{Influence of temperature}
\label{temperature}

Deformation acts in a way analogous to heating above $T_C$ since it accelerates the dynamics of the glass and gives access to high-energy ISs that had become inaccessible below $T_C$. The glass microstructure evolution under deformation, or \emph{rejuvenation} \cite{utz-prl2000}, has been opposed to \emph{aging} because the deformation reverses the aging process by increasing the energy of the glass and decreasing its stability. Also, as seen in \Fig{fig:SHEAR}(b), the steady-state IS energy in deformation is the same as in the high-temperature liquid regime. Moreover, it is known experimentally that the thermal history (annealing) affects the structure of a glass through its density fluctuations at rest \cite{levelut-jac2007,thorpe-ac2010}. This effect could be compared with the structural changes that occur when the system is submitted to a plastic deformation at a given strain rate. In addition, the plateau resulting from the cage effect \cite{berthier-jcp2002}, which arises in temporal correlation functions used to characterize the dynamics of glasses, disappears progressively with both an increasing shear rate and an increasing temperature. Finally, it is well-known experimentally that the viscosity of a supercooled metallic liquid \cite{middleman-1962,reger-leonhard-sm2000,kawamura-apl2001} or of oxide glasses \cite{simmons-jncs1998} decreases as a function of both temperature or shear rate. There is therefore a strong interplay between temperature and strain rate. From a theoretical point of view, the effect of temperature has mainly been taken into account up to now in mean-field models of plasticity, where plastic events are triggered randomly by thermal fluctuations in specific distributions of energy barriers \cite{sollich-prl1997,hebraud-prl1998,sollich-pre1998} or activated volumes \cite{falk-pre1998}. In the numerical simulation of mechanically deformed systems at finite temperature, two main difficulties must be considered: first the competition between the mechanically driven dynamics and the local dynamics of the thermostat used to maintain the temperature, and second the necessity to identify very carefully the temperature domain studied, which is a function of the applied shear rate.

\textbf{Influence of the thermostat.} This first difficulty is inherent to all MD simulations. The latter consist of solving the discretized Newton's equations of motion with different constraints, such as constant total energy (microcanonical ensemble) or constant temperature (canonical ensemble). The temperature is defined by the equipartition theorem through the average fluctuation of particle velocities:
\be
\langle \sum_{i=1}^N\frac{1}{2}m_i\delta v_i^2 \rangle = \frac{3}{2}Nk_BT.
\label{newton}
\ee

When a system is deformed plastically by MD, a thermostat must be used because the work produced by the plastic deformation is transformed into heat and the temperature will rise indefinitely. Interestingly, all thermostats involve characteristic timescales that depend on arbitrary parameters. For example, the simplest thermostats (Berendsen thermostat, rescaling of velocities) \cite{frenkel-2002} preserve Eq. \label{newton}  with a characteristic time depending on the coupling coefficient to the heat bath for Berendsen thermostat and the frequency of rescaling for the velocity rescaling. More elaborated thermostats, such as Langevin, Andersen or Nos\'e-Hoover thermostats, also involve characteristic times related to the damping coefficient for the Langevin thermostat, the probability of collision for Andersen thermostat and the dynamical coupling to fictive variables in Nos\'e-Hoover thermostat.

These timescales are important because the dynamics of the thermostat can interfere with the driven dynamics of the system. For example, a too frequent rescaling of the velocities will slow down the driven dynamics, while local heating can be unrealistically high in the opposite case. The thermostat parameters may be chosen on physical grounds if the dynamics of the dissipative processes acting in the system are known, but characterizing such processes is a major difficulty. As a result, MD simulations usually involve simply a spatially homogeneous velocity damping with an arbitrary intermediate value for the damping coefficient.

\begin{figure}
\begin{center}
\includegraphics[width=13cm]{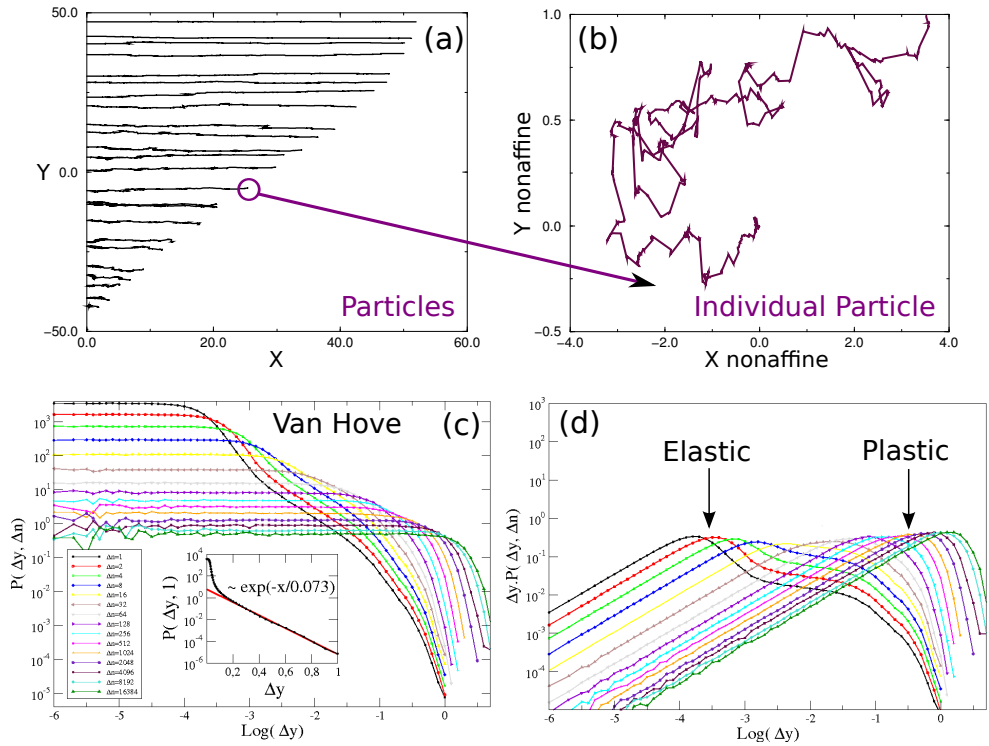}
\caption{\label{fig:VANHOVE}Stochastic analysis of the nonaffine part of the individual motion of particles in a sheared 2D LJ glass. $P(\Delta y,\Delta n)$ is the distribution of transverse motion $\Delta y$ between $\Delta n$ shear steps in QS simulations. It corresponds to the Van Hove analysis for local dynamics in glassy systems. $\Delta y.P(\Delta y,\Delta n)$ or $P(ln\Delta y, \Delta n)$ shows clearly a cross-over from a non-gaussian to a gaussian distribution with a single maximum whose position evolves like $\Delta n^{1/2}$ (corresponding to diffusive motion). The contribution of the plastic displacements is enhanced in the last figure on the right, which shows that the Gaussian distribution is due to plastic displacements. Reproduced with permission from Ref. \cite{tanguy-epje2006}.}
\end{center}
\end{figure}

\begin{figure}
\begin{center}
\includegraphics[width=13cm]{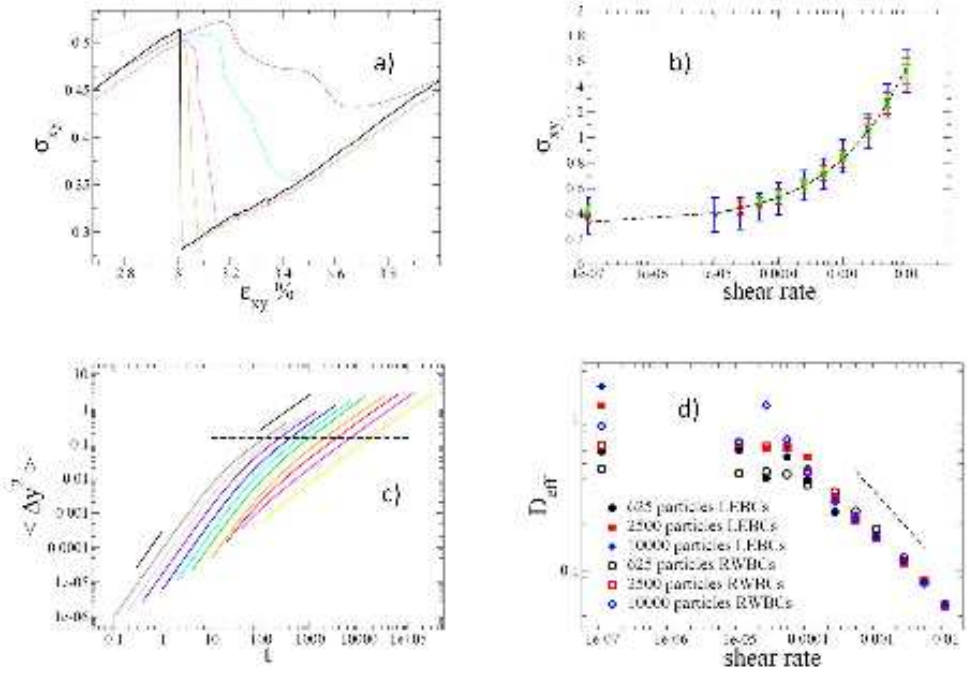}
\caption{\label{fig:ATHERMAL}(a) Stress-strain relation during the MD simulation of a sheared 2D LJ glass with Lees-Edwards boundary conditions, for different shear rates $\dot\gamma=1~10^{-5}, 2.5~10^{-5}, 5~10^{-5}, 1~10^{-4}$ and a QS protocol (black line). (b) Rheological law $\sigma_{flow}(\dot\gamma)$. The dotted line is an Herschel-Bulkley law $\sigma_{flow}=\sigma_{QS} + A \dot\gamma^{0.4}$. (c)  Transverse mean-square displacement $<\Delta y^2>$ of the particles as a function of time for different shear rates. (d) Effective diffusive coefficient $D_{eff}=<\Delta y^2>/2\Delta\gamma$. The dashed line corresponds to a power-law $D_{eff}\propto\dot\gamma^{-0.5}$. Reproduced with permission from Ref. \cite{tsamados-epje2010}.}
\end{center}
\end{figure}

\textbf{Temperature domains.} Different temperature domains can be defined, with boundaries that depend on the applied strain rate. A first limiting case is the {\it athermal regime}, which may be defined \cite{tsamados-epje2010} as the regime where the typical relative displacement between particles due to the external strain $a\dot\gamma t$ (where $a$ is the interatomic distance, $\sim \sigma_{AA}$ for a LJ potential) is larger than the typical vibration of an atom due to its local thermal activation, $\sqrt{k_BT/K}$ with $K=m\omega_D^2$ and $t=2\pi/\omega_D$, resulting in a condition on the temperature $T$:
\be
T<\frac{4\pi^2ma^2}{k_B}\dot\gamma^2 \mbox{ , or in LJ units: } T<4\pi^2\dot\gamma^2.
\label{eq:athermal}
\ee
This condition is very strict for a LJ glass ($T<40\dot\gamma^2\approx 4~10^{-7}$ for $\dot\gamma=10^{-4}$) but quite usual for a colloidal glass ($\dot\gamma>10^{-6}$ s$^{-1}$ at $T=300$ K for millimetric beads with $m=1g$). In this regime, which corresponds to $T \rightarrow 0$, Tsamados \cite{tsamados-epje2010} checked the progressive convergence to the QS regime when $\dot\gamma\rightarrow 0$, as illustrated in \Fig{fig:ATHERMAL}(a). In the athermal regime, the rheological law for the flow stress follows a Herschel-Bulkley law \cite{herschel-1926}:

\be
\sigma_0(\dot\gamma) = \sigma_{QS} + A\dot\gamma^m,
\label{eq:HerschelBulkley}
\ee
where $\sigma_{QS}$ is the flow stress in QS condition and $m = 0.4 \sim 0.5$, as shown in \Fig{fig:ATHERMAL}(b) \cite{berthier-jcp2002,rottler-pre2003,lemaitre-prl2009,tsamados-epje2010}. Similar behavior was also found experimentally in dense colloidal systems \cite{cloitre-prl2003,tang-ra2004}.

Under deformation at finite temperature, the particles exhibit a thermal as well as a mechanically-driven diffusion. Collective effects in atomic diffusion in glasses have been known for a long time \cite{ehmler-prl1998}. In the QS regime, the particles display a diffusive behavior at large strains due only to the plastic rearrangements upon external driving \cite{tanguy-epje2006,caroli-pre2007,tsamados-epje2010} as shown in \Fig{fig:VANHOVE}. In the athermal regime, the particles show first a ballistic motion followed by a diffusive behavior at longer timescales (see \Fig{fig:ATHERMAL} (d)) with a cross-over time that decreases for increasing shear rates $\dot\gamma$. Interestingly, as shown in \Fig{fig:ATHERMAL}(d), in the 2D LJ glass studied in Ref. \cite{tsamados-epje2010}, the diffusive coefficient, $D_{eff}$, defined as a function of the strain $\gamma$ rather than time (to transpose easily the definition to the QS regime) decreases with the shear rate as $D_{eff}\equiv\langle\Delta y^2\rangle/2\Delta\gamma\propto\dot\gamma^{-0.5}$ for large $\dot\gamma$, with a finite-size saturation at small $\dot\gamma$, i.e. in the QS regime. This decrease of the diffusion coefficient can be compared to the decrease of the Lindemann relaxation time $t_{MSD}$ defined by $<\Delta y^2(t_{MSD})>^{1/2}=0.1 a$, which corresponds to the typical time a particle needs to escape definitely from its initial cage (here through the plastic rearrangements due to the external driving) and is often associated with $\tau_{\alpha}$, the characteristic time for $\alpha$-relaxation (see Section \ref{cooling}). Of course, $t_{MSD}\sim \tau_{\alpha} \propto\dot\gamma^{-0.5}$ due to the $\dot\gamma$-dependence of $D_{eff}$. Also, the effective viscosity $\eta$ follows Herschel-Bulkley law with approximately the same strain-rate dependence $\eta\equiv (\sigma_{flow}-\sigma_{QS})/\dot\gamma\propto\dot\gamma^{-m'}$, with $m'=0.5 \sim 0.6$ as deduced from \Fig{fig:ATHERMAL}(b). This is in agreement with a Maxwell-like interpretation of the viscosity where $\tau_{\alpha}=\eta/\mu$, $\mu$ being the shear modulus \cite{coussot-2002}. Tsamados \cite{tsamados-epje2010} proposed an interpretation of this dependence of the relaxation time with $\dot\gamma$. The idea is that the rheological behavior of the amorphous material results from the competition between the propagation and nucleation of plastic rearrangements. The typical distance between thermally-activated nucleated sites in a 2D system is $d_n=1/(\rho_n\dot\gamma t)^{1/2}$ where $\rho_n$ is a density of nucleated sites per unit strain, depending probably on $T$. The typical distance covered by mechanically-triggered plastic rearrangements due to the diffusion of plastic activity \cite{tanguy-epje2006,shi-prl2007,manning-pre2007} from a given site is $d_p=(D_p t)^{1/2}$. If we assimilate the relaxation time to the time at which $d_n=d_p$, we obtain:
\be
\tau_{\alpha}=\frac{1}{\sqrt{\rho_n\dot\gamma D_p}}\propto \dot\gamma^{-0.5},
\ee
in agreement with the previous discussion.

In the well-defined athermal regime, the rheological law of the system is thus a Herschel-Bulkley law (Eq. \ref{eq:HerschelBulkley}) and the dynamics is entirely due to the succession of plastic rearrangements that enable to recover a diffusive memory-free behavior for the particles. Note however that finite size effects are very important at small $\dot\gamma$, i.e. in the QS regime, and they disappear only for sufficiently large strain rates, typically $\dot\gamma > 10^{-4}$ (see \Fig{fig:ATHERMAL}(d)). The diffusion coefficient then becomes a well-defined intensive parameter, with a finite non-zero value even at very small temperatures.

\begin{figure}
\begin{center}
\includegraphics[width=13cm]{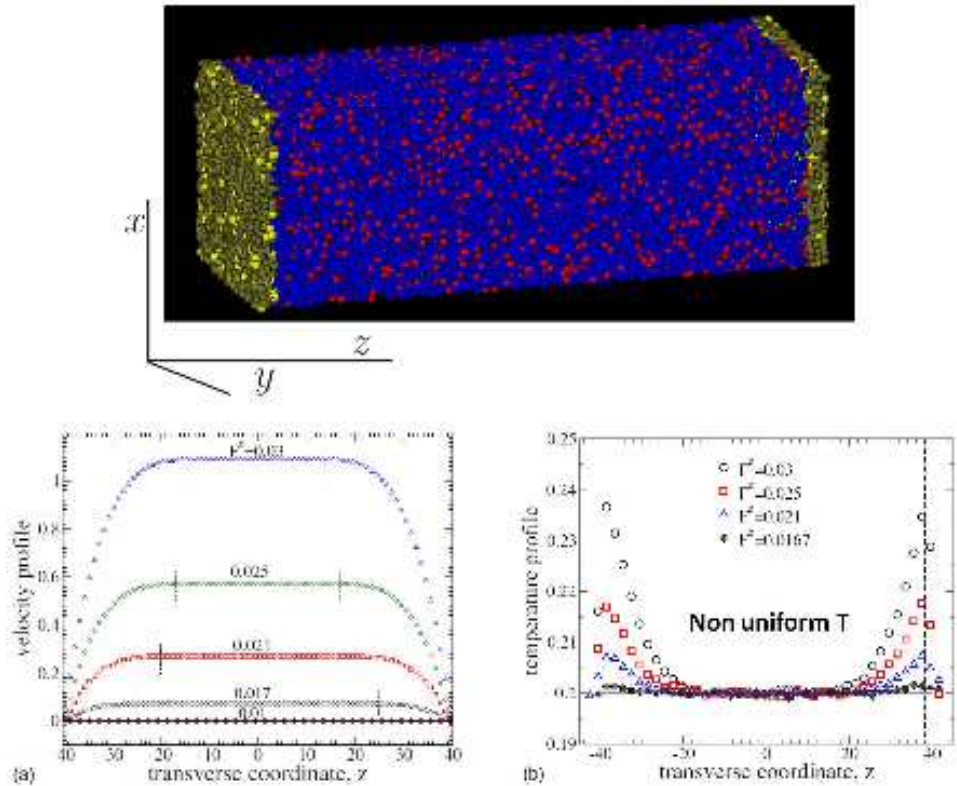}
\caption{\label{fig:NONUNIF_T} Shear velocity $v_x(z)$ and non-uniform temperature profile $T(z)$ in a Poiseuille-like flow of 3D binary LJ glass. Reproduced with permission from Ref. \cite{varnik-pre2008}.}
\end{center}
\end{figure}

In another study of the mechanical behavior of glassy materials at low temperature, Varnik \etal~\cite{varnik-pre2008} proposed to consider the case where the heat created by the plastic deformation is dissipated very efficiently by the system. This situation appears when the time needed to dissipate heat $t_d=L/c$, with $c$ the sound wave's velocity, is much smaller than the time needed to generate an energy $k_B T$ by plastic activity. The rate of heat production by plastic activity being $\sigma_{flow}\dot\gamma$, the time needed to generate the energy $k_B T$ is $t_Q=k_B T/\sigma_{flow}\dot\gamma$ and the condition $t_d \ll t_Q$ yields:
\be
\dot\gamma \ll \frac{k_B T c}{L\sigma_{flow}}.
\ee

The above relation is very different from the athermal condition in Eq. \ref{eq:athermal} and may be considered opposite because now the temperature plays a crucial role: the fact that the thermal energy is evacuated very efficiently assumes the presence of high thermal agitation.
In this situation, as illustrated in \Fig{fig:NONUNIF_T}, maintaining a uniform temperature profile inside the system is difficult, especially where the local shear rate is important, for example close to the walls where the flow is liquid-like (even if the temperature is far smaller than the melting temperature). This effect may also depend on the choice of thermostat (velocity rescaling in the direction transverse to the principal flow only \cite{varnik-pre2008}). The result we would like to emphasize on this example is that the interaction between the local temperature and local shear rate can affect the local dynamics and that local temperature and local dynamics are strongly coupled as soon as we depart from the athermal regime.

Indeed, identifying a single characteristic temperature for the rheological behavior of driven amorphous solids is very difficult. By looking at the detailed statistics of stress jumps in the stress-strain curve of a binary glass (with a Berendsen thermostat), Karmakar \etal~\cite{procaccia-pre2010} identified two main characteristic temperatures and showed that we can define separately cross-overs due to thermal fluctuations and strain rate. The cross-over due to thermal fluctuations is obtained by comparing the energy dissipated during plastic rearrangements $\langle \Delta U \rangle / N$ to the thermal energy $k_B T$. The finite-size scaling of $\langle \Delta U \rangle$ allows to define a characteristic length $\xi_2$ at which both energies coincide. For $\xi_2=L$, $T=T_{cross}(L)$ and $T>T_{cross}$ if $\xi_2<L$. If $T<T_{cross}$, thermal agitation does not affect the stress drops and the corresponding plastic rearrangements have large size dependence, while if $T>T_{cross}$, plastic rearrangements are localized with a finite extent. We should note that this assumption relates the size of a plastic event (function of $\Delta U$) to its energy barrier (only barriers of order $k_BT$ are accessible), but such relation may not be true in all systems \cite{rodney-prl2009}. The cross-over due to high strain rates allows to define another characteristic length $\xi_1$ by comparing, in a way analogous to the work of Varnik \etal~\cite{varnik-pre2008} in previous paragraph (except that here the temperature plays no role), the time needed to dissipate energy through sound propagation $t_d$, to the time needed to create plastic energy $t_P=<\Delta U>/(V.\sigma_{flow}\dot\gamma)$. The condition $t_d \ll t_P$ corresponds to $L<\xi_1$ that is to a QS process with shear-rate independent sizes of plastic rearrangements but strong system-size dependence. Karmakar  \etal~\cite{procaccia-pre2010} proposed to recover the Herschel-Bulkley law for the flow stress by looking at the $\dot\gamma$-dependence of $\xi_1$ that dominates the plastic processes at large $\dot\gamma$ (when $\xi_1<L$ or equivalently $t_d>t_P$). Finally, the comparison between the thermal effects and the shear-rate dependence can be done by comparing $\xi_1$ and $\xi_2$. For $\xi_1<\xi_2$ the shear-rate dominates the plastic processes. It corresponds to small temperatures $T<T^*$ or equivalently to large values of $\dot\gamma$. For $\xi_2>\xi_1$ ($T>T^*$), the temperature dominates the plastic processes. This description is supported by numerical results on a specific system \cite{procaccia-pre2010}. However, the different exponents obtained depend on the system-size dependence of the stress-rearrangements during plastic events. Also, it has been shown recently that the extent of plastic rearrangements strongly depends on the details of the interatomic potential used, and in particular on three-body terms \cite{fusco-pre2010}. The above analysis thus does not reflect the surprising universality of rheological laws of disordered systems but a more appropriate dynamical analysis is still lacking.

\begin{figure}
\begin{center}
\includegraphics[width=15cm]{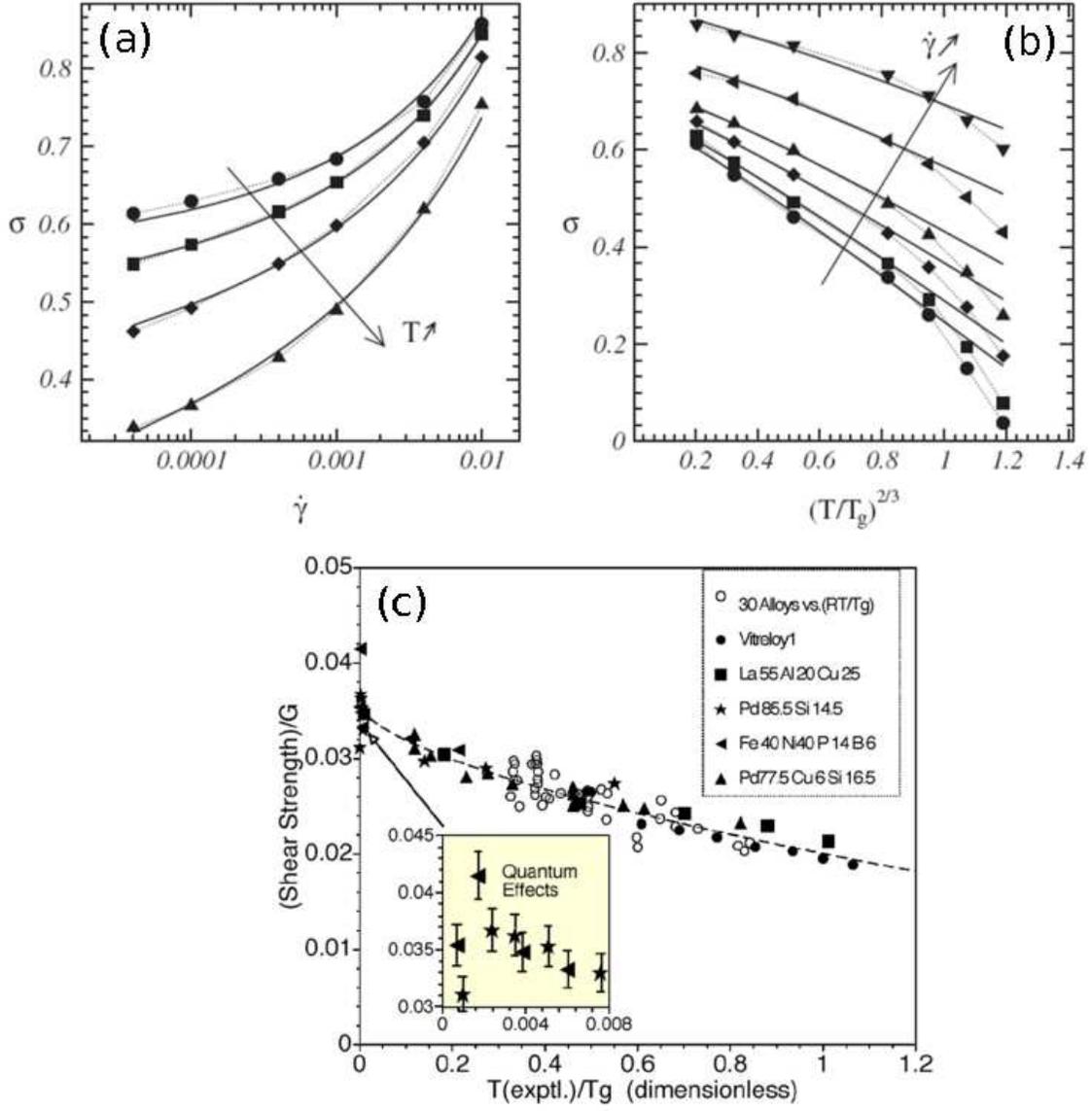}
\caption{\label{fig:SPIN_T} Temperature-dependent rheology: (a-b) Flow stress $\sigma$ as a function of shear rate $\dot\gamma$ and temperature $T$ in a 2D binary LJ glass. The continuous lines correspond to the parameterized rheological law in Eq. \ref{eq:thermalactivation}.  Reproduced with permission from Ref. \cite{caroli-prl2010}. (c) Experimental measurements of the flow stress as a function of $\dot\gamma$ for different metallic glasses close to $T_g$. The dashed line is a fit analogous to Eq. \ref{eq:thermalactivation}. Reproduced with permission from Ref. \cite{johnson-prl2005}.}
\end{center}
\end{figure}

\textbf{Thermally-activated transitions.} Another way to take temperature into account, far above the athermal regime, is to consider the global thermal escape in a mean-field description of the saddle node bifurcation preceding a plastic rearrangement \cite{caroli-prl2010}. When approaching a plastic instability, the global energy barrier presents the universal scaling form $\Delta E \propto (\gamma_c-\gamma)^{3/2}$ corresponding to a saddle node bifurcation at $\gamma_c$ \cite{maloney-prl2004a,caroli-prl2010,procaccia-pre2010}, as discussed in Section \ref{qs_def}. Considering the energy barrier as a function of $\gamma$ only, and not the other possible directions of the configuration space (this is why the model is of mean-field type), i.e. along the strain direction introduced in Section \ref{qs_def} and assuming that the system is at equilibrium along all other directions, the Kramers expression for the activation rate is:
\be
R(\gamma)=\omega\exp(-\frac{\Delta E}{k_BT}).
\ee
Assuming a constant strain rate $\dot\gamma$, the probability $P(\gamma;\gamma_0)$ that the system has not yet flipped at strain $\gamma$ starting from an initial strain $\gamma_0$ (survival probability \cite{vankampen-book}) is:
\begin{eqnarray}
P(\gamma;\gamma_0)&=&\exp\left(-\frac{1}{\dot\gamma}\int_{\gamma_0}^\gamma d\gamma' R(\gamma')\right)\\\nonumber
&=&\exp\left(-\frac{2}{3}\frac{\nu}{\dot\gamma}\left(\frac{T}{B}\right)^{5/6}\left(Q(\delta\gamma)-Q(\delta\gamma_0)\right)\right),
\end{eqnarray}
where $\delta\gamma=\gamma_c-\gamma$ and $Q(\delta\gamma)=\Gamma(5/6;B\delta\gamma^{3/2}/T)$ with $\Gamma$, the upper incomplete gamma function.
$P$ presents a very sharp transition from $P\approx 1$ to $P\approx 0$ around a strain $\gamma^*$ such that
\be
\frac{2}{3}\frac{\nu}{\dot\gamma}\left(\frac{T}{B}\right)^{5/6}Q(\delta\gamma^*)=1.
\label{eq:condition}
\ee

Considering that the flip event due to the thermal activity occurs at that threshold, the macroscopic stress should thus be of the form
\be
\sigma(\dot\gamma,T)=\sigma_0(\dot\gamma)-\mu\langle\delta\gamma^*(\dot\gamma,T)\rangle
\label{activation}
\ee
where $\sigma_0(\dot\gamma)$ is the athermal limit, $\mu$ is the shear modulus and $\langle.\rangle$ the average over structural disorder. As seen above (Eq. \ref{eq:HerschelBulkley}), $\sigma_0(\dot\gamma)$ follows a Herschel-Bulkley law $\sigma_0(\dot\gamma)=A_0+A_1 \dot\gamma^m$. The last term in the r.h.s. of Eq. \ref{activation} provides the departure from Herschel-Buckley law due to thermal activation in a mean-field approximation. It reads, after solving Eq. \ref{eq:condition} to leading order in $T/B\delta\gamma^{3/2}$. :
\be
\sigma(\dot\gamma,T)=A_0+A_1(\dot\gamma)^m-A_2T^{2/3}\left(ln\left(A_3T^{5/6}/\dot\gamma\right)\right)^{2/3}.
\label{eq:thermalactivation}
\ee
This relation between $T$ and $\dot\gamma$ is highly non trivial and shows clearly that $T$ and $\dot\gamma$ do not play similar roles that could be described through a simple scaling. \Fig{fig:SPIN_T}(a) and (b) shows fits of the flow stress for a sheared binary LJ glass using the above relation at different strain rates and temperatures, with a single set of parameters, $A_0$ through $A_3$ \cite{caroli-prl2010}, showing the high accuracy of Eq. \ref{eq:thermalactivation}. It must be noticed that the same kind of law also describes very well the rheological properties of metallic glasses, even above the glass transition temperature \cite{johnson-prl2005}, as shown in \Fig{fig:SPIN_T}(c).

The above expression does not explain the Herschel-Bulkley law obtained at very low temperature. Understanding the origin of this law probably requires a detailed description of the competition between local thermal escapes and elastically assisted propagation of plastic rearrangements, far from a simple mean field description, as mentioned in Ref. \cite{tsamados-epje2010}. Such detailed description is still lacking, but efforts in this direction are made by a detailed analysis of the evolution of energy barriers with the temperature and strain rate. It has been shown for example \cite{rodney-prl2009} that while the statistics of activation energies changes drastically upon an applied stress, only the low range of the distribution is visited during thermal dynamics. The study of the detailed evolution of the density of selected energies with temperature and strain rate, as well as the precise way of escape (localized or collective rearrangements) should contribute to a better understanding of the respective roles of temperature and strain rates. It has also been shown \cite{tanguy-epl2010} that even at zero temperature, the proximity to a plastic rearrangement (quadrupolar rearrangement or elementary shear band) affects not only the low energy distribution of activation energies, but also the low frequency eigenmodes of the system, and thus the detailed accessible rearrangements. The respective occurrence of thermal escapes and mechanical vanishing of activation barriers can thus give rise to a variety of different behaviors.

\textbf{Effective temperature.} A general way to take into account the different roles of thermal and mechanical activity is to introduce an \emph{effective temperature}.  This tempting reconciling view is possible when the average behavior of the system can be described in terms of the linear response theory, with a linear relation between the response function to a perturbation and the corresponding correlation function -- as in the usual Fluctuation-Dissipation theorem \cite{cugliandolo-pre1997}. In this situation, it has been shown with MD simulations that the fluctuations in the steady-state flow regime of sheared binary LJ glasses \cite{berthier-prl2002,berthier-jcp2002} and model foams \cite{ono-prl2002} are comparable to those in equilibrium systems maintained at an \emph{effective temperature} higher than the true temperature and function of the shear rate. More precisely, when the true temperature is above $T_C$ (supercooled regime), the effective temperature converges to the true temperature $T$ as the strain rate goes to zero, while for true temperatures below $T_C$ (glassy regime), the effective temperature remains above $T_C$, with a limiting value different from $T$ and close to $T_C$ when the strain rate goes to zero \cite{haxton-prl2007}. In addition to steady-state fluctuations, the rate of activated transitions above energy barriers \cite{ilg-epl2007} as well as the steady-state stresses \cite{haxton-prl2007} are also functions of the effective temperature $T_{eff}$ with Arrhenius dependence of the form $\exp(-\Delta E/k_BT_{eff})$. However, strong deviations from the linear behavior may appear, especially when the fluctuations become slow variables \cite{sciortino-prl2010}. Moreover, as evidenced by Eq. \ref{eq:thermalactivation}, temperature and strain rate do not play equivalent roles in thermally-activated transitions. Finally, the average energy of ISs visited in a sheared system at a given effective temperature is higher than the average IS energy of the same system maintained in equilibrium at a temperature equal to the effective temperature \cite{haxton-prl2007}. The range of validity of the effective temperature description thus still needs to be delimited clearly in out-of-equilibrium systems.

\begin{figure}
\begin{center}
\includegraphics[width=10cm]{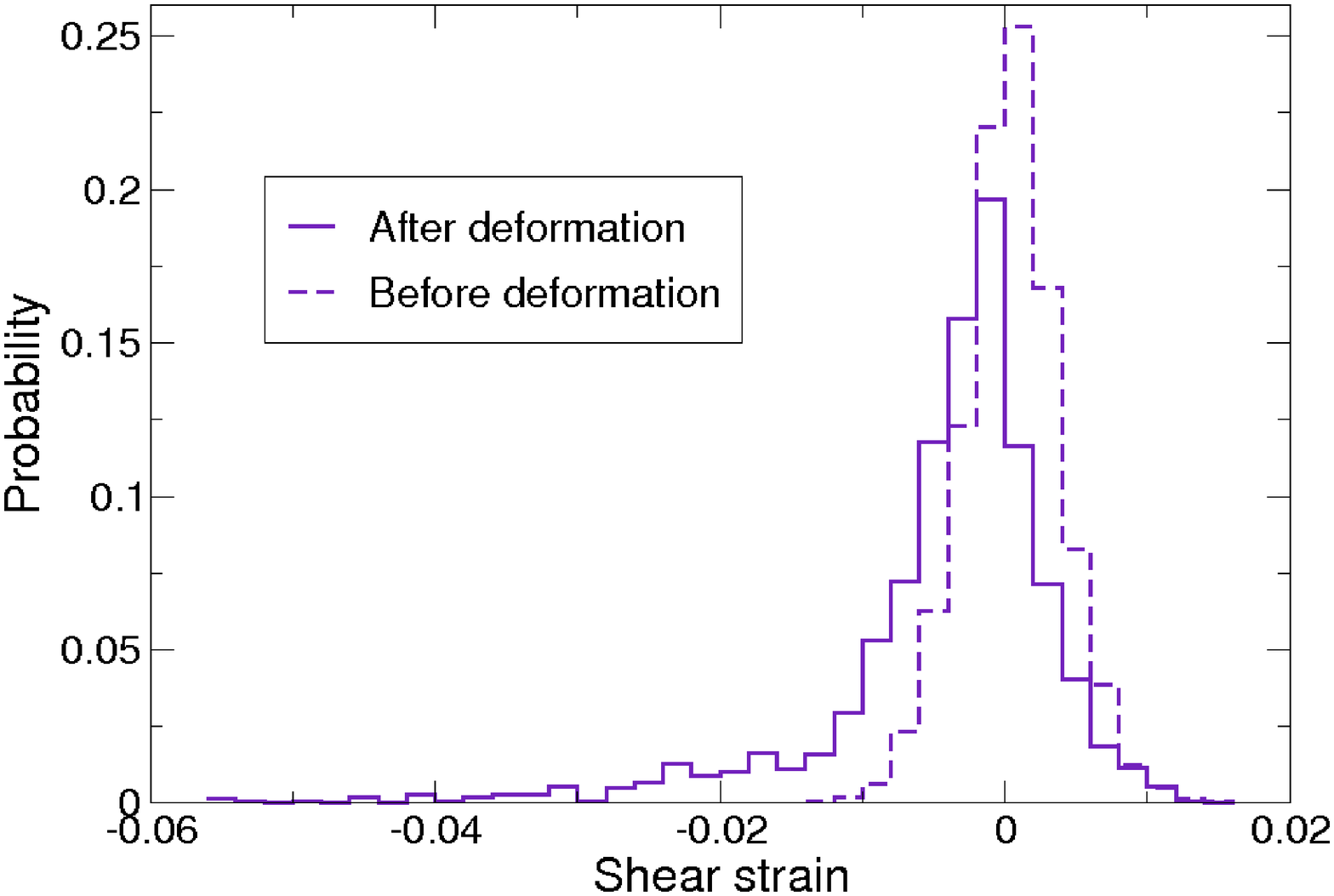}
\caption{\label{fig:POLARIZATION}Distribution of strain of thermally activated events in the configuration quenched at $2~10^{-5}$ before deformation (dashed line) and after deformation and unloading to a shear stress free state (solid line). The transitions are the same as in Figs. 3(d) and (e).}
\end{center}
\end{figure}

\textbf{Polarization.} Another reason why heating and mechanical deformation cannot be strictly equivalent is that plastic deformation can induce anisotropy in the glass microstructure whereas heating cannot. Structural anisotropy is referred to as \emph{polarization} \cite{argon-jncs1980} and is the hallmark of history-dependence in glasses. It is at the origin of several of the mechanical characteristics of glasses, such as the Baushinger effect \cite{falk-pre1998} and anelasticity, i.e. time- and temperature-dependent recovery of glasses after deformation \cite{argon-jap1968,argon-jncs1980}. Polarization has been measured experimentally in metallic glasses through x-ray diffraction spectra that become anisotropic after plastic flow in uniaxial tension \cite{suzuky-prb1987} and compression \cite{ott-am2008}. The results indicate that during deformation bonds are reorganized such that in tension, bonds in the tensile direction are cut and reform in the transverse direction, and vice versa in compression. Bond anisotropy has been reported in MD simulations of a silica glass \cite{rountree-prl2009} by considering the fabric tensor $\bf{F} = \langle \bf{n} \otimes \bf{n} \rangle$, where $\bf{n}$ is the unitary bond vector between atoms, and associated anisotropic parameter $\alpha =\sqrt{3/2\sum_{i=1}^3(\lambda_i-1/3)^2}$, where $\{\lambda_i\}$ are the eigenvalues of $\bf{F}$. Polarization is more difficult to evidence in simulated metallic glasses, presumably because the lack on angular dependence of the interatomic potentials used to model metallic glasses (see Section \ref{practicalities}) imposes fewer constraints on bond angles. The anisotropic parameter does not vary with plastic strain in metallic glasses \cite{rodney-prb2009}, but bond anisotropy was reported by computing an anisotropic pair distribution function \cite{tomida-prb1993,egami-jncs1995}. Bond orientation ordering upon shear strain has also been measured in binary LJ glasses \cite{albano-jcp2005}. In the extreme case of monodisperse amorphous systems with two body interactions and no directional bonding, crystallization can even occur \cite{mokshin-pre2008} at a sufficiently large strain whose critical value increases with strain rate. Polarization in metallic glasses is also visible through their local PEL, by considering the distribution of strains associated with thermally-activated transitions around deformed configurations \cite{rodney-prl2009,rodney-prb2009}. Examples of distributions are shown in \Fig{fig:POLARIZATION} for the most slowly quenched glass of \Fig{fig:SHEAR}. Before deformation, the distribution of strain is symmetrical around zero, while after deformation, the strain distribution is asymmetrical and contains more events with a negative strain, i.e. opposite to the initial direction of deformation, than events with a positive strain, i.e. in the same direction as the initial deformation. This asymmetry explains anelasticity since relaxation from a deformed state will tend to remove the anisotropy and will involve more events with a negative strain than a positive one, thus producing an average deformation in the direction opposite to the initial direction of deformation, as was checked by activated dynamics in a 2D LJ glass \cite{rodney-prb2009}.

\section {Mesoscopic models}
\label{meso}

From the atomistic simulations reviewed above, we know that plasticity
in glasses occurs by local plastic rearrangements, or shear
transformations (ST), that have a characteristic size $\ell_{ST}$ on
the order of a few nanometers. The cost for the high level of details
in atomistic simulations is a strong limitation in both
length- and time scales (typically a few tens of nanometers during a
few tens of nanoseconds as seen in previous Section). In order to
access larger length- and time-scales while retaining a description of
the elementary dissipative plastic processes, we wish now to develop a model for
amorphous plasticity at the mesoscopic scale that averages out
atomistic effects and accounts only for the dynamics of shear
transformations, in the same way as Dislocation Dynamics describes
crystal plasticity based on the motion, multiplication and interaction
of dislocations without explicitly accounting for atomic-scale core
effects \cite{madec-science2003}.

Generally speaking, a mesoscopic model requires four elementary ingredients: (i) a \emph{local yield criterion} for the occurrence of plastic rearrangements, (ii) an \emph{elastic coupling} to represent the reaction of the elastic matrix to the local rearrangements of the amorphous structure, (iii) an \emph{evolution rule} for the local yield criterion because a plastic rearrangement alters locally the amorphous structure, leading to either local softening or hardening of the matrix, (iv) a \emph{dynamical rule} to associate a time scale to the elementary processes. In analogy with atomistic simulations, depending on whether the temperature range considered is far below or close to the glass transition, the simulations can be conducted in the athermal quasi-static limit or time and temperature may be accounted for by using a kinetic Monte Carlo algorithm. The above elementary ingredients are reviewed below.

As an introductory illustration, directly inspired by the work of
  Dahmen {\it et al.} \cite{Dahmen-PRL97,Dahmen-PRL09} and more
  generally by depinning models of driven interfaces in random
  media \cite{Leschhorn-AnnPhys97,Kardar-PR98}, we present below a
  general equation of motion that incorporates the elementary building
  blocks mentioned above:

\be
\eta \frac{\partial \varepsilon({\bf x},t)}{\partial t} = {\cal H} \large\left\{
\sigma_{ext} + \sigma_{int}({\bf x},t)
- \sigma_{\gamma} \left[\varepsilon, {\bf x}, \{\varepsilon({\bf x},t'<t )\} \right]\large\right\}\;,
\label{eq:equation-motion}
\ee
where
\be
\sigma_{int}({\bf x},t) = \int_{0}^t dt' \int d{\bf x'} J({\bf x-x'},t-t')
[\varepsilon({\bf x'},t') - \varepsilon({\bf x},t)] \;.
\label{eq:elastic-coupling}
\ee

We are here restricted to a scalar formulation where $\varepsilon$ is
the local shear strain, $\sigma_{ext}$ is an applied shear stress,
$\sigma_{int}$ is the local shear stress accumulated at point ${\bf
  x}$ and time $t$ due to elastic stress transfer from all previous
STs since time $t=0$ (where a fully relaxed -- unstressed --
configuration is assumed) and $ \sigma_{\gamma}$ is a random pinning
stress (local yield stress) that prevents plastic slip until the local
stress ($\sigma = \sigma_{ext}+\sigma_{int}$) exceeds the local threshold. $\eta$ is an effective viscosity
that sets the characteristic relaxation rate and ${\cal H}$ is the
  Heaviside step function. This generic example shows how one may
build an equation of motion from a small number of hypothesis in terms
of a mesoscopic description of threshold criteria, elastic coupling
and dynamics, that can then be solved numerically -- or sometimes
analytically.

\subsection{Elementary ingredients of a mesoscopic model}
\label{ingredients}

\textbf{Local yield criterion.} A mesoscopic model involves a
discretization length, $\xi$, and the first question to be asked is
how to choose $\xi$ with respect to $\ell_{ST}$. This choice does not
appear clearly in the definition of most models proposed in the
literature and $\xi$ may actually be either close to or significantly larger than
$\ell_{ST}$, depending on the model considered. Related to the choice
of $\xi$ is the question of using whether a local or non-local yield
criterion for the plastic reorganizations and on what internal
variable should the criterion be based. Indeed, at the continuous
scale, all plastic criteria so far have relied on the local stress
state of the material. The well-known Tresca and von Mises criteria
for instance impose a limit, or yield surface, to
well-chosen norms of the deviatoric part of the local stress tensor (maximum deviatoric stress for Tresca, and maximum
  deviatoric elastic energy for von Mises). More elaborated criteria
can be defined to take into account the pressure dependence
\cite{Lambropoulos-SPIE98,Rottler-PRE01,KBVD-ActaMat08}. On
the other hand, at the atomistic scale, we have seen in
Section~\ref{qs_def} that a stress-based criterion for ST nucleation
is not the most relevant because plastic reorganizations are better
predicted by a lowering of the local elastic shear modulus, which is a non-local
quantity.

Relating non-local effects and elastic moduli-based criteria at the atomic scale to the well-established relevance of local stresses at the continuous scale is a very challenging task that has so far received little attention in the literature, probably due to its inherent difficulty. One way to circumvent (or reformulate) this question is to identify the relevant spatial scale $\xi$ at which mesoscopic modeling may be performed. Indeed, one may expect that non-local contributions that are significant at the reorganization
length scale $\ell_{ST}$ are averaged out in a description at a slightly larger length scale (say ~$10\ell_{ST} \approx 3$ nm in a mineral glass), while preserving the information on the dissipative reorganization due to the shear transformations.

In practice, mesoscopic models assume homogeneous linear elasticity, such that the discrete scale $\xi$ should be significantly larger than the microscopic length scale $\ell_{ST}$.  Then, mostly for the sake of simplicity, most models assume a simple (scalar) stress-based criterion:
\be
\sigma({\bf x}) = \sigma_Y({\bf x})
\label{criterion}
\ee
where $\sigma$ is the local shear stress (or shear stress invariant of the stress tensor) and $\sigma_Y $ is the
\emph{local yield stress} at location ${\bf x}$, also called slip or failure stress, i.e. a ST is triggered at ${\bf x}$ if $\sigma({\bf x}) > \sigma_Y({\bf x})$.  The local yield stress can
be chosen spatially homogeneous \cite{Picard-EPJE04,Picard-PRE05} or heterogeneous \cite{BVR-PRL02,TPVR-Meso10,Jagla-PRE07,Dahmen-PRL09}, with consequences discussed below.

An alternative way that includes thermal activation, is based on an energy approach at the ST scale \cite{BulatovArgon94a,BulatovArgon94b,BulatovArgon94c,homer-am2009,homer-prb2010}. While keeping the hypothesis of elastic homogeneity, the
authors used a lattice defined at a sub-ST scale ($\xi<\ell_{ST}$) and estimated the energetic cost of local plastic shears
involving sets of neighboring cells to recover a length scale $\sim \ell_{ST}$. The energy cost is computed as:
\begin{equation}
\Delta E = \Delta F_0 - \sigma({\bf x}) \frac{\gamma_p \Omega_0}{2},
\label{EA_bulatov}
\end{equation}
where $\Delta F_0$ is the stress-free activation free energy of a ST, biased by the work of the local stress between the initial and activated configurations, $-\sigma \Omega^*$, where $\sigma$ is the local stress resolved on the plane and direction of shear of the ST and $\Omega^* = \gamma_p \Omega_0/2$ the activation volume with $\gamma_p$ the strain associated with the ST and $\Omega_0$ the volume of the ST. In Ref. \cite{homer-am2009}, the plastic strain increment $\gamma_p$ was chosen constant, equal to 0.1 in agreement with the Lindemann criterion observed at the atomic scale (see Section \ref{qs_def}). The volume of the ST $\Omega_0$ was also constant, taken equal to 1.6 nm$^3$ to reproduce the properties of Vitreloy 1, a commercial metallic glass, i.e. contains 84 atoms, again in agreement with the atomistic simulations presented in Section \ref{qs_def}. We should note that within this approach, if an athermal yield criterion is applied, stating that plastic events occur when their activation energy vanishes, a local homogeneous stress-based yield criterion is recovered, as in \Eq{criterion}, with a yield stress $\sigma_Y = \Delta F_0/\Omega^*$.

\textbf{Evolution rule for the yield criterion.} Another crucial point
concerns the evolution of the criterion landscape under plastic
deformation. Indeed, after a reorganization has occurred, the local
structure of the glass is altered, which may change its local yield
stress. In some models proposed in the literature, the local yield
stress in \Eq{criterion} is unchanged after plastic slip
\cite{BulatovArgon94a,Picard-EPJE04,Picard-PRE05,homer-am2009,homer-prb2010}
implying a homogeneous and stationary yield stress landscape. In other models, the yield stress is renewed based on
a specific rule: the new value may be drawn from a stationary
distribution without correlation \cite{BVR-PRL02,TPVR-Meso10} or may
be systematically increased (resp. decreased) \cite{Dahmen-PRL09},
which naturally leads to hardening (resp. softening). We should note
that using a stationary distribution also leads to hardening during
the initial stage of deformation by a progressive exhaustion of the
weakest sites in the configuration \cite{TPVR-Meso10}. Two different
distributions (one for the initial configuration and the other for
renewing the yield stresses under deformation) have also been used to model aging
\cite{VR-ShearBanding11} (see below).

\textbf{Elastic coupling.} Successive shear transformations are not independent because the elastic medium surrounding the reorganizing zone reacts to the local change of conformation and acquires an additional (positive or negative) internal stress increment. The strength of this increment depends on the plastic strain associated with the ST, the size of the ST and the shear modulus \cite{TPRV-PRE08}. While ST size and shear modulus have been so far systematically chosen constant, the plastic strain has been chosen either constant \cite{BulatovArgon94a,homer-am2009} or drawn from a statistical distribution \cite{BVR-PRL02,TPVR-Meso10}.

Restricting ourselves to continuum mechanics, finding the elastic relaxation around a plastic shear transformation is similar to the plastic inclusion problem treated early-on by Eshelby \cite{Eshelby57}. We will assume here as in previous paragraphs that the mesoscopic length scale is sufficiently large to allow for a continuum and homogeneous description of elasticity. Within this approximation, there are several ways to include elastic effects in mesoscopic models:

\begin{itemize}

\item {\it Mean Field.} Ignoring the details of the elastic interaction, a first approach consists in assuming that the elastic relaxation of a reorganized region is compensated by a constant elastic stress everywhere else \cite{Dahmen-PRL09}. A statistical variant consists in drawing local elastic responses from an uncorrelated random distribution (constrained by the global balance of elastic contributions). Several distributions have been tested \cite{Lemaitre-preprint06}, from a simple Gaussian distribution to ad-hoc distributions that mimic the effect of a quadrupolar Eshelby contribution (see below). An advantage of this approach is the possibility for analytical treatment \cite{VSR-PRE04,Lemaitre-preprint06}.

\item {\it Exact numerical solution.} A second approach consists in solving numerically the equation of elastic equilibrium, thus accounting for the precise plastic strain induced by the ST and the outer boundary conditions. Elasticity can be solved using the finite element method in direct space \cite{homer-am2009} or, if the lattice is regular, using Green's function, obtained as the elastic response of an elementary cell of the lattice to a unit shear \cite{BulatovArgon94a,BVR-PRL02}, or else using Lagrange multipliers \cite{Jagla-PRE07}.

\item {\it Eshelby quadrupolar interaction.} An intermediate way consists in focusing on the dominant term of
  the elastic response at long distance. The Eshelby solution of the elastic field induced by a plastic inclusion can be developed using a multipolar expansion \cite{TPRV-PRE08}. While at short distance several terms are necessary to account for the details of the plastic reorganizations, at long distance, only the dominant term of the expansion, proportional to $1/r^d$ (where $d$ is the space dimension) survives. In two dimensions, such an analysis can be carried out using the complex potentials of Kolossov and Muskhelishvili \cite{Muskhelishvili}. It appears that only two terms survive at long distance. One is associated with the temporary dilation/contraction of a circular inclusion while the other term is associated with the local shear of a circular inclusion. This second term is responsible for the well-known quadrupolar symmetry of the elastic shear stress response:
\be \sigma_{xy}= -\frac{2\mu}{\kappa+1}\frac{ {\cal A}\gamma_p}{\pi r^2}\cos(4\theta) \label{eq:eshelby}\ee
where $\mu$ is the elastic shear modulus and $\kappa=(3-4\nu)$ for
plane strain and $\kappa=(3-\nu)/(1+\nu)$ for plane stress, $\nu$
being the Poisson's ratio. This symmetry is visible for example in the
atomistic snapshot shown in Fig. \ref{fig:multiscale}(a). Note that in
Eq. \ref{eq:eshelby}, the size of the zone under reorganization $\cal
A$ does not contribute independently but through its product with the
plastic strain, which is about twice the activation surface or activation volume in Eq. \ref{EA_bulatov}. $\gamma_p$ being on the order of $0.1$, the activation volume or surface is about a tenth of its actual size, in contrast with dislocations where $\gamma_p$ is close to 1 and activation and real surfaces and volumes are identical. While a priori simple,
the implementation of such quadrupolar solution on a discrete lattice
happens to be numerically delicate. Satisfaction of the boundary
conditions is made difficult by the long-range decay of the elastic
interaction. To circumvent potential resummation problems in direct
space, it is valuable to rewrite the interaction in Fourier space in
case of periodic boundary conditions \cite{Picard-PRE05,TPVR-Meso10}.

\end{itemize}

\textbf{Dynamical rule.} As mentioned above, there are two main
choices of dynamical rule for mesoscopic models, based on either a
quasi-static or a kinetic Monte Carlo algorithm. In the
\emph{quasi-static limit}, thermal and rate effects are ignored and
the dynamics is entirely dictated by the satisfaction of the local
yield criterion. The system can be either stress- or strain-driven
using two simple numerical protocols inherited from the field of
self-consistent criticality
\cite{Sneppen-prl92,paczuski-pre1995}. The first protocol, called
\emph{extremal dynamics} \cite{BVR-PRL02,TPVR-Meso10}, consists in
allowing one and only one event per simulation step, i.e. the weakest
site in the configuration. The applied stress (called extremal stress)
is thus adapted at each step to induce only this event and therefore accommodates a
vanishing shear rate. In the algorithm, the weakest site is first
identified and subjected to a local plastic strain, the associated
local yield stress is renewed (or not) according to the chosen rule of
evolution and the stress field is updated to account for the internal
stress change induced by the local transformation. Finally the whole
process is iterated. While the shear strain rate is reduced at its
lowest possible numerical value, the associated extremal stress
strongly fluctuates. The macroscopic yield stress is then identified
as the maximum extremal stress.

The second quasi-static protocol consists in \emph{slowly
  (quasi-statically) increasing the applied stress}
\cite{Picard-PRE05,Dahmen-PRL09}. For a given value of the external
stress, all sites satisfying their local yield criterion experience a
shear strain increment. As before, the yield criterion and internal
stress are then updated, but in contrast with before, the applied
stress is unchanged. Thus, the new configuration may contain a new set
of sites that satisfy the yield criterion, i.e. the occurrence of some
local slip may trigger other local slip events, leading to
\emph{avalanches}, as seen in the atomistic simulations in Section
\ref{qs_def}. The procedure is then iterated at constant applied stress
until the avalanche eventually stops, when a configuration is reached
where no site satisfies the yield criterion. Only then is the
external stress increased by a small amount and the whole procedure is
started again. In this approach, the number of sites experiencing slip
increases with external stress and diverges when the macroscopic
yield stress is reached \cite{Picard-PRE05}. Note that the bulk
elasticity of the material and/or the compliance of the mechanical
testing machine can be incorporated by coupling the system to a spring
\cite{Tanguy-PRE98,Zaiser-JSM07}. Tuning the value of the spring
constant then allows to drive the system either close to extremal
dynamics (high spring constant) or to a slowly increasing stress (low
spring constant).

The above protocols suffer from a clear drawback: any notion of realistic time has disappeared since the number of iterations simply counts the number of plastic rearrangements and avalanches in the system; or said in other words, while the iterations give successive configurations of the system, the time scale separating the configurations is unknown. One way to introduce a time scale and at the same time thermal effects is to employ a \emph{kinetic Monte Carlo algorithm}. This approach has been used in conjunction with the energy-based criterion in Eq. \ref{EA_bulatov}, which provides the activation energy of potentially thermally-activated STs \cite{BulatovArgon94a,homer-am2009,homer-prb2010}. From a given configuration, all potentially thermally-activated events are first determined. The activation energy of each event is calculated, from which the distribution of rates of the events is obtained using Boltzmann statistics. One event is then drawn from the distribution. Stresses, activation energies and rates are then updated and the whole procedure is iterated. This approach was developed in two \cite{BulatovArgon94a,homer-am2009} and three dimensions \cite{homer-msmse2010} and was shown to reproduce the mechanical properties of metallic glasses, including the high-temperature homogeneous flow and the low-temperature strain localization in shear bands.

Another approach to include a timescale and to account for rate
effects (but not for thermal effects) within a simple yield stress
model consists in assigning characteristic transition times for a
region to either plastically (viscously) deform or to relax back to an
elastic state. The most striking rate effect is that shear bands occur in complex yield-stress fluids only at low strain rates while at higher strain rates, homogeneous flow is recovered \cite{coussot-prl2002,lauridsen-prl2003,ovarlez-ar2009,coussot-epje2010}.
One of the first models of this family, developed to
simulate the rheology of complex fluids, is due to Picard {\it et al}
\cite{Picard-PRE05}. In this model, the displacement is discretized on
a lattice and each site can be either in an active \emph{plastic} (or
more exactly visco-plastic)
state if its local stress is larger than a yield stress or in a still
\emph{elastic} state in the opposite case. Characteristic transition
rates $\tau_{pl}^{-1}$ and $\tau_{el}^{-1}$ are then assigned for the
transition from elastic to plastic for an active site and from plastic
to elastic for an inactive site. When a site becomes "plastic",
  its strain relaxes through Maxwell dynamics, i.e. with a rate
proportional to the local stress: $\dot \gamma_p = \sigma /2 \mu \tau$ where $\mu$ is the shear modulus and $\tau$ a mechanical
relaxation time (in practice, the authors chose
$\tau_{pl}=\tau_{el}\equiv\tau$ and have thus a single timescale). The plastic strain produces an
internal stress by its associated Eshelby field. Although simple, this
elastoplastic model produces complex spatiotemporal patterns of
deformation \cite{martens-prl2011} and includes both Newtonian and
non-Newtonian regimes with a threshold strain rate
$\sigma_Y/\mu\tau$. However, we should note that $\tau_{pl}$ and $\tau_{el}$ are characteristic timescales, not directly related to microscopic processes so far.

Three model archetypes naturally emerge from the possible choices allowed by the different rules described above. All three models feature local plastic events that interact elastically. The first archetype, developed by Vandembroucq \etal~\cite{BVR-PRL02,Dahmen-PRL09,TPVR-Meso10}, which will be called \emph{depinning model}, is based on a yield stress criterion with yield stresses and plastic strains drawn from statistical distributions, an internal stress arising from the accumulation of Eshelby fields and extremal dynamics. The second model, developed initially by Bulatov and Argon \cite{BulatovArgon94a} and extended by Homer \etal~\cite{homer-am2009,homer-prb2010,homer-msmse2010}, called here the \emph{KMC model}, is based on a kinetic Monte Carlo algorithm with an energy-based yield criterion and elasticity solved by the finite element method. Here, the yield criterion is not affected by deformation and the ST plastic strain is constant. The third model, developed by Picard \etal~ \cite{Picard-EPJE04,Picard-PRE05,Bocquet-PRL09,martens-prl2011}, called a \emph{fluidity model}, is based on a constant yield stress criterion and Maxwellian viscous strain rate. It includes strain-rate effects through characteristic transition rates. It is interesting to note that the origin of disorder in the three above models is different. The depinning model includes a \emph{structural disorder} that arises from the stochastic distribution of yield stresses, while the fluidity and KMC models have no structural disorder (constant yield stress) but reflect a \emph{dynamical disorder} arising either from the Boltzmann statistics, or from the stochastic distribution of the relaxation times.

\subsection{Phenomenology}
\label{phenomenology}

We consider here three applications of mesoscopic models, namely a comparison with atomistic simulations, the interplay between aging and shear banding and avalanches in plastic flow.

\begin{figure}[htbp]
  \centering
  \includegraphics[width=50mm]{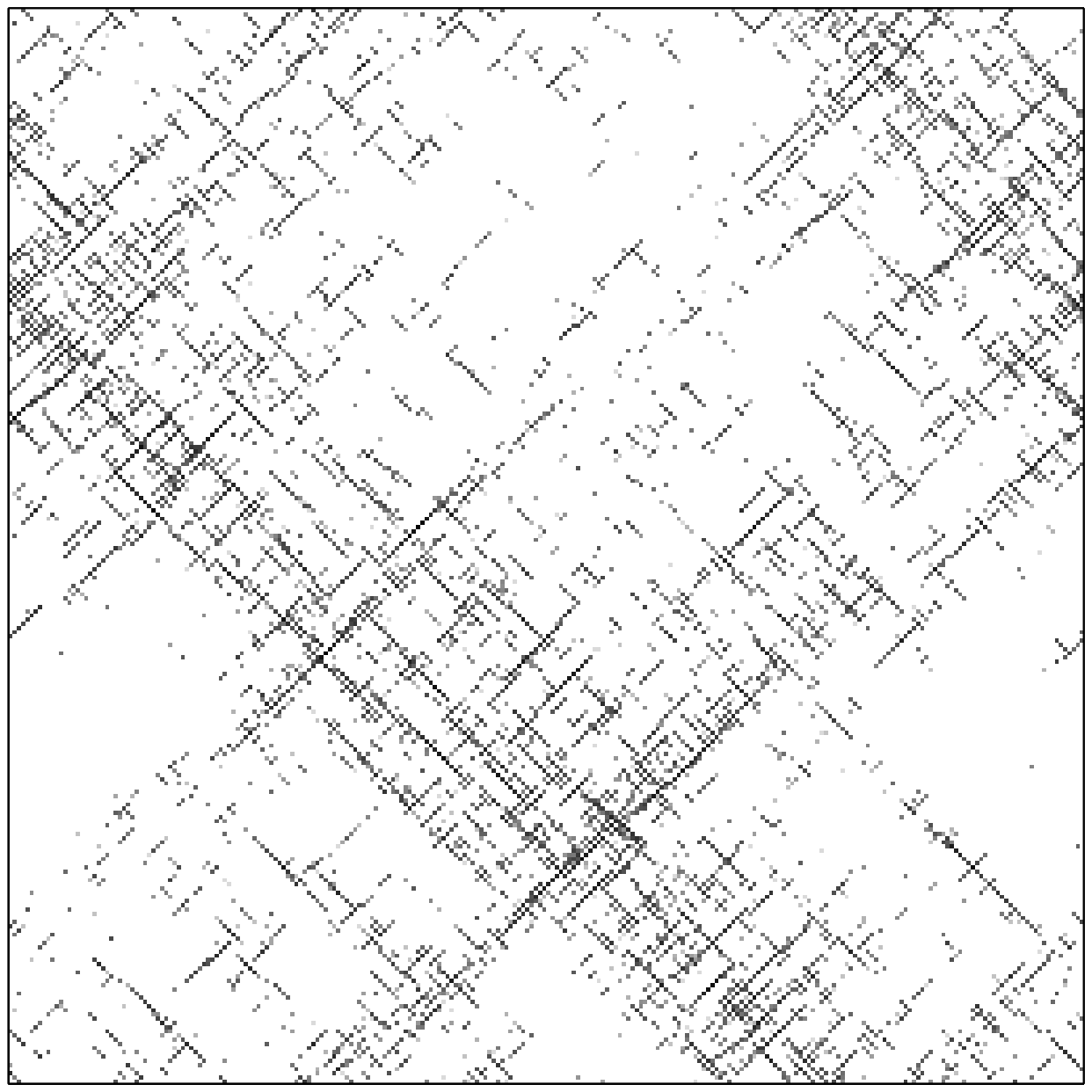}
  \includegraphics[width=49.2mm]{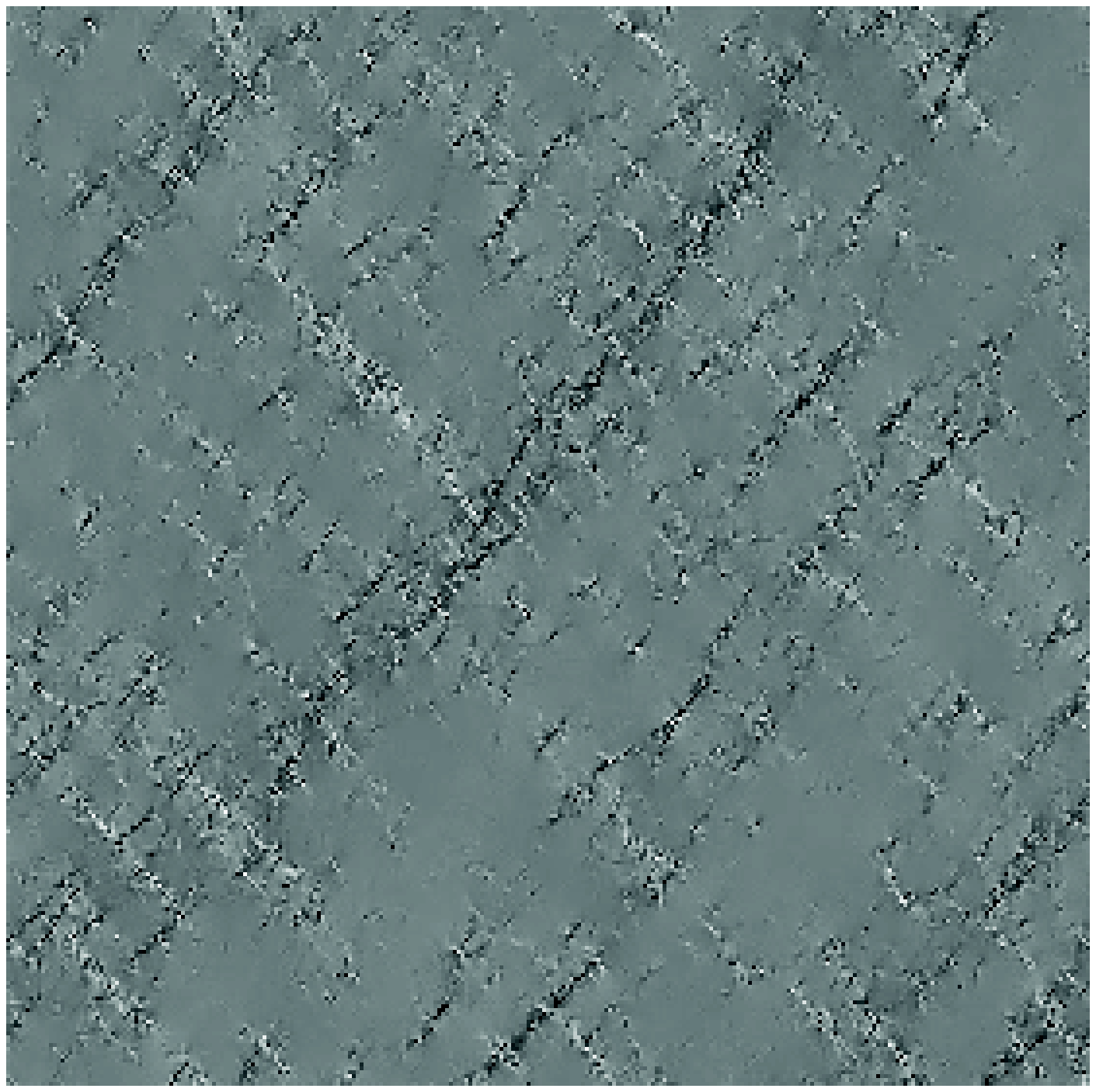}
\\
  \caption{Comparison of activity maps between a mesoscopic model (left) and an atomistic simulation (right). Left: map of cumulated plastic activity in the stationary
    regime during a deformation window $\Delta \varepsilon = 0.01$
    obtained with the mesoscopic depinning model. Reproduced from Ref. \cite{TPVR-Meso10}. Right:
    strikingly similar map of
    plastic activity (vorticity of the displacement
    field) computed on a 2D LJ binary glass under
    compression. Reproduced with permission from Ref. \cite{maloney-prl2009}.}
  \label{Talamali-Maloney}
\end{figure}

\textbf{Comparison to atomistic simulations.} \Fig{Talamali-Maloney} compares maps of plastic activities obtained with the depinning model and with atomistic simulations (2D binary LJ glass \cite{maloney-prl2009}). The similarity between the two maps is striking. In both cases, plastic strain is spread over the entire system and one can clearly distinguish elongated patterns along the directions at $\pm \pi/4$. Those correlated events are transient localization events. They are observed with all models that satisfy elasticity equilibrium  \cite{BulatovArgon94a,homer-am2009,Jagla-PRE07} and are due to the quadrupolar symmetry of the Eshelby field (Eq. \ref{eq:eshelby}).

This comparison between atomistic and mesoscopic models clearly shows
that the elementary ingredients included in the mesoscopic models,
i.e. mainly localized plastic events that interact via the Eshelby
stress field, are sufficient to reproduce the deformation pattern of
some atomic-scale glasses and to some extent the anisotropic
  character of the plastic strain power spectrum. We will see below other examples of
processes where mesoscopic and atomistic simulations agree very
well. We should note however that this comparison is qualitative and the rheological properties $\sigma(\dot\gamma)$
  and spatio-temporal description of plastic damage could be more
  discriminating for the choice of the elementary ingredients to be included in mesoscopic models.

\begin{figure}[tbp]
\centering
\includegraphics[width=0.55\textwidth]{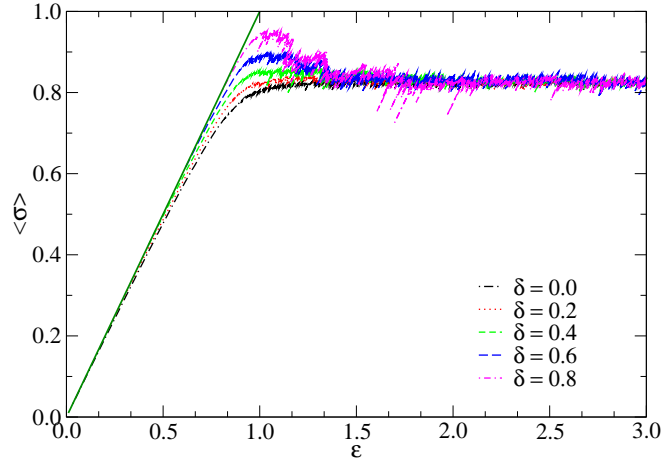}
\caption{Effect of aging on the strain/stress curve in the depinning model. The aging parameter $\delta$ is the minimum normalized yield stress in the initial glass configuration before shearing. Reproduced from Ref. \cite{VR-preprint11}.}
\label{StressPeak-vs-delta}
\end{figure}

\textbf{Localization, shear banding and aging.} Localization
  effects have been discussed early in the framework of
  mesoscopic models \cite{BVR-PRL02,Picard-PRE05}. More recent studies
  were devoted to shear banding \cite{Manning-PRE07,Manning-PRE09} and in particular its
  dependence on aging \cite{Fielding-SM09,Fielding-PRL11}. Persistent shear bands form when correlations between plastic events are stronger than the disorder in the glass. Both disorder and correlations can be of structural and dynamical origins. Correlations are strongest in case of softening, i.e. \emph{structural correlations}. There are also \emph{dynamical correlations} because when a ST is triggered, the stress increment added to neighboring sites through the Eshelby field (Eq. \ref{eq:eshelby}) may be larger than the local yield stress, leading to a succession of STs. The latter appear along elongated patterns because of the strong anisotropy of the Eshelby field. As mentioned above, there is \emph{structural disorder} for instance when yield stresses are drawn from a random distribution \cite{BVR-PRL02} or when the configuration contains an initial elastic stress field \cite{Jagla-PRE07,homer-am2009}. Disorder may also be dynamical when the glassy dynamics is itself stochastic, for instance when using a KMC algorithm. A \emph{dynamical disorder} also arises from the slow decay of the Eshelby field at long range. Indeed, the Eshelby field of a ST affects distant regions in the glass. The succession of these stress increments plays a role analogous to a temperature (or effective temperature \cite{sollich-pre1998}) that can be regarded as an uncorrelated mechanical noise \cite{Lemaitre-preprint06}.

  In summary, disorder and correlations can have structural and dynamical origins depending of the models and simulation conditions and the competition between these factors controls the formation of shear bands. In absence of structural correlations (no softening), disorder is usually stronger than dynamical correlations and no persistent shear bands form. As illustrated in Fig.~\ref{Talamali-Maloney}(a), the strain field is then characterized by an accumulation of transient localized patterns corresponding to successive small avalanches. In absence of nucleation sources (walls or local structural defects) or local softening, such patterns diffuse throughout the systems and no persistent shear band form. Only the anisotropic spatial correlation of the strain field retains a trace of the transient localization \cite{BVR-PRL02,maloney-prl2009,homer-prb2010,TPVR-Meso10}.

Persistent shear bands form when there is no structural disorder in the initial glass configuration or when structural disorder relaxes faster than the strain rate. The influence of initial conditions was demonstrated by Homer \etal~\cite{homer-am2009} using the KMC model by preparing glasses that either contained an initial density of STs (activated during a quench from high temperature) and thus contained an initial structural disorder through an internal stress field, or contained no ST and were therefore stress-free without initial disorder. Only in the latter case, shown in Fig. \ref{fig:multiscale}(b), was a persistent shear band observed. Otherwise, the deformation pattern was similar to Fig. \ref{Talamali-Maloney}(a). Another example is if the initial configuration contains a distribution of stresses that relaxes during deformation, as in the work of Jagla \cite{Jagla-PRE07}. Shear bands were observed only if the strain rate is small compared to the stress-relaxation rate, i.e. only if the configuration has time to relax to a stress-free configuration without structural disorder before plastic flow sets in.

Shear bands in complex fluids are thus limited to low strain rates because of a competition between the rate of production of structural disorder, which increases with strain rate $\dot \gamma$, and the rate of relaxation of structural disorder. Using a simple mean-field fluidity model, Coussot and Ovarlez \cite{coussot-epje2010} showed that the condition for shear band formation, using the notations introduced here, is:

\be
\tau_{el}>\tau$ \mbox{and} $\dot \gamma < \sigma_Y/\mu\tau_{el}(\sqrt{\tau_{el}/\tau}-1).
\ee

In metallic glasses, it is also known that deformation is homogeneous at high strain rates \cite{schuh-am2007} but the reason is different and is due to the fact that the strain per band is not sufficient to sustain the strain rate. Within the fluidity model, this situation may be related to the condition $\dot \gamma > \tau_{pl}^{-1}$.

\begin{figure}[tbp]
\centering
\includegraphics[width=15cm]{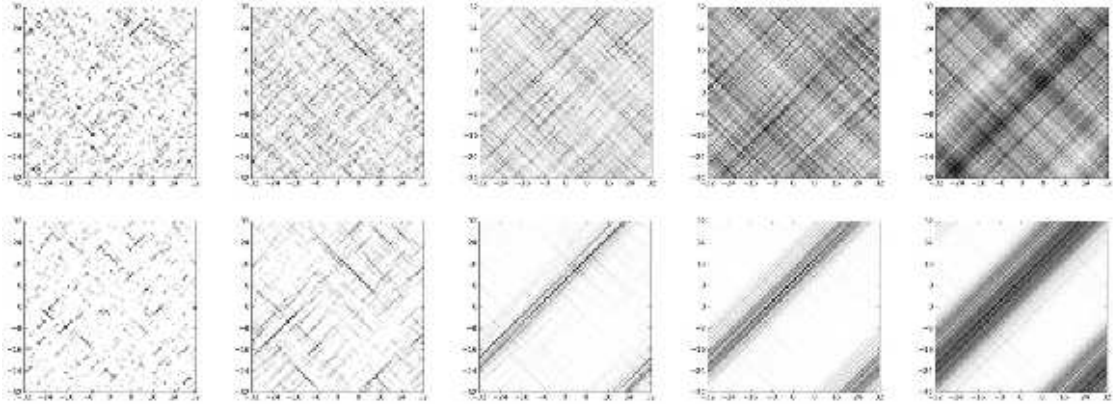}
\caption{Maps of plastic strain obtained from left to right at
strains $1/16$, $1/4$, $1$, $4$ and $16$ and
from top to bottom with an aging parameter $\delta = 0$ (top) and  $\delta = 0.5$ (bottom). Reproduced from Ref. \cite{VR-ShearBanding11}.}
\label{aging}
\end{figure}

Shear bands are prominent in aged glasses where structural relaxations and local softening are accounted for. The effect of aging and
stress relaxations have been studied using several mesoscopic models
\cite{Jagla-PRE07,Fielding-SM09,Fielding-PRL11,VR-ShearBanding11}. Within
the depinning model \cite{VR-ShearBanding11}, aging was modeled by
drawing the initial distribution of yield stresses from a statistical
distribution shifted to higher stresses compared to the distribution
used to renew the yield stress under deformation. More precisely, the
initial yield stresses were drawn from a uniform distribution between
$[\delta; 1+\delta]$ in normalized units, while under deformation, the
distribution was uniform between $[0; 1]$. The parameter $\delta$, the
minimum of the initial distribution, is called the \emph{aging
  parameter} because the yield stress is expected to increase
logarithmically with time during the aging process
\cite{rottler-prl2005,Fielding-PRL11}. The resulting stress/strain
curves as a function of $\delta$ are shown in
\Fig{StressPeak-vs-delta}, where in agreement with atomistic
simulations, as $\delta$ increases, i.e. as the glass ages, an upper
yield point develops followed by a steady-state flow state independent
of the initial configuration. Associated with the stress overshoot is
the development of a shear band shown in \Fig{aging}. The reason is
that after the first slip events, the new yield stress are drawn from
a distribution with statistically lower yield stresses, which induces
a systematic softening effect, thus leading to localization. In this
approach, once the plastic activity has concentrated along a band, the
system remains trapped for arbitrary long times while the band widens
at a logarithmic pace. A similar correlation between initial stress overshoot and shear bands
was obtained using fluidity models \cite{Fielding-SM09,Fielding-PRL11}
where the lifetime of the shear band appears bounded to the intrinsic
timescale of the model.

A notable effect on shear banding can be obtained by tuning the level
of the mechanical noise, which can be obtained by increasing the
plastic strain increment per ST \cite{VR-ShearBanding11}. Starting
from aged configurations, it appears that the higher the mechanical
noise, the faster the shear band widens and the shorter its duration. Also, shear banding is strongly affected by the
boundary conditions and is more readily obtained with fixed boundary
conditions than periodic boundary conditions \cite{Picard-PRE05}, as
in atomistic simulations (see Section~\ref{qs_def}).


Finally, we note that the way structural disorder is introduced in the above models are phenomenological and not related to any local structural defects, such as non-icosahedral environments or anomalous coordination numbers, as observed in atomistic simulations. It would thus be very
  interesting to relate the atomistic-scale structural parameters to appropriate
  mesoscopic variables, such as a relaxation time or a distribution of yield stresses.

\textbf{Avalanches. Intermittent flow.} While traditionally described in continuum mechanics by constitutive laws at
the macroscopic scale, it has progressively appeared in the last two decades that
the mechanical behavior of materials was not as smooth and regular as
anticipated. In particular crack propagation in brittle materials and plastic
flow in crystalline solids have been shown to exhibit jerky motion and scale-free spatio-temporal
correlations \cite{Miguel-Nat01,Zaiser-AdvPhys06,Bonamy-JPhysD09}.

In the context of plasticity of crystalline materials, a significant
amount of results have been obtained over the last decade (see
e.g. the comprehensive review by Zaiser about scale invariance in
plastic flow \cite{Zaiser-AdvPhys06}).  Acoustic emission measurements
performed on ice or metal monocrystals have shown a power law
distribution of the energy $P(E)\propto E^{-\kappa}$ with
$\kappa\approx 1.6$ for ice \cite{Richeton-ActaMat05} and
$\kappa\approx 1.5$ for hcp metals and alloys
\cite{Richeton-MSEA06}. The case of polycrystals is somewhat more
complex since not only a grain size related cut-off appears in the
avalanche distribution but the power law exponent is also
significantly lowered \cite{Richeton-ActaMat05}. Performing
nano-indentation measurements on nickel monocrystals, Dimiduk \etal~
found evidence for a scale-free intermittent plastic flow and
estimated $\kappa \approx 1.5-1.6$~\cite{Dimiduk-Science06}.  Very
recently analogous analysis was performed on metallic glass
samples.  Sun \etal~\cite{Wang-PRL10} measured the distributions of
stress drops occurring in the stress/strain curves for various
metallic glass samples under compression and observed scale-free
distributions with a power law exponent $\kappa \in [1.37-1.49]$.
Mean field models such as developed by Dahmen \etal~\cite{Dahmen-PRE98} give a power law distribution for the size $s$ of
avalanches $P(s)\propto s^{-3/2}$ thus with an exponent strikingly
close to the experimental results. Note that the introduction of
systematic hardening \cite{Dahmen-PRL09} induces a finite cut-off in
the initially scale-free distribution. Zaiser and Moretti~\cite{Zaiser-JSM05}
also measured a similar exponent in a dislocation-based model and
found evidence for a stiffness-induced cut-off.

While from renormalization theoretical results, Dahmen \etal~\cite{Dahmen-PRL09} argue in favor of the universality of the mean
field (MF) exponent, a rapid tour of the other models reveals a more
contrasted situation.  Starting from the closest member of the class,
Lema\^itre and Caroli \cite{Lemaitre-preprint06} obtained the MF
result when using a statistical mean field approach with a Gaussian
distribution. Indeed the average size of avalanches was shown to scale
as $\langle s \rangle \propto L^{0.5}$, consistent with a
$s^{-3/2}$-distribution. However, when using an uncorrelated random
noise reproducing the quadrupolar Eshelby field, they obtained a
clearly different scaling with $\langle s \rangle \propto L^{0.14}$. This abrupt dependence on the distribution may not
be as surprising as it may appear at first. Indeed, the latter
Eshelby-like noise happens to exhibit power-law fat tails.

Using a full Eshelby elastic interaction, Talamali \etal
~\cite{TPVR-Aval11} obtained an avalanche scale-free distribution with
an exponent $\kappa=1.25$. While significantly different from the MF
prediction, this non-trivial exponent may still be difficult to
distinguish from 1.5 experimentally. It would thus be interesting to consider other variables in order to discriminate between
  different theoretical models. Another question is whether this
difference obtained with a 2D model survives in three dimensions.

\section{Macroscopic scale}
\label{macro}

The literature on the simulation of plasticity at the macroscopic scale is more scarce, in part because glasses are brittle at the macroscopic scale, their plasticity being mostly limited to the micron-scale. However, substantial amounts of plasticity can be reached under conditions of confined plasticity, as in indentation \cite{su-am2006,charleux-jmr2006}. At the macroscopic scale, simulations are based on the finite-element method (FEM), which requires a constitutive law that relates the plastic strain rate to the state of stress and the history of deformation of the glass.

Following Spaepen \cite{Spaepen-ActaMet77}, most of the early discriptions of the viscoplastic behavior of metallic glasses
  relied on a flow rule accounting for the evolution of an internal
  state variable, the free volume. The excess free volume in metallic
  glasses is usually defined as follows. Let $V$ be the volume of the
  sample and $V_d$ the volume of the same sample with a dense
  random packing of atoms. The excess free volume, $V_f$ , is the
  difference between the two volumes, i.e., $V_f = V - V_d$ .  The flow equation for the plastic shear strain $\gamma$ derived by Spaepen writes \cite{Spaepen-ActaMet77,schuh-am2007}:

\be
\frac{\partial \gamma}{\partial t} =
\nu_0 \exp \left[
- \alpha \frac{ v^*}{v_f}\right] \cdot 2\exp\left[-\frac{\Delta G_0}{k_BT}
\right]
\sinh \left[\frac{\sigma \Omega^*}{k_B T} \right],
\ee
where $\nu_0$ is an attempt frequency that sets a characteristic timescale. The first exponential is the free-volume contribution ($v_f$ is the average free volume per atom, $v^*$ a critical volume and $\alpha$ a geometrical factor of order unity). The second exponential and the sinh term are derived from a mean-field stress-biased activation energy model similar to Eq. \ref{EA_bulatov}: $\Delta G_0$ is the shear stress-free activation enthalpy, $\sigma$ is resolved shear stress and $\Omega^*$ the activation volume (the sinh function accounts for both forward and backward shears). This flow rule is complemented by an evolution equation for the free
  volume assuming a stress-induced production term, a
  relaxation-induced annihilation term and often a diffusion term.
  The stress assisted production of free volume has a clear shear
  thinning effect. The latter effect has been shown early to induce
  localization \cite{Steif-ActaMet82} (see also the clear presentation of
  the model and its generalization as well as a linear stability
  analysis in Ref.~\cite{Nix-JMPS02}).

\begin{figure}
  \centering
\includegraphics[width=15cm]{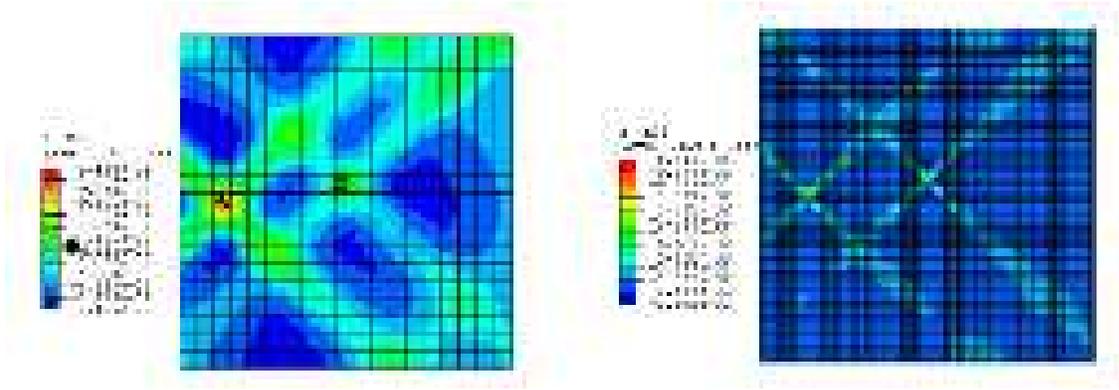}
\caption{Contour maps of $\varepsilon_{22}$ in a uni-axial traction test where the initial free volume has been slightly perturbed at two sites of the lattice. The two lattice sizes shown here illustrate the dependance of the shear band width on the sole mesh size. Reproduced with permission from Ref. \cite{YFGao-MSMSE06}}
\label{Fig-YFGao-MSMSE06}
\end{figure}

  In the very same spirit, Gao~\cite{YFGao-MSMSE06} recently
  developed an implicit finite element method for simulating
  inhomogeneous deformation and shear bands of amorphous alloys. The
  use of a numerical scheme that limits convergence problems in
  mechanically unstable systems \cite{YFGao-MSMSE04} allowed to
  follow the initiation and propagation of shear bands from local
  free volume fluctuations. Two examples are displayed in
  Fig.~\ref{Fig-YFGao-MSMSE06}. Note that the width of the shear band
  appears to be controlled only by the lattice discretization length as
  expected in a model deprived of any internal length scale.

In the recent years, such free volume based models have been
  enriched with phenomenological rate- and temperature-dependence in
  the glass transition regime in order to account for the
  technological thermoplastic forming process\cite{Anand-ActaMat08}.

While free volume has been widely used as a shear thinning
  ingredient, the natural consequence of stress-induced free volume
  expansion, \ie the introduction of an inelastic volumetric
  deformation, has been less frequently
  discussed \cite{Flores-ActaMat01,Nix-JMPS02}. A simple reason for
  leaving aside this {\it a priori} important aspect (after all, while
  dislocations are volume conserving, no such limit applies to shear
  transformations) is that in most experimental tests performed on
  metallic glasses, pressure levels remain low compared to the yield
  stress value \cite{Lowhaphandu-ScriptaMat99}. Strongly contrasting
  with such a statement is the case of indentation tests where
  pressure levels are frequently measured in GigaPascal units.

So far, two main constitutive laws have been proposed to model plasticity during indentation experiments, one in amorphous silica by Kermouche \etal~\cite{KBVD-ActaMat08,kermouche-pm2011} and the other in metallic glasses by Anand and Su for low temperatures in the shear banding regime \cite{anand-jmps2005,su-am2006} and for high temperatures in the homogeneous deformation regime \cite{anand-am2007}. The main characteristic of these constitutive laws is to account for the pressure-dependence of the plastic deformation of glasses. However, they treat this effect in different ways due to the specificities of the glasses considered. For metallic glasses, the pressure dependence is treated using a Mohr-Coulomb law \cite{Lowhaphandu-ScriptaMat99,Lee-ActaMat05}, where the yield stress in shear increases linearly with the normal stress, while for silica glasses, following early attempts based on a simple linearly
  pressure-dependent Mises criterion \cite{Lambropoulos-JACS96}, a quadratic law involving deviatoric stress and pressure, inspired from the mechanics of porous materials was employed. In both cases, the pressure dependence is related to the free-volume in the material but in the case of metallic glasses, plasticity leads to an increase of free volume and a corresponding softening of the glass, whereas in silica glasses, plasticity decreases the free volume and induces densification \cite{Rouxel-JMR05,Rouxel-ScriptaMat06} and
hardening \cite{PMMCVB-JACS06,VDCPBCM-JPCM08} of the glass. Finally, in silica glasses, only density hardening was considered (by a linear relation between the plastic limit in hydrostatic compression and the plastic strain) while no hardening in shear was accounted for, while in metallic glasses, shear softening by the increase of free volume was considered, while the Mohr-Coulomb friction coefficient was held constant, implying no density hardening (or softening). The hypothesis used in both constitutive laws are thus quite different but well-adapted to the systems considered since in both cases, the simulations were at least in qualitative agreement with experiments. In particular, in silica glasses, the authors were able to reproduce densification maps below the indentor, while in metallic glasses, the simulations showed shear band patterns, as shown in Fig. \ref{fig:multiscale}(c), in good agreement with experiments and atomic-scale simulations \cite{shi-apl2005,shi-am2007}
In the perspective of multiscale modeling, the constitutive laws could also be checked directly by MD simulations on submicrometric samples submitted to different kinds of deformations, like shear at constant pressure or hydrostatic compression.

\begin{figure}
  \centering
\includegraphics[width=15cm]{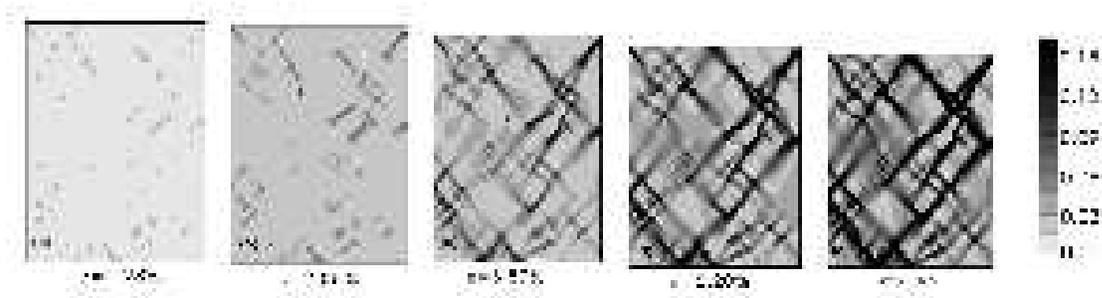}
\caption{Distribution of the effective strain at various deforming stages of a Ta particle enriched Cu-based bulk amorphous glass matrix
; (a) 1.3\%, (b) 2.6\%, (c) 3.9\%, (d) 5.2\%, and (e) 6.5\%.
 Reproduced with permission from Ref. \cite{Lee-ActaMat05}}
\label{Fig-Lee-ActaMat05}
\end{figure}

Finally, remaining in the framework of mutiscale modeling,
  an additional interesting point of comparison is given by recent
  works simulating the plastic behavior of bulk amorphous matrix
  composites \cite{Lee-ActaMat05}. In the very same spirit as
  glass-ceramics developed to improve mechanical properties of oxide
  glasses, such materials incorporate nano- or micro-particles of a
  crystalline ductile phase. As strikingly apparent in
  Fig.~\ref{Fig-Lee-ActaMat05}, the macro-scale plastic deformation
  of metal-matrix composites strongly recalls meso-scale results
  discussed above for amorphous materials. In both cases, one recovers
  similar patterns of criss-crossed shear bands.  Such a similarity is
  obviously not surprising since the very same Eshelby
  elastic interaction of quadrupolar symmetry is at work in both
  cases. Beyond the size of the plastic inclusions encountered, note
  that the main difference between these two otherwise comparable cases
  stems from the fact that shear bands are persistent in the
  metal-matrix composites since they are nucleated on well defined
  structural defects (the ductile particles) while shear
  transformations experienced at the microscopic scale in glasses are very  short-lived.

\section{Perspectives and challenges for multiscale modeling}
\label{perspective}

As mentioned in introduction, the simulations reviewed in the present article at different length and time scales were performed mostly independently. Methods and models have now reached a state of development where transfer of information from one scale to the scale above becomes possible. However, there remain challenges that are discussed in the remaining of the section.

\subsection{From atomistic to mesoscopic scales}

The striking agreement between deformation patterns found in atomistic and mesoscopic simulations, illustrated in Fig. \ref{Talamali-Maloney}, and the fact that mesoscopic simulations can account for the influence of aging and relaxation of glasses (stress-overshoot and shear banding) are clear indications that mesoscopic models are very good tools to check the main ingredients necessary to model plasticity in glasses, such as local visco-plastic events with long-range elasticity. Also, the fact that the same boundary condition effects are found in mesoscopic and atomistic simulations (shear banding favored by fixed boundary conditions) is another proof of the realism of mesoscopic models. Interestingly, both atomistic and mesoscopic models have difficulties reproducing shear banding, which is almost unavoidable at the macroscopic scale, at least at low temperature. The reason is that both atomistic and mesoscopic glass models are far less relaxed than experimental glasses and therefore do not exhibit strong enough structural relaxations and local softening which is the main source of shear bands as seen above.

Mesoscopic models have been used so far mostly phenomenologically. In order to quantitatively transfer information from the atomistic to the mesoscopic scale, several difficult challenges have to be faced. We have identified four main challenges:
\begin{itemize}
\item \emph{Length Scale.} A mesoscopic model involves a characteristic discretization length scale that must be compared to the other characteristic length scales describing the mechanical response of the system. One of these length scales is the size of the ST itself, $\ell_{ST}$. But the choice of the elastic kernel depends on the role devoted to the elastic heterogeneities. In amorphous systems, the typical size for elastic heterogeneities $\ell_{ELAST}$ is in the nanometer range \cite{ehmler-prl1998,Tanguy-PRB02}. This characteristic size has been studied only in specific systems. It has been shown for example in 2D LJ systems~\cite{tsamados-pre2009} that linear elasticity is valid only above $5$ interatomic distances, and that isotropic and homogeneous elasticity is recovered beyond $20$ interatomic distances. The experimental and numerical study of the vibrational response of different amorphous materials \cite{Monaco-science1998,Pilla-jpcm2004,Tanguy-PRB02,Tanguy-PRL06,Ruffle-prl2006,Duval-prb2007,ruffle-prl2008,Shintani-NatureMaterials2008,Champagnon-jncs2009}  show that disorder affects strongly the acoustical response of materials for wavelengths below a characteristic value $\lambda_{BP}$ depending on the pressure \cite{Amonaco-prl2006,Niss-prl2009,Starviou-prb2010,Mantisi-jpcm2010}, on the temperature \cite{Ruffle-prl2010,Baldi-prl2009}, and on the aging time \cite{Duval-jncs2006}, as well as on the precise composition of the material \cite{Fabiani-jchemphys2008}. The order of magnitude of this characteristic wavelength is the nanometer in usual glasses \cite{Tanguy-PRB02,Tanguy-PRL06,Ruffle-prl2006,Duval-prb2007}  and reflects probably a transition from weak to strong scattering of acoustic waves on strain (or elastic) heterogeneities \cite{Ruffle-prl2003,tanguy-epl2010}. It means that homogeneous elasticity cannot be valid at length scales below this characteristic value $\lambda_{BP}$ \cite{Tanguy-PRB02,Pilla-jpcm2004,Baldi-prl2009}. However, visco-plastic rearrangements in mesoscopic models will induce inhomogeneous strains~\cite{martens-prl2011} not compatible with the homogeneous elastic kernel usually chosen in the same models. The question of the origin of the nanometer scale for elastic heterogeneities in amorphous materials and the question of the scale of description for mesoscopic modeling are thus strongly related and of crucial interest to check the validity of the models. Fortunately, it seems that above the scale $\ell_{ELAST}$, homogeneous and isotropic elasticity is valid; however, the nanometer scale is precisely the most difficult size to reach experimentally. This is also why an appropriate mesoscopic modeling can be very challenging.

\item \emph{Yield criterion.} We have to understand how the ST nucleation criterion, which appears non-local and controlled by elastic moduli (or equivalently, by the stability of the Hessian matrix), may become local and based on stresses at the scale above. Understanding this point is interesting from a fundamental point of view and is also unavoidable in order to develop a physical yield criterion at the micron-scale. As seen above, yield stresses in mesoscopic models have so far been either constant or drawn from phenomenological statistical distributions (for instance uniform) without spatial correlations. However, we may expect from the disordered structure of glasses with short- and medium-range orders, that yield stresses should arise from specific distributions that reflect the level of relaxation of the glass and include some spatial correlations. Similarly, energy-based approaches use a homogeneous stress-free activation energy for STs ($\Delta F_0$), with homogeneous ST volume ($\Omega_0$) and strain increment ($\gamma_p$), while again, we expect these quantities to arise from statistical distributions. This information will have to come from atomistic simulations and will require a better understanding of the relation between the structure and properties of glasses. The first step could be to determine more precisely the nucleation criterion for STs and the relation between the local structure of the glass and its propensity for plastic deformation. One link in metallic glasses could be the fivefold coordinated atoms surrounded by icosahedra that have recently been directly correlated to hard zones under deformation in CuZr glasses \cite{peng-prl2011}, but the relation between the local structure and a local yield stress or a local activation energy has to be established. Also, such clusters are not relevant to all glasses and a better measure for the local stability of glasses is needed.

\item \emph{Time scale.} Mesoscopic models either have no time scale, as the depinning model based on extremal dynamics, or use very simplified dynamics, as the fluidity model based only on two time scales ($\tau_{pl}$ and $\tau_{el}$) that are fixed, independent of the local structure of the glass and not clearly related to any microscopic processes. These models can therefore not reproduce the slow dynamics with multiple timescales characteristic of aging glasses, as evidenced by MD simulations near the mode-coupling temperature and by the large energy range of the energy distributions found in glasses at the atomic scale (see Figs. \ref{fig:PEL}(d) and (e)). Timescales are also influenced by the temperature, both real and effective. This will require first to better understand the role of temperature and thermal activation at the atomic scale. One way could be through atomistic activated dynamics, based on kinetic Monte Carlo simulations, with distributions of activation energies determined on the fly by saddle-point search methods \cite{xu-jcp2008,rodney-prb2009}. Indeed, this technique could simulate aging or plastic deformation on experimental timescales and could be compared to the mesoscopic KMC model.

\item \emph{Softening versus hardening.} So far, the treatment of glass microstructure evolution under deformation has been treated very crudely in mesoscopic models, being either simply ignored when a constant yield criterion landscape is used or drawn from phenomenological statistical distributions. On the other hand, we expect that soft regions harden when they deform, for example because their "free volume" decreases, while strong regions soften because they increase their "free volume". Such effect is qualitatively reproduced when drawing yield stresses from statistical distributions, but we may wish to use more physical distributions. Again, this information will have to come from atomistic simulations and will require a better understanding of the structure/property relation in glasses. Depending on the microstructure of the glass (pure silica versus soda-lime glass for example), hardening can depend crucially on the geometry of the mechanical deformation (pure shear versus densification or shear at constant pressure) \cite{KBVD-ActaMat08}. Also, the origin of polarization and its link to the internal stress field produced by the plastic deformation will have to be better understood to be properly modeled in mesoscopic simulations. The basic question is whether polarization is only related to the stress field accumulated in the structure or is also encrypted in the atomic configurations. In the latter case, an anisotropic softening effect will be required (for instance a yield stress that depends on the sign of the local stress).
\end{itemize}

\subsection{From mesoscopic to macroscopic scales}

There is less connection between mesoscopic models and macroscopic constitutive laws. In particular, the two constitutive laws mentioned above are mostly based on the pressure-dependence of the mechanical response of glasses, an effect neglected in all mesoscopic models so far. Also, the constitutive laws are based on the notion of free-volume and a relation between free volume, yield stress and pressure dependence is assumed. But, if the pressure-dependence of the mechanical properties of glasses has been demonstrated at the atomic scale \cite{schuh-naturemat2003,lund-am2003,ogata-im2006}, the notion of free volume remains unclear, in particular in metallic glasses where only very little volume change is observed under and after deformation. A first step could therefore be to identify an internal variable that could be characterized at the atomic scale and linked to the mechanical properties, for example related to the distribution of yield stresses. That internal variable will certainly have to be tensorial and not just scalar in order to account for the dependence on pressure and polarization. This information could then be imported in a mesoscopic model that would serve to develop a realistic constitutive law at the macroscopic scale.

\section{Summary and conclusion}
\label{conclusion}

Glasses are characterized at the atomistic scale by broad energy spectra and heterogeneities that are consequences of the hierarchical multifunnel structure of their PEL. The latter explains most of the dynamical properties of supercooled liquids and glasses: (i) the strong influence of the quench rate on final glass state, (ii) dynamical heterogeneities, (iii) slow aging process in glassy state and (iv) anelastic recovery after deformation. Plastic deformation in glasses also involves heterogeneities, in the form of local rearrangements that interact elastically, as evidenced by both atomistic and mesoscopic simulations, and we have seen an analogy between string-like events in supercooled liquids and shear transformations in deformed glasses that are both localized rearrangements that allow to system to transit between metabasins.

Glasses relaxed more slowly have more time to explore deeper regions of their PEL. They are more stable thermodynamically (lower potential energy) and kinetically (higher activation energies). They have also a higher mechanical strength with a longer elastic regime and larger elastic moduli. These effects arise naturally in atomistic simulations and are reproduced phenomenologically in mesoscopic models, for instance by increasing the initial level of yield stresses. However, a more quantitative description of this effect at the micron scale is still lacking, because it requires a deeper understanding of the dynamics of relaxation and aging in glasses. This is related to a better understanding of the dissipative processes at small scale that are far to be understood in general, as soon as they involve electron-phonon couplings. Moreover, the effect of bond directionality, already known in crystalline plasticity, is far from being quantitatively understood, as well as the corresponding relation between local structure and plastic damage in amorphous materials, and eventually its mesoscopic description.

Instabilities in the plastic regime are observed at all scales and depend on the way the simulations are carried out. The simplest models, both at the atomistic and mesoscopic scales, do not lead to persistent shear bands. Indeed, in atomistic simulations, unless slow quench rates on MD timescales are used to prepare glasses, or strong interatomic bond directionality, the glass will only undergo transient localizations in the form of elementary shear bands. Similarly, at the mesoscale, we have seen that correlations and disorder can have different origins (dynamical and structural) and that a model without softening, i.e. without structural correlations, but with structural disorder (initial distribution of internal stresses or with stochastic yield stresses) show transient localization patterns, but no permanent shear bands. The other condition to form shear bands is when the glass contains no structural disorder, which explains the strain rate effect of shear banding: shear bands form only when the strain rate is slow enough that the rate of production of structural disorder (which increase with the strain rate) is slower than the relaxation rate of structural disorder. Shear bands are observed in relaxed glasses that exhibit softening, which requires to apply a slow quench at atomistic scale or to include softening effects at the mesoscopic scale. Another way to generate permanent shear bands is to run simulations on covalent systems with strong three-body interactions (bond directionality) at low pressure, because a constant pressure contributes to homogenize the deformation. A better understanding of softening and relaxations at the atomistic and meso-scales is a major issue, which requires to better understand the physics of STs (their nucleation criterion and relation to local softening or hardening). At the macroscopic scale, softening can be included in constitutive laws but the question is what is the relevant internal variable and its corresponding dynamics to represent faithfully the dynamics of STs and of their interactions and relation to softening. At the macroscopic scale, reproducing history-dependence may require to generalize the notions of effective temperature and free volume to transform them into tensorial quantities to reproduce anisotropy and the fact that the symmetry of the plastic strain tensor gets embedded in glass microstructure, leading to anelasticity. Less-relaxed glasses that present a larger density of nucleation sites for plasticity (although the nature of these sites remains elusive, as discussed above) have a lower strength but a higher ductility. This observation is at the basis of several strategies to produce ductile metallic glasses, such as prestraining \cite{zhang-natmat2006,lee-sm2008} or ion irradiation \cite{raghavan-sm2010}.

Finally, the question of the temperature dependence of the mechanical response is still largely an open question. Different regimes have been explored, especially the athermal regime and the mean-field regime at higher temperatures. But a relevant description of thermal processes and their competition with mechanical instabilities needs a deeper understanding of the evolution of activation barriers and memory effects in mechanically driven systems and the question of the competition between the dynamics of the numerically chosen thermostat and the local dynamics of the mechanical instabilities makes this question difficult to solve at the moment.

We end with a few open questions that we deem important to answer in order to make significant progress. At the atomic scale, can we identify a geometrical characteristic of weak zones in metallic glasses, more general than coordination defects or special structural environments like icosahedra? Related to temperature, is it possible to quantify properly the respective nucleation and diffusion from a nucleated center of plastic rearrangements? How does the size of the plastic rearrangements depend on the shear rate? What is responsible for the nucleation of large scale rearrangements like elementary shear bands, and do these large scale rearrangements resist thermal activation? Do these large scale events imply non-uniform temperatures, and local melting that would favor crack propagation? What are the characteristic timescales for structural and mechanical relaxations that must be included in mesoscopic models?
All these questions need a more realistic description of the local dissipative processes, that are strongly system-dependent. At the mesoscale, is it possible to find a characteristic length scale for self-consistent mesoscopic models? Which internal variable will account for the specific microscopic structure of the system (for instance, bond directionality), and corresponding relaxations? At the macro-scale, is it possible to derive a general constitutive law from microscopic modeling that can take into account the competition between shear and densification, with a complete description of the mechanical history? How does it depend on the specific composition of a glass? The above questions are but a few among a vast list of topics for future research that will benefit from the continuous progress in simulation techniques and experiments, and will hopefully lead to a general understanding and description of glassy mechanics consistent across length- and time-scales.

\section{References}

\bibliography{msmse}

\begin{thebibliography}{100}

\bibitem{albano-jcp2005}
F.~Albano and M.L. Falk.
\newblock Shear softening and structure in a simulated three-dimensional binary
  glass.
\newblock {\em J. Chem. Phys.}, 122:154508, 2005.

\bibitem{allen-book}
M.P. Allen and D.~Tildesley.
\newblock {\em Computer simulation of liquids}.
\newblock Clarendon Press, Oxford, 1987.

\bibitem{Anand-ActaMat08}
L.~Anand and D.~Hennann.
\newblock A constitutive theory for the mechanical response of amorphous metals
  at high temepratures spanning the glass transition temperature: application
  to microscale thermoplastic forming.
\newblock {\em Acta Mater.}, 56:3290--3305, 2008.

\bibitem{anand-jmps2005}
L.~Anand and C.~Su.
\newblock A theory for amorphous viscoplastic materials undergoing finite
  deformations, with application to metallic glasses.
\newblock {\em J. Mech. Phys. Solids}, 53:1362--1396, 2005.

\bibitem{anand-am2007}
L.~Anand and C.~Su.
\newblock A constitutive theory for metallic glasses at high homologous
  temperatures.
\newblock {\em Acta Mater.}, 55:3735--3747, 2007.

\bibitem{angelani-prl2000}
L.~Angelani, R.~Di~Leonardo, G.~Ruocco, A.~Scala, and F.~Sciortino.
\newblock Saddles in the energy landscape probed by supercooled liquids.
\newblock {\em Phys. Rev. Lett.}, 85:5356--5359, 2000.

\bibitem{angelani-pre2000}
L.~Angelani, G.~Parisi, G.~Ruocco, and G.~Viliani.
\newblock Potential energy landscape and long-time dynamics in a simple model
  glass.
\newblock {\em Phys. Rev. E}, 61:1681--1691, 2000.

\bibitem{appignanesi-prl2006}
G.A. Appignanesi, J.A. Rodriguez~Fris, R.A. Montani, and W.~Kob.
\newblock Democratic particle motion for metabasin transitions in simple glass
  formers.
\newblock {\em Phys. Rev. Lett.}, 96:057801, 2006.

\bibitem{argon-jap1968}
A.S. Argon.
\newblock Delayed elasticity in inorganic glasses.
\newblock {\em J. Appl. Phys.}, 39:4080--4086, 1968.

\bibitem{argon-am1979}
A.S. Argon.
\newblock Plastic deformation in metallic glasses.
\newblock {\em Acta Metall.}, 27:47--58, 1979.

\bibitem{argon-mse1979}
A.S. Argon and H.Y. Kuo.
\newblock Plastic flow in a disordered bubble raft.
\newblock {\em Mat. Sci. Eng.}, 39:101--109, 1979.

\bibitem{argon-jncs1980}
A.S. Argon and H.Y. Kuo.
\newblock Free energy spectra for inelastic deformation of five metallic glass
  alloys.
\newblock {\em J. Non-Cryst. Sol.}, 37:241--266, 1980.

\bibitem{ashby-sm2006}
M.F. Ashby and A.L. Greer.
\newblock Metallic glasses as structural materials.
\newblock {\em Scripta Mat.}, 54:321--326, 2006.

\bibitem{bacon-book}
D.~Bacon, Y.~Osetsky, and D.~Rodney.
\newblock Dislocation-obstacle interactions at the atomic level.
\newblock In J.~Hirth and L.~Kubin, editors, {\em Dislocations in Solids},
  volume~15, chapter~88, pages 1--90. Elsevier, 2009.

\bibitem{bailey-prb2004}
N.P. Bailey, J.~Schiotz, and K.W. Jacobsen.
\newblock Simulation of cu-mg metallic glass: Thermodynamics and structure.
\newblock {\em Phys. Rev. B}, 69:144205, 2004.

\bibitem{bailey-prl2007}
N.P. Bailey, J.~Schiotz, A.~Lema\^itre, and K.W. Jacobsen.
\newblock Avalanche size scaling in sheared three-dimensional amorphous solid.
\newblock {\em Phys. Rev. Lett.}, 98:095501, 2007.

\bibitem{Baldi-prl2009}
G.~Baldi, A.~Fontana, G.~Monaco, L.~Orsingher, S.~Rols, F.~Rossi, and B.~Ruta.
\newblock Connection between the boson peak and elastic properties in silicate
  glasses.
\newblock {\em Phys. Rev. Lett.}, 102:195502, 2009.

\bibitem{BVR-PRL02}
J.-C. Baret, D.~Vandembroucq, and S.~Roux.
\newblock An extremal model of amorphous plasticity.
\newblock {\em Phys. Rev. Lett.}, 89:195506, 2002.

\bibitem{barrat-review2010}
J.L Barrat and A.~Lema\^itre.
\newblock Heterogeneities in amorphous systems under shear.
\newblock In L.~Berthier, G.~Biroli, J.P. Bouchaud, L.~Cipelletti, and W.~van
  Saarloos, editors, {\em Dynamical Heterogeneities in Glasses, Colloids, and
  Granular Media}. Oxford University Press, 2010.

\bibitem{Bazant1996}
M.Z. Bazant and E.~Kaxiras.
\newblock Modeling of covalent bonding in solids by inversion of cohesive
  energy curves.
\newblock {\em Phys. Rev. Lett.}, 77:4370--4373, 1996.

\bibitem{EDIP1997}
M.Z. Bazant, E.~Kaxiras, and J.F. Justo.
\newblock Environment-dependent interatomic potential for bulk silicon.
\newblock {\em Phys. Rev. B}, 56:8542--8552, 1997.

\bibitem{berthier-jcp2002}
L.~Berthier and J.L. Barrat.
\newblock Nonequilibrium dynamics and fluctuation-dissipation relation in a
  sheared fluid.
\newblock {\em J. Chem. Phys.}, 116:6228--6242, 2002.

\bibitem{berthier-prl2002}
L.~Berthier and J.L. Barrat.
\newblock Shearing a glassy material: numerical test of nonequilibrium
  mode-coupling approaches and experimental proposals.
\newblock {\em Phys. Rev. Lett.}, 89:095702, 2002.

\bibitem{berthier-arXiv2010}
L.~Berthier and G.~Biroli.
\newblock A theoretical perspective on the glass transition and nonequilibrium
  phenomena in disordered materials.
\newblock arxiv:1011.2578, 2010.

\bibitem{kob-book}
K.~Binder and W.~Kob.
\newblock {\em Glassy materials and disordered solids: an introduction to their
  statistical mechanics}.
\newblock World Scientific, Singapore, 2005.

\bibitem{Bocquet-PRL09}
L.~Bocquet and A.~Ajdari.
\newblock Kinetic theory of plastic flow in soft glassy materials.
\newblock {\em Phys. Rev. Lett.}, 103:036001, 2009.

\bibitem{Bonamy-JPhysD09}
D.~Bonamy.
\newblock Intermittency and roughening in the failure of brittle heterogeneous
  materials.
\newblock {\em J. Phys. D}, 42:2114014, 2009.

\bibitem{broderix-prl2000}
K.~Broderix, K.K. Bhattacharya, A.~Cavagna, A.~Zippelius, and I.~Giardina.
\newblock Energy landscape of a lennard-jones liquid: statistics of stationary
  points.
\newblock {\em Phys. Rev. Lett.}, 85:5360--5364, 2000.

\bibitem{buchner-prl2000}
S.~B\"uchner and A.~Heuer.
\newblock Metastable states as a key to the dynamics of the supercooled
  liquids.
\newblock {\em Phys. Rev. Lett.}, 84:2168--2171, 2000.

\bibitem{BulatovArgon94a}
V.~V. Bulatov and A.~S. Argon.
\newblock A stochastic model for continuum elasto-plastic behavior. {I}.
  numerical approach and strain localization.
\newblock {\em Model. Simul. Mater. Sci. Eng.}, 2:167, 1994.

\bibitem{BulatovArgon94b}
V.~V. Bulatov and A.~S. Argon.
\newblock A stochastic model for continuum elasto-plastic behavior. {II}. a
  study of the glass transition and structural relaxation.
\newblock {\em Model. Simul. Mater. Sci. Eng.}, 2:185, 1994.

\bibitem{BulatovArgon94c}
V.~V. Bulatov and A.~S. Argon.
\newblock A stochastic model for continuum elasto-plastic behavior. {III}.
  plasticity in ordered versus disordered solids.
\newblock {\em Model. Simul. Mater. Sci. Eng.}, 2:203, 1994.

\bibitem{cances-jcp2009}
E.~Canc\`es, F.~Legoll, M.C. Marinica, K.~Minoukadeh, and F.~Willaime.
\newblock Some improvements of the activation-relaxation technique method for
  finding transition pathways on potential energy surfaces.
\newblock {\em J. Chem. Phys.}, 130:114711, 2009.

\bibitem{Carre2008}
A.~Carré, J.~Horbach, S.~Ispas, and W.~Kob.
\newblock New fitting scheme to obtain effective potential from car-parrinello
  molecular dynamics simulations: application to silica.
\newblock {\em Eur. Phys. Lett.}, 82:17001, 2008.

\bibitem{Champagnon-jncs2009}
B.~Champagnon, L.~Wondraczek, and T.~Deschamps.
\newblock Boson peak, structural inhomogeneity, light scattering and
  transparency of silicate glasses.
\newblock {\em J. Non-Cryst. Sol.}, 355:712, 2009.

\bibitem{charleux-jmr2006}
L.~Charleux, S.~Gravier, M.~Verdier, M.~Fivel, and J.J. Blandin.
\newblock Indentation plasticity of amorphous and partially crystallized
  metallic glasses.
\newblock {\em J. Mater. Res.}, 22:525--532, 2006.

\bibitem{caroli-prl2010}
J.~Chattoraj, C.~Caroli, and A.~Lema\^itre.
\newblock Universal, additive effect of temperature on the rheology of
  amorphous solids.
\newblock {\em Phys. Rev. Lett.}, 105:266001, 2010.

\bibitem{chen-apl2006}
G.L. Chen, X.J. Liu, X.D. Hui, H.K. Hou, K.F. Yao, C.T. Liu, and J.~Wadsworth.
\newblock Molecular dynamic simulations and atomic structures of amorphous
  materials.
\newblock {\em Appl. Phys. Lett.}, 88:203115, 2006.

\bibitem{cheng-am2009}
Y.Q. Cheng, A.J. Cao, and E.~Ma.
\newblock Correlation between the elastic modulus and the intrinsic plastic
  behavior of metallic glasses: the roles of atomic configuration and alloy
  composition.
\newblock {\em Acta Mater.}, 57:3253--3267, 2009.

\bibitem{cheng-am2008}
Y.Q. Cheng, A.J. Cao, H.W. Sheng, and E.~Ma.
\newblock Local order influences initiation of plastic flow in metallic glass:
  effect of alloy composition and sample cooling history.
\newblock {\em Acta Mater.}, 56:5263--5275, 2008.

\bibitem{cloitre-prl2003}
M.~Cloitre, R.~Borrega, F.~Monti, and L.~Leibler.
\newblock Glassy dynamics and flow properties of soft colloidal pastes.
\newblock {\em Phys. Rev. Lett.}, 90:068303, 2003.

\bibitem{coslovich-jcp2007}
D.~Coslovich and G.~Pastore.
\newblock Understanding fragility in supercooled lennard-jones mixtures. i.
  locally preferred structures.
\newblock {\em J. Chem. Phys.}, 127:124504, 2007.

\bibitem{coussot-2002}
P.~Coussot and J.L. Grossiord, editors.
\newblock {\em Understanding rheology - From blood circulation to concrete
  hardening}, Paris, 2002. EDP Sciences.

\bibitem{coussot-epje2010}
P.~Coussot and G.~Ovarlez.
\newblock Physical origin of shear-banding in jammed systems.
\newblock {\em Eur. Phys. J. E}, 33:183--188, 2010.

\bibitem{coussot-prl2002}
P.~Coussot, J.S. Raynaud, F.~Bertrand, P.~Moucheront, J.P. Guilbaud, H.T.
  Huynh, S.~Jarny, and D.~Lesueur.
\newblock Coexistence of liquid and solid phases in flowing soft-glassy
  materials.
\newblock {\em Phys. Rev. Lett.}, 88:218301, 2002.

\bibitem{cugliandolo-pre1997}
L.F. Cugliandolo, J.~Kurchan, and L.~Peliti.
\newblock Energy flow, partial equilibration, and effective temperatures in
  systems with slow dynamics.
\newblock {\em Phys. Rev. E}, 55:3898--3914, 1997.

\bibitem{Dahmen-PRE98}
K.~Dahmen, D.~Erta\c{s}, and Y.~Ben-Zion.
\newblock Gutemberg-richter and characteristic earthquake behavior
  insimple-mean-field models of heterogeneous faults.
\newblock {\em Phys. Rev. E}, 58:1494--1501, 1998.

\bibitem{Dahmen-PRL09}
K.~A. Dahmen, Y.~Ben-Zion, and J.~T. Uhl.
\newblock Micromechanical model for deformation in solids with universal
  predictions for stress-strain curves and slip avalanches.
\newblock {\em Phys. Rev. Lett.}, 102:175501, 2009.

\bibitem{thorpe-ac2010}
A.M.R. de~Graff and M.F. Thorpe.
\newblock The long-wavelength limit of the structure factor of amorphous
  silicon and vitreous silica.
\newblock {\em Acta Cryst.}, A66:22, 2010.

\bibitem{debenedetti-nature2001}
P.G. Debenedetti and F.H. Stillinger.
\newblock Supercooled liquids and the glass transition.
\newblock {\em Nature}, 410:259--267, 2001.

\bibitem{delogu-prl2008}
F.~Delogu.
\newblock Indentification and characterization of potential shear
  transformation zones in metallic glasses.
\newblock {\em Phys. Rev. Lett.}, 100:255901, 2008.

\bibitem{demkowicz-prl2004}
M.J. Demkowicz and A.S. Argon.
\newblock High-density liquidlike component facilitates plastic flow in a model
  amorphous silicon system.
\newblock {\em Phys. Rev. Lett.}, 93:025505, 2004.

\bibitem{demkowicz-prb2005}
M.J. Demkowicz and A.S. Argon.
\newblock Autocatalytic avalanches of unit inelastic shearing events are the
  mechanism of plastic deformation in amorphous silicon.
\newblock {\em Phys. Rev. B}, 72:245206, 2005.

\bibitem{denny-prl2003}
R.A. Denny, D.R. Reichman, and J.P. Bouchaud.
\newblock Trap model and slow dynamics in supercooled liquids.
\newblock {\em Phys. Rev. Lett.}, 90:025503, 2003.

\bibitem{devincre-science2008}
B.~Devincre, T.~Hoc, and L.~Kubin.
\newblock Dislocation mean free paths and strain hardening of crystals.
\newblock {\em Science}, 320:1745--1748, 2008.

\bibitem{Dimiduk-Science06}
D.M. Dimiduk, C.~Woodward, and R.~LeSarand~M.D. Uchic.
\newblock Scale free intermittent flow in crystal plasticity.
\newblock {\em Science}, 26:1188--1190, 2006.

\bibitem{doliwa-pre2003b}
B.~Doliwa and A.~Heuer.
\newblock Energy barriers and activated dynamics in a supercooled lennard-jones
  liquid.
\newblock {\em Phys. Rev. E}, 67:031506, 2003.

\bibitem{doliwa-pre2003a}
B.~Doliwa and A.~Heuer.
\newblock Hopping in a supercooled lennard-jones liquid: metabasins, waiting
  time distribution, and diffusion.
\newblock {\em Phys. Rev. E}, 67:030501, 2003.

\bibitem{doliwa-prl2003}
B.~Doliwa and A.~Heuer.
\newblock What does the potential energy landscape tell us about the dynamics
  of supercooled liquids and glasses?
\newblock {\em Phys. Rev. Lett.}, 91:235501, 2003.

\bibitem{donati-prl1998}
C.~Donati, J.F. Douglas, W.~Kob, S.J. Plimpton, P.H. Poole, and S.C. Glotzer.
\newblock Stringlike cooperative motion in a supercooled liquid.
\newblock {\em Phys. Rev. Lett.}, 80:2338--2342, 1998.

\bibitem{donati-pre1999}
C.D. Donati, S.C. Glotzer, P.H. Poole, W.~Kob, and S.J. Plimpton.
\newblock Spatial correlations of mobility and immobility in a glass-forming
  lennard-jones liquid.
\newblock {\em Phys. Rev. E}, 60:3107, 1999.

\bibitem{doye-jcp2002}
J.P.K. Doye and D.J. Wales.
\newblock Saddle points and dynamics of lennard-jones clusters, solids and
  supercooled liquids.
\newblock {\em J. Chem. Phys.}, 116:3777--3788, 2002.

\bibitem{duan-prb2005}
G.~Duan, D.~Xu, Q.~Zhang, G.~Zhang, T.~Cagin, W.L. Johnson, and W.A.
  Goddard~III.
\newblock Molecular dynamics study of the binary cu46zr54 metallic glass
  motivated by experiments: Glass formation and atomic-level structure.
\newblock {\em Phys. Rev. B}, 71:224208, 2005.

\bibitem{Duval-jncs2006}
A.~Duval, S.~Etienne, G.~Simeoni, and A.~Mermet.
\newblock Physical aging effect on the boson peak and heterogeneous
  nanostructure of a silicate glass.
\newblock {\em J. Non-Cryst. Sol.}, 352:4525, 2006.

\bibitem{Duval-prb2007}
A.~Duval, A.~Mermet, and L.~Saviot.
\newblock Boson peak and hybridization of acoustic modes with vibrations of
  nanometric heterogeneities in glasses.
\newblock {\em Phys. Rev. B}, 75:024201, 2007.

\bibitem{dyre-prl1987}
J.C. Dyre.
\newblock Master-equation approach to the glass transition.
\newblock {\em Phys. Rev. Lett.}, 58:792--795, 1987.

\bibitem{ediger-jcp1996}
M.D. Ediger, C.A. Angell, and S.R. Nagel.
\newblock Supercooled liquids and glasses.
\newblock {\em J. Chem. Phys.}, 100:13200--13212, 1996.

\bibitem{egami-jncs1995}
T.~Egami, W.~Dmowski, P.~Kosmetatos, M.~Boord, T.~Tomida, E.~Oikawa, and
  A.~Inoue.
\newblock Deformation induced bond orientational order in metallic glasses.
\newblock {\em J. Non-Cryst. Sol.}, 192\&193:591--594, 1995.

\bibitem{ehmler-prl1998}
H.~Ehmler, A.~Heesemann, K.~R\"atzke, and F.~Faupel.
\newblock Mass dependence of diffusion in a supercooled metallic melt.
\newblock {\em Phys. Rev. Lett.}, 80:4919--4922, 1998.

\bibitem{Eshelby57}
J.~D. Eshelby.
\newblock The determination of the elastic field of an ellipsoidal inclusion,
  and related problems.
\newblock {\em Proc. Roy. Soc. A}, 241:376, 1957.

\bibitem{Fabiani-jchemphys2008}
E.~Fabiani, A.~Fontana, and U.~Buchenau.
\newblock Neutron scattering study of the vibrations in silica and germania.
\newblock {\em J. Chem. Phys.}, 128:244507, 2008.

\bibitem{falk-pre1998}
M.L. Falk and J.S. Langer.
\newblock Dynamics of viscoplastic deformation in amorphous solids.
\newblock {\em Phys. Rev. E}, 57:7192--7205, 1998.

\bibitem{falk-epj2010}
M.L. Falk and C.E. Maloney.
\newblock Simulating the mechanical response of amorphous solids using
  atomistic methods.
\newblock {\em Eur. Phys. J. B}, 75:405--413, 2010.

\bibitem{Fielding-SM09}
S.~M. Fielding, M.~E. Cates, and P.~Sollich.
\newblock Shear banding, aging and noise dynamics in soft glassy materials.
\newblock {\em Soft Matter}, 5:2378--2382, 2009.

\bibitem{Dahmen-PRL97}
D.~S. Fisher, K.~Dahmen, and Y.~Ben-Zion.
\newblock Statistics of earthquakes in simple models of heterogeneous faults.
\newblock {\em Phys. Rev. Lett.}, 78:4885--4888, 1997.

\bibitem{Flores-ActaMat01}
K.~M Flores and R.~H. Dauskardt.
\newblock Mean stress effects on flow localization and failure in a bulk
  metallic glass.
\newblock {\em Acta Mater.}, 49:2527--2537, 2001.

\bibitem{frenkel-2002}
D.~Frenkel and B.~Smith.
\newblock {\em Understanding molecular simulation: from algorithms to
  applications}.
\newblock Academic Press, San Diego, 2002.

\bibitem{fusco-pre2010}
C.~Fusco, T.~Albaret, and A.~Tanguy.
\newblock Role of local order in the small-scale plasticity of model amorphous
  materials.
\newblock {\em Phys. Rev. E}, 82:066116, 2010.

\bibitem{ganesh-prb2006}
P.~Ganesh and M.~Widom.
\newblock Signature of nearly icosahedral structures in liquid and supercooled
  liquid copper.
\newblock {\em Phys. Rev. B}, 74:134205, 2006.

\bibitem{ganesh-prb2008}
P.~Ganesh and M.~Widom.
\newblock Ab initio simulations of geometrical frustration in supercooled
  liquid fe and fe-based metallic glass.
\newblock {\em Phys. Rev. B}, 77:014205, 2008.

\bibitem{YFGao-MSMSE06}
Y.~F. Gao.
\newblock An implicit finite element method for simulating inhomogeneous
  deformation and shear bands of amorphous alloys based on the free-volume
  model.
\newblock {\em Model. Simul. Mater. Sci. Eng.}, 14:1329--1345, 2006.

\bibitem{YFGao-MSMSE04}
Y.~F. Gao and A.~F. Bower.
\newblock A simple technique for avoiding convergebce problem in finite element
  simulations of crack nucleation and growth on cohesive interfaces.
\newblock {\em Model. Simul. Mater. Sci. Eng.}, 12:453--463, 2006.

\bibitem{Pizzagalli2003}
J.~Godet, L.~Pizzagalli, S.~Brochard, and P.~Beauchamp.
\newblock Comparison between classical potentials and ab initio moethods for
  silicon under large shear.
\newblock {\em J. Phys.: Cond. Mat.}, 15:6943--6953, 2003.

\bibitem{goldstein-jcp1969}
M.~Goldstein.
\newblock Viscous liquids and the glass transition: a potential energy barrier
  picture.
\newblock {\em J. Chem. Phys.}, 51:3728--3739, 1969.

\bibitem{gotze-rpp1992}
W.~G\"otze and L.~Sj\"ogren.
\newblock Relaxation processes in supercooled liquids.
\newblock {\em Rep. Prog. Phys.}, 55:241--376, 1992.

\bibitem{gu-apl2008}
X.J. Gu, S.J. Poon, G.J. Shiflet, and M.~Widom.
\newblock Mechanical properties, glass transition temperature, and bond
  enthalpy trends of high metalloid fe-based bulk metallic glasses.
\newblock {\em Appl. Phys. Lett.}, 92:161910, 2008.

\bibitem{guckenheimer-book}
J.~Guckenheimer and P.~Homes.
\newblock {\em Nonlinear oscillations, dynamical systems, and bifurcation of
  vector fields, 3rd Ed.}
\newblock Springer-Verlag, New York, 1997.

\bibitem{han-pre2007}
X.J. Han and H.~Teichler.
\newblock Liquid-to-glass transition in bulk glass-forming cu60ti20zr20 alloy
  by molecular dynamics simulations.
\newblock {\em Phys. Rev. E}, 75:061501, 2007.

\bibitem{harmon-prl2007}
J.S. Harmon, M.D. Demetriou, W.L. Johnson, and K.~Samwer.
\newblock Anelastic to plastic transition in metallic glass-forming liquids.
\newblock {\em Phys. Rev. Lett.}, 99:135502, 2007.

\bibitem{Hauch1999}
J.A. Hauch, D.~Holland, M.P. Marder, and H.L. Swinney.
\newblock Dynamic fracture in single crystal silicon.
\newblock {\em Phys. Rev. Lett.}, 82:3823--3826, 1999.

\bibitem{haxton-prl2007}
T.K. Haxton and A.J. Liu.
\newblock Activated dynamics and effective temperature in a steady state
  sheared glass.
\newblock {\em Phys. Rev. Lett.}, 99:195701, 2007.

\bibitem{hebraud-prl1998}
P.~H\'ebraud and F.~Lequeux.
\newblock Mode-coupling theory for the pasty rheology of soft glassy materials.
\newblock {\em Phys. Rev. Lett.}, 81:2934--2937, 1998.

\bibitem{heilmaier-jmpt2001}
M.~Heilmaier.
\newblock Deformation behavior of zr-based metllic glasses.
\newblock {\em J. Mat. Proc. Tech.}, 117:374--380, 2001.

\bibitem{herschel-1926}
W.H. Herschel and R.~Bulkley.
\newblock Konsistenzmessungen von gummi-benzollosungen.
\newblock {\em Kolloid Zeitschrift}, 39:291, 1926.

\bibitem{heuer-prl1997}
A.~Heuer.
\newblock Properties of a glass-forming system as derived from its potential
  energy landscape.
\newblock {\em Phys. Rev. Lett.}, 78:4051--4055, 1997.

\bibitem{heuer-jpcm2008}
A.~Heuer.
\newblock Exploring the potential energy landscape of glass-forming systems:
  from inherent structures via metabasins to macroscopic transport.
\newblock {\em J. Phys.: Cond. Mat.}, 20:373101, 2008.

\bibitem{hirth-1982}
J.P. Hirth and J.~Lothe.
\newblock {\em Theory of dislocations}.
\newblock Wiley, New York, 1982.

\bibitem{homer-prb2010}
E.R. Homer, D.~Rodney, and C.A. Schuh.
\newblock Kinetic monte carlo study of activated states and correlated
  shear-transformation-zone activity during the deformation of an amorphous
  metal.
\newblock {\em Phys. Rev. E}, 81:064204, 2010.

\bibitem{homer-am2009}
E.R. Homer and C.A. Schuh.
\newblock Mesoscale modeling of amorphous metals by shear transformation zone
  dynamics.
\newblock {\em Acta Mater.}, 57:2823--2833, 2009.

\bibitem{homer-msmse2010}
E.R. Homer and C.A. Schuh.
\newblock Three-dimensional shear transformation zone dynamics model for
  amorphous metals.
\newblock {\em Model. Simul. Mater. Sci. Eng.}, 18:065009, 2010.

\bibitem{Huang2003}
L.~Huang and J.~Kieffer.
\newblock Molecular dynamics study of cristobalite silica using a charge
  transfer three-body potential: phase transformation and structural disorder.
\newblock {\em J. Chem. Phys.}, 118:1487--1498, 2003.

\bibitem{Nix-JMPS02}
R.~Huang, Z.~Suo, J.~H. Prevost, and W.~D. Nix.
\newblock Inhomogeneous deformation in metallic glasses.
\newblock {\em J. Mech. Phys. Solids}, 50:1011--1027, 2002.

\bibitem{ilg-epl2007}
P.~Ilg and J.L. Barrat.
\newblock Driven activation vs. thermal activation.
\newblock {\em Eur. Phys. Lett.}, 79:26001, 2007.

\bibitem{inoue-am2000}
A.~Inoue.
\newblock Stabilization of metallic supercooled liquid and bulk amorphous
  alloys.
\newblock {\em Acta Mater.}, 48:279--306, 2000.

\bibitem{Ispas2002}
S.~Ispas, M.~Benoit, P.~Jund, and R.~Jullien.
\newblock Structural properties of glassy and liquid tetrasilicate, comparison
  between ab initio and classical molecular-dynamics simulations.
\newblock {\em J. Non-Cryst. Sol.}, 307-310:946--955, 2002.

\bibitem{Ispas2005}
S.~Ispas, N.~Zotov, S.~de~Wispaleare, and W.~Kob.
\newblock Vibrational properties of a sodium tetrasilicate glass: ab initio
  versus classical force field.
\newblock {\em J. Non-Cryst. Sol.}, 351:1144--1150, 2005.

\bibitem{Jagla-PRE07}
E.~A. Jagla.
\newblock Strain localization driven by structural relaxation in sheared
  amorphous solids.
\newblock {\em Phys. Rev. E}, 76:046119, 2007.

\bibitem{jakse-prl2003}
N.~Jakse and A.~Pasturel.
\newblock Local order of liquid and supercooled zirconium by ab initio
  molecular dynamics.
\newblock {\em Phys. Rev. Lett.}, 91:195501, 2003.

\bibitem{Rouxel-ScriptaMat06}
Hui Ji, V.~Keryvin, T.~Rouxel, and T.~Hammouda.
\newblock Densification of window glass under very high pressure and its
  relevance to vickers indentation.
\newblock {\em Scripta Mat.}, 55:1159--1162, 2006.

\bibitem{johari-jcp1970}
G.P. Johari and M.~Goldstein.
\newblock Viscous liquids and the glass transition. ii. secondary relaxations
  in glasses of rigid molecules.
\newblock {\em J. Chem. Phys.}, 53:2372--2389, 1970.

\bibitem{johnson-mrs1999}
W.L. Johnson.
\newblock Bulk glass-forming metallic alloys: science and technology.
\newblock {\em Mater. Res. Soc. Bull.}, 24:42--56, 1999.

\bibitem{johnson-prl2005}
W.L. Johnson and K.~Samwer.
\newblock A universal criterion for plastic yielding of metallic glasses with a
  $(t/tg)^{3/2}$ temperature dependence.
\newblock {\em Phys. Rev. Lett.}, 95:1995501, 2005.

\bibitem{jonsson-prl1988}
H.~J\'onsson and H.C. Andersen.
\newblock Isosahedral ordering in the lennard-jones liquid and glass.
\newblock {\em Phys. Rev. Lett.}, 60:2295--2298, 1988.

\bibitem{Kardar-PR98}
M.~Kardar.
\newblock Nonequilibrium dynamics of interfaces and lines.
\newblock {\em Phys. Rep.}, 301:85--112, 1998.

\bibitem{procaccia-pre2010}
S.~Karmakar, E.~Lerner, and I.~Procaccia.
\newblock Statistical physics of elasto-plastic steady states in amorphous
  solids.
\newblock {\em Phys. Rev. E}, 82:055103, 2010.

\bibitem{kawamura-apl2001}
Y.~Kawamura and A.~Inoue.
\newblock Newtonian viscosity of supercooled liquid in a pd40ni40p20 metallic
  glass.
\newblock {\em Appl. Phys. Lett.}, 76:1114--1116, 2001.

\bibitem{kawasaki-prl2007}
T.~Kawasaki, T.~Araki, and H.~Tanaka.
\newblock Correlation between dynamic heterogeneity and medium-range order in
  two-dimensional glass-forming liquids.
\newblock {\em Phys. Rev. Lett.}, 99:215701, 2007.

\bibitem{kazimirov-prb2008}
V.Y. Kazimirov, D.~Louca, M.~Widom, X.J. Gu, S.J. Poon, and G.J. Shiflet.
\newblock Local organization and atomic clustering in multicomponent amorphous
  steels.
\newblock {\em Phys. Rev. B}, 78:054112, 2008.

\bibitem{KBVD-ActaMat08}
G.~Kermouche, E.~Barthel, D.~Vandembroucq, and P.~Dubujet.
\newblock Mechanical modelling of indentation-induced densification in
  amorphous silica.
\newblock {\em Acta Mater.}, 56:3222--3228, 2008.

\bibitem{Khonik-jap2009}
V.~Khonik, Y.P. Mitrofanov, S.A. Lyakhov, D.A. Khoviv, and R.A. Kouchakov.
\newblock Recovery of structural relaxation in aged metallic glass as
  determined by high-precision in situ shear modulus measurements.
\newblock {\em J. Appl. Phys.}, 105:123521, 2009.

\bibitem{klement-nature1960}
W.~Klement, R.H. Willens, and P.~Duwez.
\newblock Noncrystalline structure in solidified gold-silicon alloys.
\newblock {\em Nature}, 187:869--870, 1960.

\bibitem{kob-pre1995}
W.~Kob and H.C. Andersen.
\newblock Testing mode-coupling theory for a supercooled binary lennard-jones
  mixture: the van hove correlation function.
\newblock {\em Phys. Rev. E}, 51:4626--4641, 1995.

\bibitem{kob-prl1997}
W.~Kob, C.~Donati, S.J. Plimpton, P.H. Poole, and S.C. Glotzer.
\newblock Dynamical heterogeneities in a supercooled lennard-jones liquid.
\newblock {\em Phys. Rev. Lett.}, 79:2827--2831, 1997.

\bibitem{kobayashi-jpsj1980}
S.~Kobayashi, K.~Maeda, and S.~Takeuchi.
\newblock Computer simulation of atomic structure of cu57zr43 amorphous alloy.
\newblock {\em J. Phys. Soc. Japan}, 48:1147--1152, 1980.

\bibitem{kobayashi-am1980}
S.~Kobayashi, K.~Maeda, and S.~Takeuchi.
\newblock Computer simulation of deformation of amorphous cu57zr43.
\newblock {\em Acta Metall.}, 28:1641--1652, 1980.

\bibitem{kushima-jcp2009}
A.~Kushima, X.~Lin, J.~Li, J.~Eapen, J.C. Mauro, X.~Qian, P.~Diep, and S.~Yip.
\newblock Computing the viscosity of supercooled liquids.
\newblock {\em J. Chem. Phys.}, 130:224504, 2009.

\bibitem{lacks-prl2001}
D.J. Lacks.
\newblock Energy landscape and the non-newtonian viscoity of liquids and
  glasses.
\newblock {\em Phys. Rev. Lett.}, 87:225502, 2001.

\bibitem{Lambropoulos-JACS96}
J.~C. Lambropoulos, S.~Xu, and T.~Fang.
\newblock Constitutive law for the densification of fused silica, with
  applications in polishing and microgrinding.
\newblock {\em J. Am. Ceram. Soc.}, 79:1441--1452, 1996.

\bibitem{lancon-epl1986}
F.~Lan\c{c}on, L.~Billard, and P.~Chaudhari.
\newblock Thermodynamical properties of a two-dimensional quasicrystal from
  molecular dynamics calculations.
\newblock {\em Europhys. Lett.}, 2:625--629, 1986.

\bibitem{lauridsen-prl2003}
J.~Lauridsen, G.~Chanan, and M.~Dennin.
\newblock Velocity profiles in slowly sheared bubble rafts.
\newblock {\em Phys. Rev. Lett.}, 93:018303, 2004.

\bibitem{Lee-ActaMat05}
J.~C. Lee, Y.~C. Kim, J.~P. Ahn, and H.~S. Kim.
\newblock Enhanced plasticity in bulk amorphous matrix composites: macroscopic
  and microscopic viewpoint studies.
\newblock {\em Acta Mater.}, 53:129--139, 2005.

\bibitem{lee-sm2008}
S.C. Lee, C.M. Lee, J.W. Yang, and J.C. Lee.
\newblock Microstructural evolution of an elastostatically compressed amorphous
  alloy and its influence on the mechanical properties.
\newblock {\em Scripta Mat.}, 58:591--594, 2008.

\bibitem{lemaitre-prl2009}
A.~Lema\^itre and C.~Caroli.
\newblock Rate-dependent avalanche size in athermally sheared amorphous solids.
\newblock {\em Phys. Rev. Lett.}, 065501:2009, 103.

\bibitem{Lemaitre-preprint06}
A.~Lema{\^i}tre and C.~Caroli.
\newblock Dynamical noise and avalanches in quasi-static flow of amorphous
  materials.
\newblock arxiv:0609689v1, 2006.

\bibitem{lemaitre-pre2007}
A.~Lema\^itre and C.~Caroli.
\newblock Plastic response of a two-dimensional amorphous solid to quasistatic
  shear: transverse particle diffusion and phenomenology of dissipative events.
\newblock {\em Phys. Rev. E}, 76:036104, 2007.

\bibitem{caroli-pre2007}
A.~Lema\^itre and C.~Caroli.
\newblock Plastic response of a two-dimensional amorphous solid to quasistatic
  shear: Transverse particle diffusion and phenomenology of dissipative events.
\newblock {\em Phys. Rev. E}, 76:036104, 2007.

\bibitem{leonforte-phd}
F.~L\'eonforte.
\newblock {\em Vibrations et microm\'ecanique de mat\'eriaux amorphes
  mod\`{e}les}.
\newblock PhD thesis, Universit\'e Lyon 1, 2006.

\bibitem{Tanguy-PRB05}
F.~Leonforte, R.~Boissiere, A.~Tanguy, J.~P. Wittmer, and J.-L. Barrat.
\newblock Continuum limit of amorphous elastic bodies iii: three-dimensional
  systems.
\newblock {\em Phys. Rev. B}, 72:224206, 2005.

\bibitem{Tanguy-PRL06}
F.~Leonforte, A.~Tanguy, and J.-L. Barrat.
\newblock Inhomogeneous elastic response of silica glass.
\newblock {\em Phys. Rev. Lett.}, 97:055501, 2006.

\bibitem{Tanguy-PRB02}
F.~Leonforte, A.~Tanguy, J.~P. Wittmer, and J.-L. Barrat.
\newblock Continuum limit of amorphous elastic bodies i: a finite-size study of
  low-frequency harmonic vibrations.
\newblock {\em Phys. Rev. B}, 66:174205, 2002.

\bibitem{Leschhorn-AnnPhys97}
H.~Leschhorn, T.~Nattermann, S.~Stepanow, and L.H. Tang.
\newblock Driven interface depinning in a disordered medium.
\newblock {\em Ann. Phys. (Leipzig)}, 6:1, 1997.

\bibitem{levelut-jac2007}
C.~Levelut, R.~Le~Parc, A.~Faivre, R.~Bruning, B.~Champagnon, V.~Martinez, J.P.
  Simon, F.~Bley, and J.L. Hazemann.
\newblock Density fluctuations in oxide glasses investigated by small-angle
  x-ray scattering.
\newblock {\em J. Appl. Cryst.}, 40:S512--S516, 2007.

\bibitem{lewandowski-nmat2006}
J.J. Lewandowski and A.L. Greer.
\newblock Temperature rise at shear bands in metallic glasses.
\newblock {\em Nature Mat.}, 5:15--18, 2006.

\bibitem{Lowhaphandu-ScriptaMat99}
P.~Lowhaphandu, S.~L Montgomery, and J.~J. Lewandowski.
\newblock Effects of superimposed hydrostatic pressure on flow and fracture of
  a zr–ti–ni–cu–be bulk amorphous alloy.
\newblock {\em Scripta Mat.}, 41:19--24, 1999.

\bibitem{lund-am2003}
A.C. Lund and C.A. Schuh.
\newblock Yield surface of a simulated metallic glass.
\newblock {\em Acta Mater.}, 51:5399--5411, 2003.

\bibitem{madec-science2003}
R.~Madec, B.~Devincre, L.~Kubin, T.~Hoc, and D.~Rodney.
\newblock The role of collinear interaction in dislocation-induced hardening.
\newblock {\em Science}, 301:1879--1882, 2003.

\bibitem{malandro-prl1998}
D.L. Malandro and D.J. Lacks.
\newblock Molecular-level mechanical instabilities and enhanced self-diffusion
  in flowing liquids.
\newblock {\em Phys. Rev. Lett.}, 81:5576--5580, 1998.

\bibitem{malandro-jcp1999}
D.L. Malandro and D.J. Lacks.
\newblock Relationships of shear-induced changes in the potential energy
  landscape to the mechanical properties of ductile glasses.
\newblock {\em J. Chem. Phys.}, 110:4593--4601, 1999.

\bibitem{malek-pre2000}
R.~Malek and N.~Mousseau.
\newblock Dynamics of lennard-jones clusters: a characterization of the
  activation-relaxation technique.
\newblock {\em Phys. Rev. E}, 62:7723, 2000.

\bibitem{maloney-prl2004b}
C.~Maloney and A.~Lema\^itre.
\newblock Subextensive scaling in the athermal, quasistatic limit of amorphous
  matter in plastic shear flow.
\newblock {\em Phys. Rev. Lett.}, 93:016001, 2004.

\bibitem{maloney-prl2004a}
C.~Maloney and A.~Lema\^itre.
\newblock Universal breakdown of elasticity at the onset of material failure.
\newblock {\em Phys. Rev. Lett.}, 93:195501, 2004.

\bibitem{maloney-pre2006}
C.E. Maloney and A.~Lema\^itre.
\newblock Amorphous systems in athermal, quasistatic shear.
\newblock {\em Phys. Rev. E}, 74:016118, 2006.

\bibitem{maloney-prl2009}
C.E. Maloney and M.O. Robbins.
\newblock Anisotropic power law strain correlations in sheared amorphous 2d
  solids.
\newblock {\em Phys. Rev. Lett.}, 102:225502, 2009.

\bibitem{Manning-PRE09}
M.L. Manning, E.G. Daub, J.S. Langer, and J.M. Carlson.
\newblock Rate-dependent shear bands in a shear-transformation-zone model of
  amorphous solids.
\newblock {\em Phys. Rev. E}, 79:016110, 2009.

\bibitem{manning-pre2007}
M.L. Manning, J.S. Langer, and J.M. Carlson.
\newblock Strain localization in a shear transformation zone model for
  amorphous solids.
\newblock {\em Phys. Rev. E}, 76:056106, 2007.

\bibitem{Manning-PRE07}
M.L. Manning, J.S. Langer, and J.M. Carlson.
\newblock Strain localization in a shear transformation zone model for
  amorphous solids.
\newblock {\em Phys. Rev. E}, 76:056106, 2007.

\bibitem{Mantisi-jpcm2010}
B.~Mantisi, S.~Adichtchev, S.~Sirotkin, L.~Rafaelly, L.~Wondraczek, H.~Behrens,
  N.V. Surovtsev, A.~Pillonnet, E.~Duval, B.~Champagnon, and A.~Mermet.
\newblock Non-debye normalization of the glass vibrational density of states in
  mildly densified silicate glasses.
\newblock {\em J. Phys.: Cond. Mat.}, 22:025402, 2010.

\bibitem{martens-prl2011}
K.~Martens, L.~Bocquet, and J.L. Barrat.
\newblock Connecting diffusion and dynamical heterogeneities in actively
  deformed amorphous systems.
\newblock {\em Phys. Rev. Lett.}, 106:156001, 2011.

\bibitem{mayr-prl2006}
S.G. Mayr.
\newblock Activation energy of shear transformation zones: a key for
  understanding rheology of glasses and liquids.
\newblock {\em Phys. Rev. Lett.}, 97:195501, 2006.

\bibitem{mendelev-pm2009}
M.I. Mendelev, M.J. Kramer, R.T. Ott, D.J. Sordelet, D.~Yagodin, and P.~Popel.
\newblock Development of suitable interatomic potentials for simulation of liq
  and amorphous cu-zr alloys.
\newblock {\em Philos. Mag.}, 89:967--987, 2009.

\bibitem{mendelev-jap2007a}
M.I. Mendelev, D.J. Sordelet, and M.J. Kramer.
\newblock Using atomistic computer simulations to analyze x-ray diffraction
  data from metallic glasses.
\newblock {\em J. Appl. Phys.}, 102:043501, 2007.

\bibitem{Micoulaut2006}
M.~Micoulaut, Y.~Guissani, and B.~Guillot.
\newblock Simulated structural and thermal properties of glassy and liquid
  germania.
\newblock {\em Phys. Rev. E}, 73:031504, 2006.

\bibitem{middleman-1962}
S.~Middleman.
\newblock {\em The flow of high polymers}.
\newblock Academic Press, New York, 1962.

\bibitem{middleton-prb2001}
T.F. Middleton and D.J. Wales.
\newblock Energy landscape of some model glass formers.
\newblock {\em Phys. Rev. B}, 64:024205, 2001.

\bibitem{Miguel-Nat01}
M.C. Miguel, A.~Vespignani, S.~Zapperi, J.~Weiss, and J.-R. Grasso.
\newblock Intermittent dislocation flow in viscoplastic deformation.
\newblock {\em Nature}, 410:667--671, 2001.

\bibitem{miller-jmps2008}
R.E. Miller and D.~Rodney.
\newblock On the nonlocal nature of dislocation nucleation during
  nanoindentation.
\newblock {\em J. Mech. Phys. Solids}, 56:1203--1223, 2008.

\bibitem{miller-msmse2009}
R.E. Miller and E.B. Tadmor.
\newblock A unified framework and performance benchmark of fourteen multiscale
  atomistic/continuum coupling methods.
\newblock {\em Model. Simul. Mater. Sci. Eng.}, 17:053001, 2009.

\bibitem{mokshin-pre2008}
A.~Mokshin and J.L. Barrat.
\newblock Shear induced crystallization of an amorphous system.
\newblock {\em Phys. Rev. E}, 77:021505, 2008.

\bibitem{Amonaco-prl2006}
A.~Monaco, A.I. Chumakov, G.~Monaco, W.A. Crichton, A.~Meyer, L.~Comez,
  D.~Fioretto, J.~Korecki, and R.~Rüffer.
\newblock Effect of densification on the density of vibrational states of
  glasses.
\newblock {\em Phys. Rev. Lett.}, 97:135501, 2006.

\bibitem{Fielding-PRL11}
R.~L. Moorcroft, M.~E. Cates, and S.~M. Fielding.
\newblock Age-dependent transient shear banding in soft glasses.
\newblock {\em Phys. Rev. Lett.}, 106:055502, 2011.

\bibitem{Muskhelishvili}
N.I. Muskhelishvili.
\newblock {\em Some basic problems of the mathematical theory of elasticity}.
\newblock Kluwer, Groningen, 1953.

\bibitem{Niss-prl2009}
K.~Niss, B.~Begen, B.~Frick, J.~Ollivier, A.~Beraud, A.~Sokolov, and
  S.~Alba-Simionesco.
\newblock Influence of pressure on the boson peak: stronger than elastic medium
  transformation.
\newblock {\em Phys. Rev. Lett.}, 99:055502, 2007.

\bibitem{nogaret-jnm2008}
T.~Nogaret, D.~Rodney, M.~Fivel, and C.~Robertson.
\newblock Clear band formation simulated by dislocation dynamics: role of
  helical turns and pile-ups.
\newblock {\em J. Nucl. Mat.}, 380:22--29, 2008.

\bibitem{ogata-im2006}
S.~Ogata, F.~Shimizu, J.~Li, M.~Wakeda, and Y.~Shibutani.
\newblock Atomistic simulation of shear localization in cu-zr bulk metallic
  glass.
\newblock {\em Intermetallics}, 14:1033--1037, 2006.

\bibitem{ono-prl2002}
I.K. Ono, C.S. O'Hern, D.J. Durian, S.A. Langer, A.J. Liu, and S.R. Nagel.
\newblock Effective temperatures of a driven system near jamming.
\newblock {\em Phys. Rev. Lett.}, 89:095703, 2002.

\bibitem{ott-am2008}
R.T. Ott, M.~Heggen, M.~Feuerbacher, E.S. Park, D.H. Kim, M.J. Kramer, M.F.
  Besser, and D.J. Sordelet.
\newblock Anelastic strain and structural anisotropy in homogeneously deformed
  cu64.5zr35.5 metallic glass.
\newblock {\em Acta Mater.}, 56:5575--5583, 2008.

\bibitem{ovarlez-ar2009}
G.~Ovarlez, S.~Rodts, X.~Chateau, and P.~Coussot.
\newblock Phenomenology and physical origin of shear localization and shear
  banding in complex fluids.
\newblock {\em Acta Rheol.}, 48:831--844, 2009.

\bibitem{paczuski-pre1995}
M.~Paczuski, S.~Maslov, and P.~Bak.
\newblock Avalanche dynamics in evolution, growth, and depinning models.
\newblock {\em Phys. Rev. E}, 53:414--443, 1995.

\bibitem{park-sm2007}
K.W. Park, J.~Jang, M.~Wakeda, Y.~Shibutani, and J.C. Lee.
\newblock Atomic packing density and its influence on the properties of cu-zr
  amorphous alloys.
\newblock {\em Scripta Mat.}, 57:805--808, 2007.

\bibitem{pedersen-prl2010}
U.R. Pedersen, T.B. Schr{\o}der, J.C. Dyre, and P.~Harrowell.
\newblock Geometry of slow structural fluctuations in a supercooled binary
  alloy.
\newblock {\em Phys. Rev. Lett.}, 104:105701, 2010.

\bibitem{peng-prl2011}
H.L. Peng, M.Z. Li, and W.H. Wang.
\newblock Structural signature of plastic deformation in metallic glasses.
\newblock {\em Phys. Rev. Lett.}, 106:135503, 2011.

\bibitem{peng-apl2010}
H.L. Peng, M.Z. Li, W.H. Wang, C.Z. Wang, and K.M. Ho.
\newblock Effect of local structures and atomic packing on glass forming
  ability in cuxzr100-x metallic glasses.
\newblock {\em Appl. Phys. Lett.}, 96:021901, 2010.

\bibitem{kermouche-pm2011}
A.~Perriot, E.~Barthel, G.~Kermouche, G.~Quérel, and D.~Vandembroucq.
\newblock On the plastic deformation of soda-lime glass - a cr$^{3+}$
  luminescence study of densification.
\newblock {\em Philos. Mag.}, 91:1245--1255, 2011.

\bibitem{PMMCVB-JACS06}
A.~Perriot, V.~Martinez, C.~Martinet, B.~Champagnon, D.~Vandembroucq, and
  E.~Barthel.
\newblock Raman micro-spectroscopic map of plastic strain in indented amorphous
  silica.
\newblock {\em J. Am. Ceram. Soc.}, 89:596--601, 2006.

\bibitem{phillips-msmse1999}
R.~Phillips, D.~Rodney, V.~Shenoy, E.~Tadmor, and M.~Ortiz.
\newblock Hierarchical models of plasticity: dislocation nucleation and
  interaction.
\newblock {\em Model. Simul. Mater. Sci. Eng.}, 7:769, 1999.

\bibitem{Picard-EPJE04}
G.~Picard, A.~Ajdari, F.~Lequeux, and L.~Bocquet.
\newblock Elastic consequences of a single plastic event: A step towards the
  microscopic modeling of the flow of yield stress fluids.
\newblock {\em Eur. Phys. J. E}, 15:371--381, 2004.

\bibitem{Picard-PRE05}
G.~Picard, A.~Ajdari, F.~Lequeux, and L.~Bocquet.
\newblock Slow flows of yield stress fluids: complex spatio-temporal behaviour
  within a simple elasto-plastic model.
\newblock {\em Phys. Rev. E}, 71:010501(R), 2005.

\bibitem{Pilla-jpcm2004}
O.~Pilla, A.~Fontana, J.R. Gonçalvez, M.~Montagna, F.~Rossi, G.~Viliani,
  L.~Angelani, G.~Ruocco, G.~Monaco, and F.~Sette.
\newblock The low energy excess of vibrational sates in a-sio2: the role of
  transverse dynamics.
\newblock {\em J. Phys.: Cond. Mat.}, 16:8519, 2004.

\bibitem{raghavan-sm2010}
R.~Raghavan, K.~Boopathy, R.~Ghisleni, M.A. Pouchon, U.~Ramamurty, and
  J.~Michler.
\newblock Ion irradiation enhances the mechanical performance of metallic
  glasses.
\newblock {\em Scripta Mat.}, 62:462--465, 2010.

\bibitem{reger-leonhard-sm2000}
A.~Reger-Leonhard, M.~Heilmaier, and J.~Eckert.
\newblock Newtonian flow of zr55cu30al10ni5 bulk metallic glassy alloys.
\newblock {\em Scripta Mat.}, 43:459--464, 2000.

\bibitem{Richeton-MSEA06}
T.~Richeton, P.~Dobron, F.~Chmelik, J.~Weiss, and F.~Louchet.
\newblock On the critical character of plasticity in metallic single crystals.
\newblock {\em Mater. Sci. Eng. A}, 424:190--195, 2006.

\bibitem{Richeton-ActaMat05}
T.~Richeton and J.~Weiss.
\newblock Dislocation avalanches: Role of temperature, grain size and strain
  hardening.
\newblock {\em Acta Mater.}, 53:4463--4471, 2006.

\bibitem{rodney-prl2009}
D.~Rodney and C.~Schuh.
\newblock Distribution of thermally activated plastic events in a flowing
  glass.
\newblock {\em Phys. Rev. Lett.}, 102:235503, 2009.

\bibitem{rodney-prb2009}
D.~Rodney and C.A. Schuh.
\newblock Yield stress in metallic glasses: the jamming-unjamming transition
  studied through monte carlo simulations based on the activation-relaxation
  technique.
\newblock {\em Phys. Rev. B}, 80:184203, 2009.

\bibitem{Rottler-PRE01}
J.~Rottler and M.~O. Robbins.
\newblock Yield conditions for deformation of amorphous polymer glasses.
\newblock {\em Phys. Rev. E}, 64:051801, 2001.

\bibitem{rottler-pre2003}
J.~Rottler and M.O. Robbins.
\newblock Shear yielding of amorphous glassy solids: Effect of temperature and
  strain rate.
\newblock {\em Phys. Rev. E}, 68:011507, 2003.

\bibitem{rottler-prl2005}
J.~Rottler. and M.O. Robbins.
\newblock Unified description of aging and rate effects in yield of glassy
  solids.
\newblock {\em Phys. Rev. Lett.}, 95:225504, 2005.

\bibitem{rountree-prl2009}
C.L. Rountree, D.~Vandembroucq, M.~Talamali, E.~Bouchaud, and S.~Roux.
\newblock Plasticity-induced structural anisotropy of silica glass.
\newblock {\em Phys. Rev. Lett.}, 102:195501, 2009.

\bibitem{Ruffle-prl2010}
B.~Ruffl\'e, S.~Ayrinhac, E.~Courtens, R.~Vacher, and M.~Foret.
\newblock Scaling the t-dependent boson peak of vitreous silica with the
  high-frequency bulk modulus derived from brillouin scattering data.
\newblock {\em Phys. Rev. Lett.}, 104:067402, 2010.

\bibitem{Ruffle-prl2003}
B.~Ruffl\'e, M.~Foret, E.~Courtens, R.~Vacher, and G.~Monaco.
\newblock Observation of the onset of strong scattering on high frequency
  acoustic phonons in densified silica glass.
\newblock {\em Phys. Rev. Lett.}, 90:095502, 2003.

\bibitem{Ruffle-prl2006}
B.~Ruffl\'e, G.~Guimbretiere, E.~Courtens, and R.~Vacher.
\newblock Glass specific behavior in the damping of acoustic-like vibrations.
\newblock {\em Phys. Rev. Lett.}, 96:045502, 2006.

\bibitem{ruffle-prl2008}
B.~Ruffl\'e, D.A. Parshin, E.~Courtens, and R.~Vacher.
\newblock Boson peak and its relation to acoustic attenuation in glasses.
\newblock {\em Phys. Rev. Lett.}, 100:015501, 2008.

\bibitem{sciortino-prl2010}
J.~Russo and F.~Sciortino.
\newblock How do self-assembling polymers and gels age compared to glasses?
\newblock {\em Phys. Rev. Lett.}, 104:195701, 2010.

\bibitem{sastry-nature1998}
S.~Sastry, P.G. Debenedetti, and F.H. Stillinger.
\newblock Signatures of distinct dynamical regimes in the energy landscape of a
  glass-forming liquid.
\newblock {\em Nature}, 393:554--557, 1998.

\bibitem{schroder-jcp2000}
T.B. Schroder, S.~Sastry, J.C. Dyre, and S.C. Glotzer.
\newblock Crossover to potential energy landscape dominated dynamics in a model
  glass-forming liquid.
\newblock {\em J. Chem. Phys.}, 112:9834--9840, 2000.

\bibitem{johnson-prl2004}
J.~Schroers and W.L. Johnson.
\newblock Ductile bulk metallic glass.
\newblock {\em Phys. Rev. Lett.}, 93:255506, 2004.

\bibitem{schuh-am2007}
C.A. Schuh, T.C. Hufnagel, and U.~Ramamurty.
\newblock Mechanical behavior of amorphous alloys.
\newblock {\em Acta Mater.}, 55:4067--4109, 2007.

\bibitem{schuh-naturemat2003}
C.A. Schuh and A.C. Lund.
\newblock Atomistic basis for the plastic yield criterion of metallic glasses.
\newblock {\em Nature Mat.}, 2:449--452, 2003.

\bibitem{sciortino-prl1999}
F.~Sciortino, W.~Kob, and P.~Tartaglia.
\newblock Inherent structure entropy of supercooled liquids.
\newblock {\em Phys. Rev. Lett.}, 83:3214--3218, 1999.

\bibitem{Monaco-science1998}
F.~Sette, M.~Krisch, C.~Masciovecchio, G.~Ruocco, and G.~Monaco.
\newblock Dynamics of glasses and glass-forming liquids studied by inelastic
  x-ray scattering.
\newblock {\em Science}, 280:1550--1555, 1998.

\bibitem{sheng-nature2006}
H.W. Sheng, W.K. Luo, F.M. Alamgir, J.M. Bai, and E.~Ma.
\newblock Atomic packing and short-to-medium-range order in metallic glasses.
\newblock {\em Nature}, 439:419--425, 2006.

\bibitem{shenoy-jmps1999}
V.~Shenoy, R.~Miller, E.~Tadmor, D.~Rodney, R.~Phillips, and M.~Ortiz.
\newblock An adaptive finite element approach to atomic-scale mechanics : the
  quasicontinuum method.
\newblock {\em J. Mech. Phys. Solids}, 47:611--642, 1999.

\bibitem{shi-prl2005}
Y.~Shi and M.L. Falk.
\newblock Strain localization and percolation of stable structure in amorphous
  solids.
\newblock {\em Phys. Rev. Lett.}, 95:095502, 2005.

\bibitem{shi-apl2005}
Y.~Shi and M.L. Falk.
\newblock Structural transformation and localization during simulated
  nanoindentation of a noncrystalline metal film.
\newblock {\em Appl. Phys. Lett.}, 86:011914, 2005.

\bibitem{shi-prb2006}
Y.~Shi and M.L. Falk.
\newblock Atomic-scale simulations of strain localization in three-dimensional
  model amorphous solids.
\newblock {\em Phys. Rev. B}, 73:214201, 2006.

\bibitem{shi-am2007}
Y.~Shi and M.L. Falk.
\newblock Stress-induced structural transformation and shear banding during
  simulated nanoindentation of a metallic glass.
\newblock {\em Acta Mater.}, 55:4317--4324, 2007.

\bibitem{shi-prl2007}
Y.~Shi, M.B. Katz, H.~Li, and M.L. Falk.
\newblock Evaluation of the disorder temperature and free-volume formalisms via
  simulations of shear banding in amorphous solids.
\newblock {\em Phys. Rev. Lett.}, 98:185505, 2007.

\bibitem{tanaka-natphys2006}
H.~Shintani and H.~Tanaka.
\newblock Frustration on the way to crystallization in glass.
\newblock {\em Nature Phys.}, 2:200--206, 2006.

\bibitem{Lambropoulos-SPIE98}
A.~Shorey, K.~Xin, K.~H. Chen, and J.~C. Lambropoulos.
\newblock Deformation of fused silica : nanoindentation and densification.
\newblock {\em Proc. SPIE}, 3424:72--81, 1998.

\bibitem{simmons-jncs1998}
J.H. Simmons, R.~Ochoa, K.D. Simmons, and J.J. Mills.
\newblock Non-newtonian viscous flow in soda-lime-silica glass at forming and
  annealing temperatures.
\newblock {\em J. Non-Cryst. Sol.}, 105:313, 1988.

\bibitem{Shintani-NatureMaterials2008}
H.~Sintani and H.~Tanaka.
\newblock Universal link between the boson peak and transverse phonons in
  glass.
\newblock {\em Nature Mat.}, 7:870, 2008.

\bibitem{Sneppen-prl92}
K.~Sneppen.
\newblock Self-organized pinning and interface growth in a random medium.
\newblock {\em Phys. Rev. Lett.}, 69:3539--3542, 1992.

\bibitem{sollich-pre1998}
P.~Sollich.
\newblock Rheological constitutive equation for a model of soft glassy
  materials.
\newblock {\em Phys. Rev. E}, 58:738--759, 1998.

\bibitem{sollich-prl1997}
P.~Sollich, F.~Lequeux, P.~H\'ebraud, and M.E. Cates.
\newblock Rheology of soft glassy materials.
\newblock {\em Phys. Rev. Lett.}, 78:2020--2023, 1997.

\bibitem{Spaepen-ActaMet77}
F.~Spaepen.
\newblock Microscopic mechanism for steady-state inhomogeneous flow in metallic
  glasses.
\newblock {\em Acta Metall.}, 25:407--415, 1977.

\bibitem{srolovitz-prb1981}
D.~Srolovitz, T.~Egami, and V.~Vitek.
\newblock Radial distribution function and structural relaxation in amorphous
  solids.
\newblock {\em Phys. Rev. B}, 24:6936--6944, 1981.

\bibitem{srolovitz-am1983}
D.~Srolovitz, V.~Vitek, and T.~Egami.
\newblock An atomistic study of deformation of amorphous metals.
\newblock {\em Acta Metall.}, 31:335--352, 1983.

\bibitem{Starviou-prb2010}
E.~Starviou, C.~Raptis, and K.~Syassen.
\newblock Effects of pressure on the boson peak of tellurite (teo2)(1-x)(zno)x
  glasses: evidence of an elastic glass to glass transition.
\newblock {\em Phys. Rev. B}, 81:174202, 2010.

\bibitem{Steif-ActaMet82}
P.~S. Steif, F.~Spaepen, and J.~W. Hutchinson.
\newblock Strain localization in amorphous metals.
\newblock {\em Acta Metall.}, 30:447--455, 1982.

\bibitem{stevenson-natphys2006}
J.D. Stevenson, J.~Schmalian, and P.G. Wolynes.
\newblock The shapes of cooperatively rearranging regions in glass-forming
  liquids.
\newblock {\em Nature Phys.}, 2:268--274, 2006.

\bibitem{stillinger-science1995}
F.H. Stillinger.
\newblock A topographic view of supercooled liquids and glass formation.
\newblock {\em Science}, 267:1935--1939, 1995.

\bibitem{stillinger-science1984}
F.H. Stillinger and T.A. Weber.
\newblock Packing structures and transitions in liquids and solids.
\newblock {\em Science}, 225:983--989, 1984.

\bibitem{Stillinger1985}
F.H. Stillinger and T.A. Weber.
\newblock Computer simulation of local order in condensed phases of silicon.
\newblock {\em Phys. Rev. B}, 31:5262, 1985.

\bibitem{su-am2006}
C.~Su and L.~Anand.
\newblock Plane strain indentation of a zr-based metallic glass: experiments
  and numerical simulation.
\newblock {\em Acta Mater.}, 54:179--189, 2006.

\bibitem{Wang-PRL10}
B.A. Sun, H.B. Yu, W.~Jiao, D.Q. Zhao, and W.H. Wang.
\newblock Plasticity of ductile metallic glasses: A self organized critical
  state.
\newblock {\em Phys. Rev. Lett.}, 105:035501, 2010.

\bibitem{suzuky-prb1987}
Y.~Suzuki, J.~Haimovich, and T.~Egami.
\newblock Bond-orientational anisotropy in metallic glasses observed by x-ray
  diffraction.
\newblock {\em Phys. Rev. B}, 35:2162--2168, 1987.

\bibitem{TPRV-PRE08}
M.~Talamali, V.~Pet\"aj\"a, S.~Roux, and D.~Vandembroucq.
\newblock Path independent integrals to identify two-dimensional localized
  plastic events.
\newblock {\em Phys. Rev. E}, 78:016109, 2008.

\bibitem{TPVR-Meso10}
M.~Talamali, V.~Pet\"aj\"a, D.~Vandembroucq, and S.~Roux.
\newblock Strain localization and anisotropic correlations in a mesoscopic
  model of amorphous plasticity.
\newblock arxiv:1005.2463, 2010.

\bibitem{TPVR-Aval11}
M.~Talamali, V.~Pet\"aj\"a, D.~Vandembroucq, and S.~Roux.
\newblock Avalanches, precursors and finite size fluctuations in a mesoscopic
  model of amorphous plasticity.
\newblock arxiv:1103.5017, 2011.

\bibitem{talati-epl2009}
M.~Talati, T.~Albaret, and A.~Tanguy.
\newblock Atomistic simulations of elastic and plastic properties in amorphous
  silicon.
\newblock {\em Europhys. Lett.}, 86:66005, 2009.

\bibitem{tanaka-nmat2010}
H.~Tanaka, T.~Kawasaki, H.~Shintani, , and K.~Watanabe.
\newblock Critical-like behaviour of glass-forming liquids.
\newblock {\em Nature Mat.}, 9:324--331, 2010.

\bibitem{tang-ra2004}
H.S. Tang and D.M. Kalyon.
\newblock Estimation of the parameters of herschel-bulkley fluid under wall
  slip using a combination of capillary and squeeze flow viscometers.
\newblock {\em Rheologica Acta}, 43:80--88, 2004.

\bibitem{Tanguy-PRE98}
A.~Tanguy, M.~Gounelle, and S.~Roux.
\newblock From individual to collective pinning: effect of long-range elastic
  interactions.
\newblock {\em Phys. Rev. E}, 58:1577--1590, 1998.

\bibitem{tanguy-epje2006}
A.~Tanguy, F.~Leonforte, and J.L. Barrat.
\newblock Plastic response of a 2d lennard-jones amorphous solid: detailed
  analyses of the local rearrangements at very slow strain rate.
\newblock {\em Eur. Phys. J. E}, 20:355--364, 2006.

\bibitem{tanguy-epl2010}
A.~Tanguy, B.~Mantisi, and M.~Tsamados.
\newblock Vibrational modes as a predictor for plasticity in a model glass.
\newblock {\em Europhys. Lett.}, 90:16004, 2010.

\bibitem{tarjus-jpcm2005}
G.~Tarjus, S.A. Kivelson, Z.~Nussinov, and P.~Viot.
\newblock The frustration-based approach of supercooled liquids and the glass
  transition: a review and critical assessment.
\newblock {\em J. Phys.: Cond. Mat.}, 17:R1143--R1182, 2005.

\bibitem{Tersoff1988}
J.~Tersoff.
\newblock Empirical interatomic potential for silicon with improved elastic
  propertie.
\newblock {\em Phys. Rev. B}, 38:9902--9905, 1988.

\bibitem{tomida-prb1993}
T.~Tomida and T.~Egami.
\newblock Molecular-dynamics study of structural anisotropy and anelasticity in
  metallic glasses.
\newblock {\em Phys. Rev. B}, 48:3048--3057, 1993.

\bibitem{tsamados-epje2010}
M.~Tsamados.
\newblock Plasticity and dynamical heterogeneity in driven glassy materials.
\newblock {\em Eur. Phys. J. E}, 32:165--181, 2010.

\bibitem{tsamados-pre2009}
M.~Tsamados, A.~Tanguy, C.~Goldenberg, and J.L. Barrat.
\newblock Local elasticity map and plasticity in a model lennard-jones glass.
\newblock {\em Phys. Rev. E}, 80:026112, 2009.

\bibitem{tsamados-epje2008}
M.~Tsamados, A.~Tanguy, F.~L\'eonforte, and J.L. Barrat.
\newblock On the study of local-stress rearrangements during quasi-static
  plastic shear of a model glass: do local-stress components contain enough
  information?
\newblock {\em Eur. Phys. J. E}, 26:283--293, 2008.

\bibitem{Tsuneyuki1988}
S.~Tsuneyuki, M.~Tsukada, H.~Aoki, and Y.~Matsui.
\newblock First-principles interatomic potential of silica applied to molecular
  dynamics.
\newblock {\em Phys. Rev. Lett.}, 61:869--872, 1988.

\bibitem{utz-prl2000}
M.~Utz, P.~Debenedetti, and F.H. Stillinger.
\newblock Atomistic simulation of aging and rejuvenation in glasses.
\newblock {\em Phys. Rev. Lett.}, 84:1471--1475, 2000.

\bibitem{valiquette-prb2003}
F.~Valiquette and N.~Mousseau.
\newblock Energy landscape of relaxed amorphous silicon.
\newblock {\em Phys. Rev. B}, 68:125209, 2003.

\bibitem{Beest1990}
B.W.H. van Beest, G.J. Kramer, and R.A. van Santen.
\newblock Force fields for silicas and aluminophosphates based on ab initio
  calculations.
\newblock {\em Phys. Rev. Lett.}, 64:1955, 1990.

\bibitem{vankampen-book}
N.~Van~Kampen.
\newblock {\em Stochastic processes in physics and chemistry}.
\newblock North-Holland, Amsterdam, 1992.

\bibitem{VDCPBCM-JPCM08}
D.~Vandembroucq, T.~Deschamps, C.~Coussa, A.~Perriot, E.~Barthel,
  B.~Champagnon, and C.~Martinet.
\newblock Density hardening plasticity and mechanical aging of silica glass
  under pressure: A raman spectroscopic study.
\newblock {\em J. Phys.: Cond. Mat.}, 20:485221, 2008.

\bibitem{VR-ShearBanding11}
D.~Vandembroucq and S.~Roux.
\newblock Mechanical noise dependent aging and shear banding behavior of a
  mesoscopic model of amorphous plasticity.
\newblock arxiv:1104.4863, 2011.

\bibitem{VR-preprint11}
D.~Vandembroucq and S.~Roux.
\newblock Shear banding behavior in glasses: a mesoscopic approach.
\newblock {\em unpublished}, 2011.

\bibitem{VSR-PRE04}
D.~Vandembroucq, R.~Skoe, and S.~Roux.
\newblock Universal fluctuations of depinning forces, application to finite
  temperature behavior.
\newblock {\em Phys. Rev. E}, 70:051101, 2004.

\bibitem{varnik-prl2003}
F.~Varnik, L.~Bocquet, J.L. Barrat, and L.~Berthier.
\newblock Shear localization in a model glass.
\newblock {\em Phys. Rev. Lett.}, 90:095702, 2003.

\bibitem{varnik-pre2008}
F.~Varnik and D.~Raabe.
\newblock Profile blunting and flow blockage in a yield-stress fluid: A
  molecular dynamics study.
\newblock {\em Phys. Rev. E}, 77:011504, 2008.

\bibitem{viasnoff-prl2002}
V.~Viasnoff and F.~Lequeux.
\newblock Rejuvenation and overaging in a colloidal glass under shear.
\newblock {\em Phys. Rev. Lett.}, 89:065701, 2002.

\bibitem{wahnstrom-pra1991}
G.~Wahnstr\"om.
\newblock Molecular-dynamics study of a supercooled two-component lennard-jones
  system.
\newblock {\em Phys. Rev. A}, 44:3752--3764, 1991.

\bibitem{wales-book}
D.J. Wales.
\newblock {\em Energy landscapes}.
\newblock Cambridge University Press, 2003.

\bibitem{wales-nature1998}
D.J. Wales, M.A. Miller, and T.R. Walsh.
\newblock Archetypal energy landscapes.
\newblock {\em Nature}, 394:758--760, 1998.

\bibitem{widmer-prl2006}
A.~Widmer-Cooper and P.~Harrowell.
\newblock Predicting the long-time dynamic heterogeneity in a supercooled
  liquid on the basis of short-time heterogeneities.
\newblock {\em Phys. Rev. Lett.}, 96:185701, 2006.

\bibitem{widmer-naturephys2008}
A.~Widmer-Cooper, H.~Perry, P.~Harrowell, and D.R. Reichman.
\newblock Irreversible reorganization in a supercooled liquid originates from
  localized soft modes.
\newblock {\em Nature Phys.}, 4:711--715, 2008.

\bibitem{xu-jcp2008}
L.~Xu and G.~Henkelman.
\newblock Adaptive kinetic monte carlo for first-principles accelerated
  dynamics.
\newblock {\em J. Chem. Phys.}, 129:114104, 2008.

\bibitem{Rouxel-JMR05}
S.~Yoshida, T.~Rouxel, and J.C. Sangleboeuf.
\newblock Quantitative evaluation of indentation-induced densification in
  glass.
\newblock {\em J. Mater. Res.}, 55:1159--1162, 2006.

\bibitem{Zaiser-AdvPhys06}
M.~Zaiser.
\newblock Scale invariance in plastic flow of crystalline solids.
\newblock {\em Adv. Phys.}, 55:185--245, 2006.

\bibitem{Zaiser-JSM05}
M.~Zaiser and P.~Moretti.
\newblock Fluctuation phenomena in crystal plasticity-a continuum model.
\newblock {\em J. Stat. Mech.}, P08004:79--97, 2005.

\bibitem{Zaiser-JSM07}
M.~Zaiser and P.~Nikitas.
\newblock Slip avalanches in crystal plasticity: scaling of the avalanche
  cut-off.
\newblock {\em J. Stat. Mech.}, P04013:1--11, 2007.

\bibitem{zhang-natmat2006}
Y.~Zhang, W.H. Wang, and A.L. Greer.
\newblock Making metallic glasses plastic by control of residual stress.
\newblock {\em Nature Mat.}, 5:857--860, 2006.

\bibitem{zhang-am2003}
Z.F. Zhang, J.~Eckert, and L.~Schultz.
\newblock Difference in compressive and tensile fracture mechanisms of
  zr59cu20al10ni8ti3 bulk metallic glass.
\newblock {\em Acta Mater.}, 51:1167--1179, 2003.

\bibitem{zink-prb2006}
M.~Zink, K.~Samwer, W.L. Johnson, and S.G. Mayr.
\newblock Plastic deformation of metallic glasses: size of shear transformation
  zones from molecular dynamics simulations.
\newblock {\em Phys. Rev. B}, 73:172203, 2006.

\end{thebibliography}

\end{document}